\documentclass[aps,nofootinbib,superscriptaddress, showpacs,preprintnumbers,  nofootinbibt,twocolumn]{revtex4-2}
\usepackage{epsfig}
\usepackage{multirow}
\usepackage{eurosym}
\usepackage{xcolor}
\usepackage{dcolumn}
\usepackage{bm}
\usepackage{enumerate}
\usepackage{float}
\usepackage{epstopdf}
\usepackage{amsmath}
\usepackage{bm}
\usepackage{amsfonts}
\usepackage{amssymb}
\usepackage{graphicx}
\usepackage{alphalph,mathtools}
\usepackage{etoolbox}
\usepackage{color}
\usepackage{booktabs}
\usepackage{footnote}
\usepackage{makecell,tabularx}
\usepackage{hyperref}
\hypersetup{colorlinks,citecolor=blue}
\hypersetup{colorlinks=true,linkcolor=magenta,filecolor=magenta,
    urlcolor=blue}

\def\be{\begin{equation}}
    \def\ee{\end{equation}}
\def\bea{\begin{eqnarray}}
    \def\eea{\end{eqnarray}}

\begin{document}

    \title{Data Analysis of three parameter models of deceleration parameter in FLRW Universe}
    \author{Amine Bouali}
    \email{a1.bouali@ump.ac.ma} \affiliation{Laboratory of Physics of
        Matter and Radiation, Mohammed I University, BP 717, Oujda,
        Morocco.}
    \author{Himanshu Chaudhary}
    \email{himanshuch1729@gmail.com} \affiliation{Department of
        Applied Mathematics, Delhi Technological University, Delhi-110042,
        India,} \affiliation{Pacif Institute of Cosmology and Selfology (PICS) Sagara, Sambalpur 768224, Odisha, India} 
        \affiliation{Department of Mathematics, Shyamlal College,
        University of Delhi, Delhi-110032, India.}
    \author{Ujjal Debnath}
    \email{ujjaldebnath@gmail.com} \affiliation{Department of
        Mathematics, Indian Institute of Engineering Science and
        Technology, Shibpur, Howrah-711 103, India.}
    \author{Alok Sardar}
    \email{alokmath94@gmail.com} \affiliation{Department of
        Mathematics, Indian Institute of Engineering Science and
        Technology, Shibpur, Howrah-711 103, India.}
    \author{G.Mustafa}
    \email{gmustafa3828@gmail.com} \affiliation{Department of Physics, Zhejiang Normal University, Jinhua 321004, People’s Republic of China,}
    \affiliation{New Uzbekistan University, Mustaqillik ave. 54, 100007 Tashkent, Uzbekistan,}

    \begin{abstract}
        Constraining the dark energy deceleration parameter is one of the fascinating topics in the recent cosmological paradigm. This work aims to reconstruct the dark energy using parametrization of the deceleration
        parameter in a flat FLRW universe filled with radiation, dark
        energy, and pressure-less dark matter. Thus, we have
        considered four well-motivated parameterizations of $q(z)$, which
        can provide the evolution scenario from the deceleration to
        acceleration phase of the Universe. We have evaluated the
        expression of the corresponding Hubble parameter of each
        parametrization by imposing it into the Friedmann equation. We
        have constrained the model parameter through $H(z)$,
        Pantheon, baryons acoustic oscillation (BOA), and Cosmic Microwave Background (CMB) dataset. Next, we
        have estimated the best-fit values of the model parameters by
        using Monte Carlo Markov Chain (MCMC) technique and implementing
        $H(z)$+ BAO+SNIa+CMB dataset. Then we analyzed the cosmographic
        parameter, such as deceleration, jerk, and snap parameters, graphically by employing the best-fit values of the model
        parameter. Moreover, we have analyzed statefinder and $Om$ diagnostics
        parameters for each scenario to discriminate various dark energy
        models. Using the information criteria, the viability of the models have examined. In the end, we have analogized our outcomes with the standard $\Lambda$CDM
        model to examine the viability of our models.
    \end{abstract}


    \maketitle
    \tableofcontents


    \section{Introduction}\label{Introduction}

    The confirmation of the cosmic accelerated expansion of the
    Universe through different
    surveys~\cite{1:1998fmf,2:1996grv,3:1998vns}, opened an emerging field
    of study in modern cosmology. To disclose the physical
    mechanism behind the such accelerated expansion is a challenging task
    for modern researchers. In light of this, several cosmological
    models have been implemented to alleviate this phenomenon based
    mainly on two different approaches. The first approach is the
    modification of the gravitational part of the Einstein field
    equation-so called modified gravity~\cite{4:2003aw,5:2003ft}, and
    the other is the existence of an exotic fluid having positive
    energy density and negative pressure, named dark energy (DE). This work will mainly concentrate on the second approach and
    suppose that the DE is responsible for the universe's accelerated expansion~\cite{6:2003cyd,7:2004lze,8:2005xhg}. It is also
    interesting to note that DE violates strong energy
    conditions. This violation of energy generates anti-gravitational
    effects for which the transition from deceleration to acceleration takes place~\cite{9:2005xb}. So far, numerous DE
    models have been suggested to describe the present-day cosmic
    accelerated expansion~\cite{10:1999gb,11:2002ji,12:2002gy,13,14}.
    Among them, the $\Lambda$CDM ($\Lambda$-cold dark matter) model is
    considered as simplest and widely accepted DE model. Despite its
    success, it sufferers from some other problems, namely
    \textit{coincidence} problem, \textit{fine-tuning} problem, and
    \textit{age} problem~\cite{13,15,16:1988cp}. To solve this issue,
    it is quite natural for physicists to figure out another
    alternative DE model that can describe the present status of the
    Universe. The scalar field models, like quintessence, phantom,
    k-essence, and so on, contribute to the understanding of the origins and nature of DE. Despite these designs of dark energy, cosmic accelerated expansion is still a problem in physics today. \\

    Several theoretical approaches have been developed to describe the cosmic accelerated expansion phenomenon, but none of these is known as the appropriate one. The most recent direction to explore the accelerating Universe at the phenomenological level is the DE parametrization of the equation of state parameter. The main idea in this approach is to consider a specific evolution scenario instead of considering any DE model a priori and then determine the nature of the exotic component that is triggering cosmic acceleration. It is known as the model-independent approach, which depends on estimating model parameters from existing observational datasets. But such an approach has some drawbacks: (i) most of the parametrization suffers from divergence problem, (ii) the parametrization technique would have missed subtle results about the true nature of dark energy due to the assumed parametric form. Various studies have been executed to describe the cosmic acceleration of the Universe through viable parametrization of EOS (see Refs.\cite{17:2005pa,18-Rivera:2019aol,19:2020rho,chaudhary2023constraints}). Inspired by the parametrization of EOS, the parametrization of the deceleration parameter has been implemented and studied extensively in the literature. Since the Universe evolves from earlier deceleration to late time acceleration. For this reason, any cosmological model should have a transition from deceleration phase to acceleration phase of expansion to explain the whole evolution of the Universe. Thus, the deceleration parameter plays a crucial role which is defined by $q=-\frac{a\ddot{a}}{\dot{a}^2}$, where $a(t)$ is the usual scale factor. The sign of $q$ decides whether the Universe is accelerating i.s. $(q<0$) or decelerating i.s. $q>0$. Recently, several theoretical models have been developed to analyze the entire evolutionary history of the Universe through parametrization of $q(z)$ as a function of scale factor $(a(t))$ or time $(t)$ or redshift $(z)$~\cite{20:2012ya,21,22:2008mt,23:2004lze,24:2007gvk,25:2009zza,26,27:2013lya,28,29:2006gs,30:2001mx,31:2015ali,39,bouali,bouali2023model,bouali2023data,ref1,ref2,ref3,ref4,ref5,ref6,ref7}. The advantage of the parametrization of deceleration parameter is that it can provide finite results without considering any particular gravitational theory. However, such approaches also have disadvantages similar to the parametrization of EOS. Since we do not yet have any concrete and satisfactory theoretical model of the Universe which can describe the whole evolutionary history of the Universe. So, the idea of adopting a parametric approach may be a preliminary step towards the expansion history of the Universe.\\
    Recently, Mamon et al.~\cite{38} studied a special form of deceleration parameter and obtained the best-fit values using $\chi^2$ minimization technique with available observational data. They also analyzed the evolution of the jerk parameter for the considered parametrized model. Gadbali, Mandal, and Sahoo~\cite{39} have explored a specific parametrization of deceleration parameter in the context of $f(Q)$ gravity theory and constrained the model parameter by using Bayesian analysis with observational data. Motivated by the above work, we explored four suitable parametrizations of deceleration parameter to determine the best viable model compared to $\Lambda$CDM (see Ref.~\cite{41:2023rdx}). The present work is indeed an extension of our previous work. Here, we have assumed three well-motivated parametrizations (model 1-model3) of deceleration parameter containing three unknown parameters. We also introduced a new parametrization of the deceleration parameter, containing three unknown parameters. This work mainly focuses on constraining the model parameters using various observational datasets. In particular, we have chosen to use $H(z)$ dataset consisting of 57 measurements, Pantheon dataset consisting of 1048 measurements, 17 measurements of BAO and CMB Distant Prior  to get the best-fit values of the model parameters. We adopt Monte-Carlo Markov Chain (MCMC) analysis to fix the model parameters on the recently released data. We perform observational data analysis by $\chi^2$-minimization technique on the view of $H(z)$+SNIa+BAO+CMB dataset. This analysis provides us the bounds of arbitrary parameters $q_0$, $q_1$, and $q_2$ within $1\sigma$, $2\sigma$ confidence levels. In the end, we performed a graphical analysis of cosmographic parameters like deceleration, jerk, and snap parameters for all models. In addition to this, we have analyzed $r-s$, $q-r$ planes, and $Om$ diagnostics parameters for all the considered models.\\
    The paper is structured as follows: section \ref{sec2} is assigned for the basic equations for the FLRW universe. In section \ref{sec3}, we have adopted four parametrizations (model 1-model 3) and a new parametrization of the deceleration parameter. Then we obtained the Hubble solution for each parametrization. In section \ref{sec4}, the model parameters have been constrained using various datasets and place the best-fit values of the model parameters by implementing the MCMC method with observation data. In section \ref{sec5}, we perform a cosmographic survey. The $r-s$, $q-r$ planes and $Om$ diagonostic parameter is also addressed in section \ref{sec6} $\&$ \ref{sec7}. In section \ref{sec8}, we present the information criteria for our models. The results of each model have been comprehensively discussed in Section \ref{results}. In section \ref{sec9}, we have reported our conclusions based on the findings of our work.

    \section{Basic Equations of FLRW Model}\label{sec2}

    The line element of the spatially flat Friedmann-Lemaitre-Robertson-Walker (FLRW) universe is assumed as

    \begin{eqnarray}
        ds^2=-dt^2 +a^2(t)\left[dr^2+r^2\left(d\theta^2 + sin^2 \theta
        d\phi^2 \right)\right]
    \end{eqnarray}

    with $a(t)$ as the scale factor, as usual. The Friedmann equations
    are taken as
    \begin{equation}\label{F1}
        H^2=\frac{8\pi G}{3}~\rho
    \end{equation}
    and
    \begin{equation}\label{F2}
        \dot{H}=-4\pi G(\rho+p)
    \end{equation}
    here, $H=\dot{a}/a$ is the Hubble parameter. We suppose that the universe is composed of radiation, dark matter (DM), and dark energy (DE).
    Then total energy density $\rho$ and total pressure $p$ becomes $\rho=\rho_{r}+\rho_{m}+\rho_{d}$ and
    $p=p_{r}+p_{m}+p_{d}$. Now we consider that the radiation, DM, and DE are separately conserved. Thus, one can write
    \begin{equation}\label{rad}
        \dot{\rho}_{r}+3H(\rho_{r}+p_{r})=0,
    \end{equation}
    \begin{equation}\label{DM}
        \dot{\rho}_{m}+3H(\rho_{m}+p_{m})=0
    \end{equation}
    and
    \begin{equation}\label{DE}
        \dot{\rho}_{d}+3H(\rho_{d}+p_{d})=0
    \end{equation}
    For radiation, $p_{r}=\frac{1}{3}\rho_{r}$, then from equation
    (\ref{rad}) we have $\rho_{r}=\rho_{r0}a^{-4}$. If we assume pressure-less DM (i.e., $p_{m}=0$), from equation
    (\ref{DM}) we get $\rho_{m}=\rho_{m0}a^{-3}$.\\

    Now, the deceleration parameter can be written as
    \begin{eqnarray}
        q=-1-\frac{\dot{H}}{H^2}
    \end{eqnarray}
    So the corresponding deceleration parameter for DE has the
    expression
    \cite{41:2023rdx,42:2022jbw}
    \begin{eqnarray}
        q_{d}=-1-\frac{\dot{H_{d}}}{H_{d}^2}
    \end{eqnarray}
    where $H_{d}$ is the Hubble rate corresponding to dark energy. So from equations (\ref{F1}) and (\ref{F2}), we can write
    \begin{equation}\label{F11}
        H_{d}^2=\frac{8\pi G}{3}~\rho_{d}
    \end{equation}
    and
    \begin{equation}\label{F22}
        \dot{H}_{d}=-4\pi G(\rho_{d}+p_{d})
    \end{equation}

    Using equations \eqref{F11}, \eqref{F22} and \eqref{DE}, the fluid energy density yields
    \begin{equation} \label{rho}
        \rho_{d}=\rho_{d0}~ e^{\int \frac{2(1+q_{d})}{1+z}dz}
    \end{equation}
    where $\rho_{d0}$ represents the present value of the density
    parameter, and $z$ is the redshift parameter described as
    $1+z=\frac{1}{a}$ (presently, $a_{0}=1$).\\

    Defining the dimensionless density parameters as $\Omega_{r0}=\frac{8\pi G\rho_{r0}}{3H_{0}^{2}}$,
    $\Omega_{m0}=\frac{8\pi G\rho_{m0}}{3H_{0}^{2}}$ and
    $\Omega_{d0}=\frac{8\pi G\rho_{d0}}{3H_{0}^{2}}$, then from equation
    (\ref{F1}), we have the Hubble parameter as:
    \begin{eqnarray}\label{H}
        H^{2}(z) =H_{0}^{2}\left[\Omega_{r0}(1+z)^{4}+\Omega_{m0}(1+z)^{3} \right. \nonumber\\
       \left. + (1-\Omega_{r0}-\Omega_{m0})~ e^{\int \frac{2(1+q_{d})}{1+z}dz}\right]
    \end{eqnarray}


    \section{Parameterized deceleration parameter}\label{sec3}

    In this section, we consider some parameterized deceleration parameter
    analogs of some well-established parametric models of the equation
    of state parameter and calculated the corresponding Hubble
    parameter in terms of redshift $z$.

    \subsection{Model 1}

    The Alam-Sahni-Saini-Starobinsky (ASSS) model for parametrized
    equation of state parameter has been studied in
    \cite{54:2003fg,55}. The equivalent ASSS type parametrization of
    deceleration parameter has been introduced in \cite{43,44} and is
    given by
    \begin{equation}\label{Ma}
        q_{d}(z)=-1+\frac{q_1(1+z)+2q_2
            (1+z)^2}{3\left[q_0+q_1(1+z)+q_2(1+z)^2\right]}
    \end{equation}
    with $q_{0}$, $q_{1}$ and $q_{2}$ are constants. Then the energy
    density reads

    \begin{equation}\label{rho Ma}
        \rho_d=\rho_{d0}~\left\{q_0+q_1(1+z)+q_2 (1+z)^2\right\}^{2/3}
    \end{equation}

    From equation (\ref{H}), we obtain

    \begin{equation}
        \begin{aligned}
            H^{2}(z)& =H_{0}^{2}[\Omega_{r0}(1+z)^{4}+\Omega_{m0}(1+z)^{3}\\
            & +(1-\Omega_{r0}-\Omega_{m0})\\
            &({q_0+q_1(1+z)+q_2(1+z)^2})^{2/3})]
        \end{aligned}
    \end{equation}

    \subsection{Model 2}

    The Pade-II model for parametrized equation of state parameter has
    been explored in \cite{57,58,59}. The equivalent Pade-II type
    parametrization of deceleration parameter has been introduced in
    \cite{43,44} which is taken as
    \begin{equation}\label{PII}
        q_{d}(z)=\frac{q_0+q_1 log(1+z)}{1+q_2 log(1+z)}
    \end{equation}
    where $q_{0}$, $q_{1}$ and $q_{2}$ are constants. Then the energy
    density yields

    \begin{equation}\label{rho PII}
        \rho_d=\rho_{d0}~(1+z)^{(q_1+q_2)}~\left\{1+q_2~log(1+z)\right\}^{\frac{2(q_0
                q_2-q_1)}{{q_2}^2}}
    \end{equation}

    From equation (\ref{H}), we obtain
    \begin{equation}
        \begin{aligned}
            H^{2}(z)& =H_{0}^{2}[\Omega_{r0}(1+z)^{4}+\Omega_{m0}(1+z)^{3}\\
            & +(1-\Omega_{r0}-\Omega_{m0})\\
            &(1+z)^{(q_1+q_2)} ({1+q_{2}log(1+z)})^{\frac{2(q_0
                    q_2-q_1)}{(q_2)^{2}}}]
        \end{aligned}
    \end{equation}

    \subsection{Model 3}

    The Pade-I model for parametrized equation of state parameter has
    been investigated in \cite{57,58,59}. The equivalent Pade-I type
    parametrization of deceleration parameter has been introduced in
    \cite{43,44}, which takes the form:

    \begin{equation}\label{PI}
        q_{d}(z)=\frac{q_0 z+q_1 (1+z)}{1+q_2 (1+z)}
    \end{equation}

    where $q_{0}$, $q_{1}$ and $q_{2}$ are constants. The energy density becomes

    \begin{equation}\label{rho PI}
        \rho_d=\rho_{d0}~(1+z)^{2(1-q_0)}~(1+q_2+q_2z)^{\frac{2(q_0+q_1+q_0
                q_2)}{q_2}}
    \end{equation}
    From equation (\ref{H}), we obtain

    \begin{equation}
        \begin{aligned}
            H^{2}(z)& =H_{0}^{2}[\Omega_{r0}(1+z)^{4}+\Omega_{m0}(1+z)^{3}\\
            & +(1-\Omega_{r0}-\Omega_{m0})\\
            &(1+z)^{2(1-q_0)} (1+q_2+q_2z)^{\frac{2(q_0+q_1+q_0 q_2)}{q_2}}]
        \end{aligned}
    \end{equation}

    \subsection{New Model}
    We propose a new model of deceleration parameter given as
    \begin{equation}\label{New}
        q_{d}(z)=q_0+\frac{2+(1+z)^{3}}{q_{1}+q_{2}(1+z)^{3}}
    \end{equation}
    where $q_{0}$ and $q_{1}$ are constants. Thus, the energy
    density \eqref{rho} reads
    \begin{equation}\label{rho New}
        \rho_d=\rho_{d0}~(1+z)^{3(1+q_0+\frac{2}{q_{1}})}[q_{1}+q_{2}(1+z)^{3}]^{\frac{(q_{1}-2 q_{2})}{q_{1}q_{2}}}
    \end{equation}

    From equation (\ref{H}), we obtain

    \begin{equation}
        \begin{aligned}
            H^{2}(z)& =H_{0}^{2}[\Omega_{r0}(1+z)^{4}+\Omega_{m0}(1+z)^{3}\\
            & +(1-\Omega_{r0}-\Omega_{m0})
            ~(1+z)^{3(1+q_0+\frac{2}{q_{1}})}\\
            &[ q_{1}+q_{2}(1+z)^{3}]^{\frac{(q_{1}-2 q_{2})}{q_{1}q_{2}}}]
        \end{aligned}
    \end{equation}

    \section{Data description with Results }\label{sec4}

    The investigation of multiple parameterizations of the deceleration
parameter serves a valuable purpose in understanding the late-time cosmic evolution more
comprehensively. By exploring a range of parameterizations, we gain insights into the underlying
theoretical framework that governs the cosmic acceleration. The motivation for
this approach lies in the fact that different theoretical models may exhibit distinct evolutionary
behaviors of the deceleration parameter. By examining various parameterizations,
we can assess the consistency and compatibility of these models with observational data,
ultimately leading us towards a better understanding of the fundamental physics driving
the accelerated expansion of the universe.

\subsection{Data description}\label{DD}
    Throughout this part, we will use three distinct
    observational datasets to limit our model parameters. 
    We utilized
    the H(z) datasets of 57 measurements, the Pantheon dataset of
    1048, 17 measurements of BAO, and CMB Distant Prior to achieving the optimal value for the proposed model
    parameters. To construct the MCMC
    \cite{28emcee}, we used the open-source tools Polychord
    \cite{29polychord} and GetDist \cite{30getdist}.The total $\chi^2$
    function of the combination $H(z)$ + Pantheon + BAO + CMB and define
    as

    \begin{equation}
        \chi^{2}_{tot}=\chi_{H(z)}^{2}+{\chi}_{SNIa}^2+{\chi}_{BAO}^2++{\chi}_{CMB}^2.
    \end{equation}

    \subsubsection{$H(z)$ Dataset}
    Numerous observational datasets must be used to achieve significant constraints on the model parameters. We employ the $H(z)$
    measurements in our investigation to constrain the model
    parameters. In general, the Hubble parameter can be determined either by estimating the BAO in the radial direction of galaxy clustering \cite{H(z)} or using the differential age technique, which also provides the redshift dependency of the Hubble parameter as

    \begin{equation}
        H(z)=-\frac{1}{1+z} \frac{d z}{d t},
    \end{equation}
    here $d z$/$d t$ is computed using two moving galaxies in a
    proportionate manner. To estimate the model's parameters, the study
    takes into account 57 Hubble measurements, which are spans
    throughout the redshift range of $0.07 \leqslant z \leqslant 2.42$.
    To compare the model's theoretical predictions with observation,
    we use the chi-square function.

    \begin{equation}
        \chi_{HZ}^{2}\\=\sum_{i=1}^{57} \frac{\left[H_{t
                h}\left(z_{i},\right)-H_{o b
                s}\left(z_{i}\right)\right]^{2}}{\sigma_{H\left(z_{i}\right)}^{2}},
    \end{equation}

    where $H_{\text {th }}$ and $H_{o b s}$ denote the model prediction and
    observed value of Hubble rate, respectively. Also, $\sigma_{H\left(z_{i}\right)}$ characterizes the standard error at the redshift $z_{i}$. The Hubble function numerical values for
    the appropriate redshifts are shown in
    \cite{bouali2023cosmological}.

    \subsubsection{type Ia supernovae (SNIa)}
    The comic accelerated expansion is
    determined by measuring type Ia supernovae (SNIa). So
    far, SNIa has proven to be one of the most robust successful methods
    for studying the nature of dark energy. In recent years, several
    supernova data sets have been established~\cite{Pan1,Pan2,Pan3,Pan4,Pan5}. The Pantheon sample has lately
    been updated~\cite{pantheon+}. The former dataset contains 1048
    spectroscopically verified SNIa spanning in the redshift range of
    $0<z< 2.3$. SNIa are also astronomical objects that act as
    standard candles for determining relative distances. As a
    consequence, SN Ia samples are combined with the distance modulus
    $\mu=m-M$, where $m$ indicates a certain object's apparent
    magnitude of a specific SNIa. The chi-square of the SNIa
    measurements is given by

    \begin{equation}
        {\chi}_{SN}^2={\Delta \mu}^{T}\hspace{0.1cm}.\hspace{0.1cm}{\bf C}_{SN}^{-1}\hspace{0.1cm}.\hspace{0.1cm}{\Delta \mu} .
    \end{equation}

    ${\bf C}_{SN}$ is represented by a covariance matrix, and
    ${\Delta \mu}=\mu_{obs}-\mu_{th}$, where $\mu_{obs}$ signifies the
    measured distance modulus of a certain SNIa, meanwhile the
    theoretical distance modulus is represented as $\mu_{th}$, and
    calculated as

    \begin{equation}
        \mu_{th}(z)=5\log_{10} \frac{D_{L}(z)}{(H_{0}/c) Mpc} +25,
    \end{equation}

    Here $H_{0}$ signifies the current Hubble rate, and $c$ reflects
    the speed of light. In a flat FLRW Universe, the
    luminosity distance, $D_{L}$, is given by:

    \begin{equation}
        D_L(z)=(1+z)H_0\int_{0}^{z}\frac{dz^{\prime}}{H\left(z^{\prime}\right)}.
    \end{equation}

    Because we limit the model's free parameters at the same time,
    i.e., by using the Pantheon sample, and hence

    \begin{eqnarray}
        {\chi}_{SN}^2&=&{\Delta \mu}^{T}\times {\bf C}_{Pantheon}^{-1}\nonumber \times {\Delta \mu} .
    \end{eqnarray}

    \subsubsection{Baryon Acoustic Oscillations (BAO)}
    We picked 17 measurements of BAO (please see Table 1 of this work
    \cite{benisty2021testing}) measures from
    \cite{bao1,bao2,bao3,bao4,bao5,bao6,bao7,bao8,bao9,bao10,bao11,bao12}
    the greatest BAO dataset of (333) measurements because considering
    the whole catalog of BAO might result in a very considerable
    inaccuracy because of data correlations; so we chose a
    representative subset to minimize errors. Transverse BAO
    experiments produce measurements. of $D_H(z) / r_d=c /
    H(z)r_d$Â along with a co-moving angular diameter
    distance~\cite{bao13,bao14}.

    \begin{equation}
        D_M=\frac{c}{H_0} S_k\left(\int_0^z \frac{d
            z^{\prime}}{E\left(z^{\prime}\right)}\right) \text {, }
    \end{equation}

    with

    \begin{equation}
        S_k(x)= \begin{cases}\frac{1}{\sqrt{\Omega_k}} \sinh
            \left(\sqrt{\Omega_k} x\right) & \text { if } \quad \Omega_k>0 \\
            x & \text { if } \quad \Omega_k=0 \\ \frac{1}{\sqrt{-\Omega_k}}
            \sin \left(\sqrt{-\Omega_k} x\right) & \text { if } \quad
            \Omega_k<0 .\end{cases}
    \end{equation}

    Considering the angular diameter distance $D_A=$ $D_M /(1+z)$
    and the $D_V(z)/r_d$. This corresponds to the combination of the
    BAO peaked coordinates and the sound horizon $r_d$ at the drag
    epoch. Furthermore, we could immediately derive "line-of-sight"
    (or "radial") observations from the Hubble parameter.

    \begin{equation}
        D_V(z) \equiv\left[z D_H(z) D_M^2(z)\right]^{1/3} \text {. }
    \end{equation}

    \subsubsection{Cosmic Microwave Background (CMB) }\label{CMB}

    The CMB Distant Prior measurements are taken from Ref.~\cite{chen2019distance}. Using distance priors, you can gain helpful information about the CMB power spectrum in two ways: the acoustic scale $l_A$ represents the temperature power spectrum of the CMB in the transverse direction, and the "shift parameter" $R$ affects the temperature spectrum of the CMB along the line-of-sight path to determine peak heights as follows:
    \begin{equation}
        l_A=(1+z_d) \frac{\pi D_A(z)}{r_s},
    \end{equation}

    \begin{equation}
        \quad R(z)=\frac{\sqrt{\Omega_m} H_0}{c}(1+z_d) D_A(z)
    \end{equation}

    These are the observables that are reported~\cite{chen2019distance}: $R_z=1.7502 \pm 0.0046, \quad l_A=301.471 \pm 0.09, \quad
    n_s=0.9649 \pm 0.0043$ and $r_s$ is an independent parameter, with
    an associated covariance matrix (see Ref.~\cite{chen2019distance}). The points represent the inflationary
    observables as well as the CMB epoch expansion rate. Besides the CMB points, we also consider other data from the late universe. The result is a successful test of the model in
    relation to the data.
    
    \begin{figure}[!htb]
        \begin{minipage}{0.49\textwidth}
            \centering
            \includegraphics[scale=0.3]{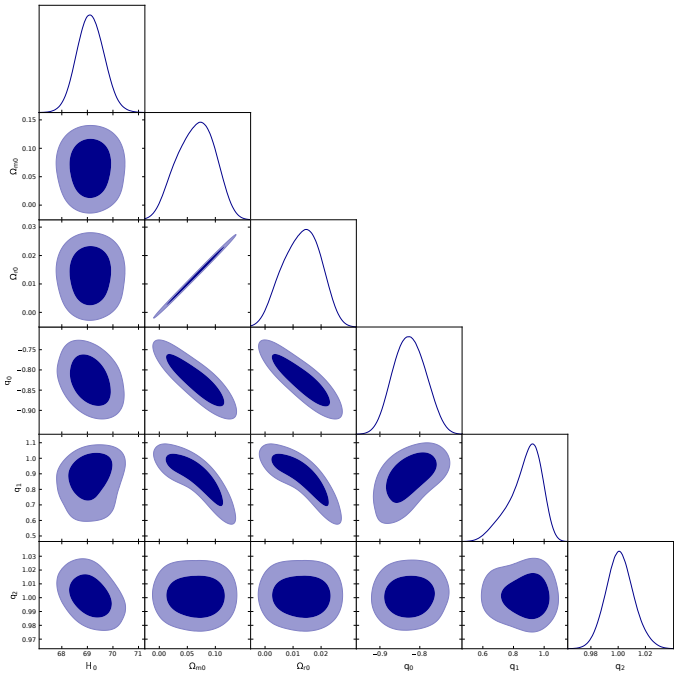}
            \caption{Plot of MCMC confidence contours at
                1$\sigma$ and 2$\sigma$ for Model 1.}\label{MCMC1}
        \end{minipage}
    \end{figure}
    \begin{figure}[!htb]
        \begin{minipage}{0.49\textwidth}
            \centering
            \includegraphics[scale=0.3]{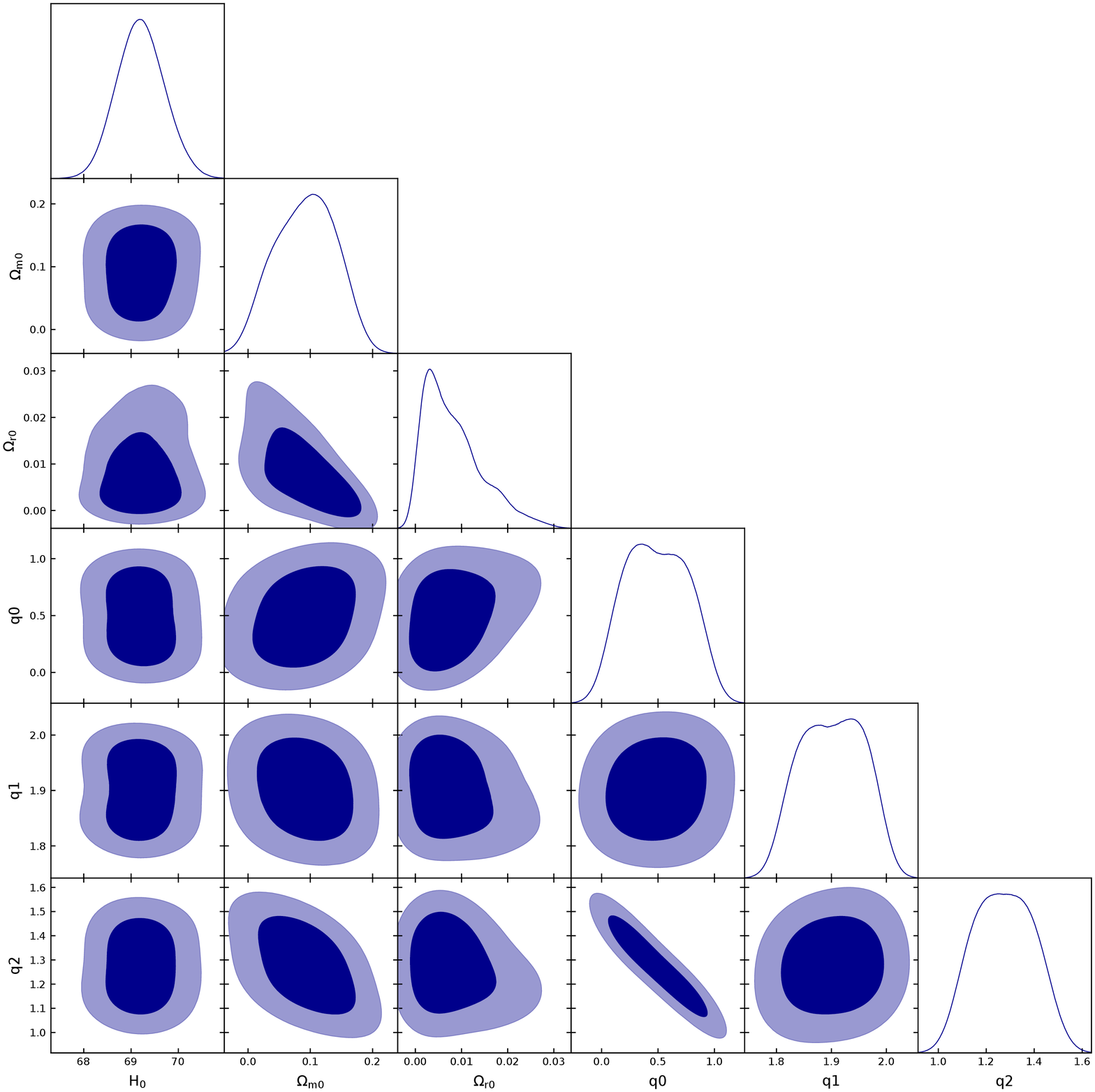}
            \caption{Plot of MCMC confidence contours at
                1$\sigma$ and 2$\sigma$ for Model 2.}\label{MCMC2}
        \end{minipage}
    \end{figure}
    \begin{figure}[!htb]
        \begin{minipage}{0.49\textwidth}
            \centering
            \includegraphics[scale=0.3]{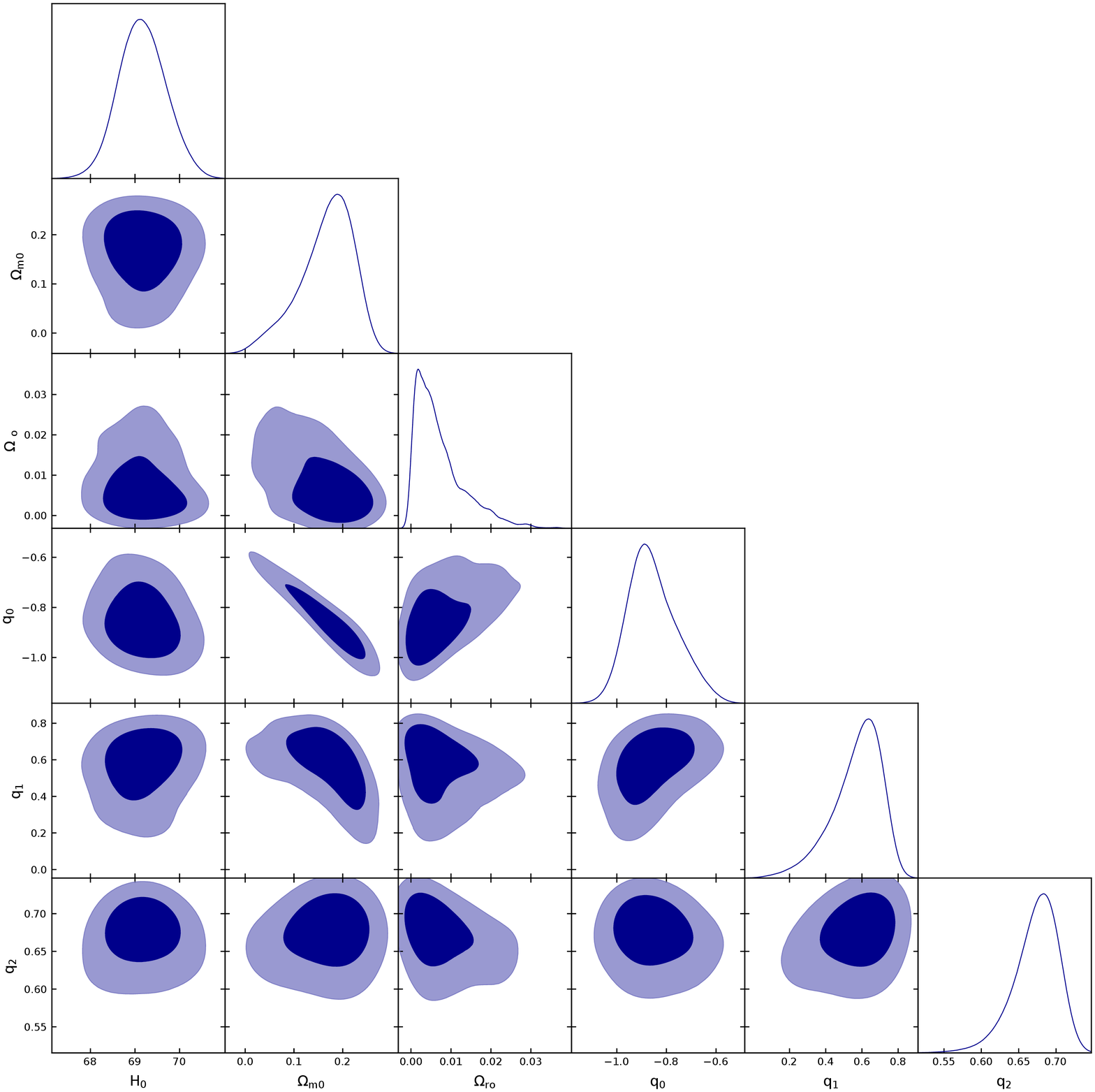}
            \caption{Plot of MCMC confidence contours at
                1$\sigma$ and 2$\sigma$ for Model 3.}\label{MCMC3}
        \end{minipage}
    \end{figure}
    \begin{figure}[!htb]
        \begin{minipage}{0.49\textwidth}
            \centering
            \includegraphics[scale=0.3]{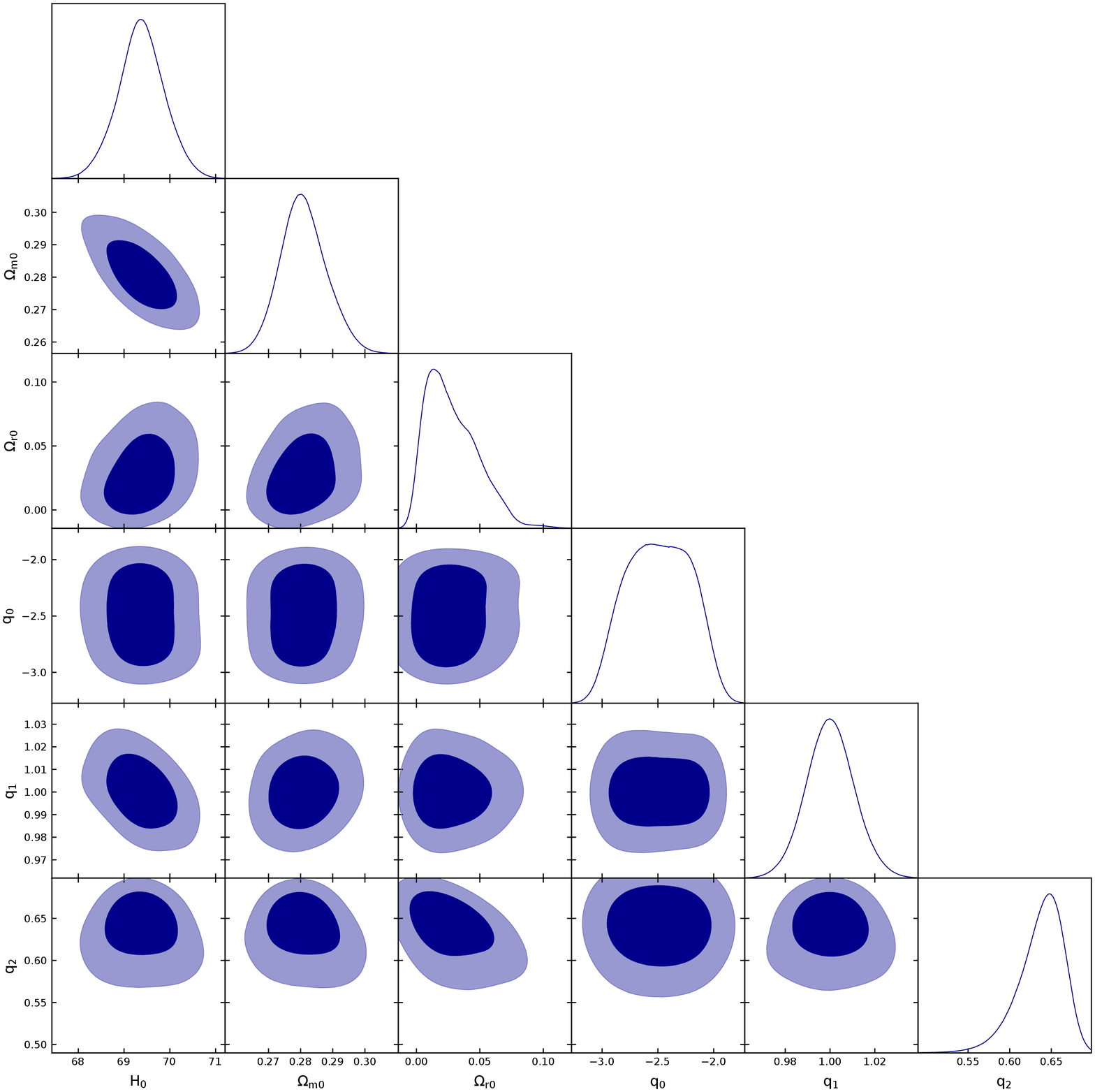}
            \caption{Plot of MCMC confidence contours at
                1$\sigma$ and 2$\sigma$ for Model 4.}\label{MCMC4}
        \end{minipage}
    \end{figure}

\begin{table}[H]
\begin{center}
\begin{tabular}{|c|c|c|c|}
\hline
\multicolumn{4}{|c|}{MCMC Results} \\
\hline
Model & Priors & Parameters & Best fit Value \\
\hline
$  \Lambda$CDM Model &$[50.,100.]$& $H_0$ &$69.854848_{-1.259100}^{+1.259100}$ \\
& $[0.,1.]$ &$\Omega_{\mathrm{m0}}$ & $0.268654_{-0.012822}^{+0.012822}$ \\
\hline
Model 1 & $[50.,100.]$ &$H_0 $ &$69.121060_{-0.517902}^{+0.517902}$ \\
& $[0.,1.]$ &$\Omega_{\mathrm{m0}}$ & $0.065585_{-0.038524}^{+0.038524}$ \\
& $[0.,1.]$ &$\Omega_{\mathrm{r0}}$ & $0.013117_{-0.007705}^{+0.007705}$ \\
&$[-1.5,-0.5.]$ &$q_0$ &$-0.825763_{-0.041951}^{+0.041951}$ \\
& $[0.5,1.5]$&$q_1$ &$0.879053 _{-0.111601}^{+0.111601}$  \\
&$[0.5,1.5]$ &$q_2$ &$1.001254 _{-0.010014}^{+0.010014}$  \\
\hline
Model 2 & $[50.,100.]$&$H_0$ &$69.203662_{-0.529034}^{+0.529034}$ \\
&$[0.,1.]$ &$\Omega_{\mathrm{m0}}$ & $0.192287_{-0.058696}^{+0.058696}$ \\
&$[0.,1.]$ &$\Omega_{\mathrm{r0}}$ & $ 0.008128_{-0.006286}^{+0.006286}$ \\
&$[0.,1.]$ &$q_0$ &$0.489155_{-0.311871}^{+0.311871}$ \\
&$[1.5,2.5]$ &$q_1$ &$1.902027_{-0.066852}^{+0.066852}$  \\
&$[1.,1.4]$ &$q_2$ &$1.274424_{-0.143687}^{+0.143687}$ \\
\hline
Model 3 &$[50.,100.]$ &$H_0$ &$69.789290_{-0.649419}^{+0.469419}$ \\
&$[0.,1.]$ &$\Omega_{\mathrm{m}}$ & $0.205031_{-0.084481}^{+0.084481}$ \\
&$[0.,1.]$ &$\Omega_{\mathrm{r0}}$ & $0.00175_{-0.007645}^{+0.007645}$ \\
&$[-1.2,-0.4]$ &$q_0$ &$-0.780911_{-0.060833}^{+0.060833}$ \\
&$[0.,1.]$ &$q_1$ &$0.610530_{-0.067778}^{+0.067778}$  \\
& $[0.,1.]$&$q_2$ &$0.680228_{-0.285537}^{+0.285537}$ \\
\hline
Model 4 & $[50.,100.]$&$H_0$ &$69.391442_{-0.478121}^{+0.478121}$ \\
&$[0.,1.]$ &$\Omega_{\mathrm{m0}}$ & $0.280604_{-0.006564}^{+0.006564}$ \\
& $[0.,1.]$&$\Omega_{\mathrm{r0}}$ & $0.023950_{-0.017250}^{+0.017250}$ \\
&$[-3.5,-2.5.]$ &$q_0$ &$-2.975495_{-0.764170}^{+0.764170}$ \\
& $[0.5,1.5]$&$q_1$ &$1.000462_{-0.009284}^{+0.009284}$  \\
& $[0.,1.]$&$q_2$ &$0.646235_{-0.023499}^{+0.023499}$ \\
\hline
\end{tabular}
\caption{Summary of the MCMC results using $H(z)$ + BAO + SNIa + CMB dataset} \label{table1}
\end{center}
\end{table}

    \subsection{Observational, and theoretical comparisons of the Hubble functions}
    Following extracting the best-fit value of the free parameter of
    each Model, one could also contrast the model predictions against
    the observational data and also the $\Lambda$CDM model,
    correspondingly.

    \subsubsection{Comparison with the Hubble data points.} 
\begin{figure}[!htb]
        \begin{minipage}{0.49\textwidth}
            \centering
            \includegraphics[scale=0.4]{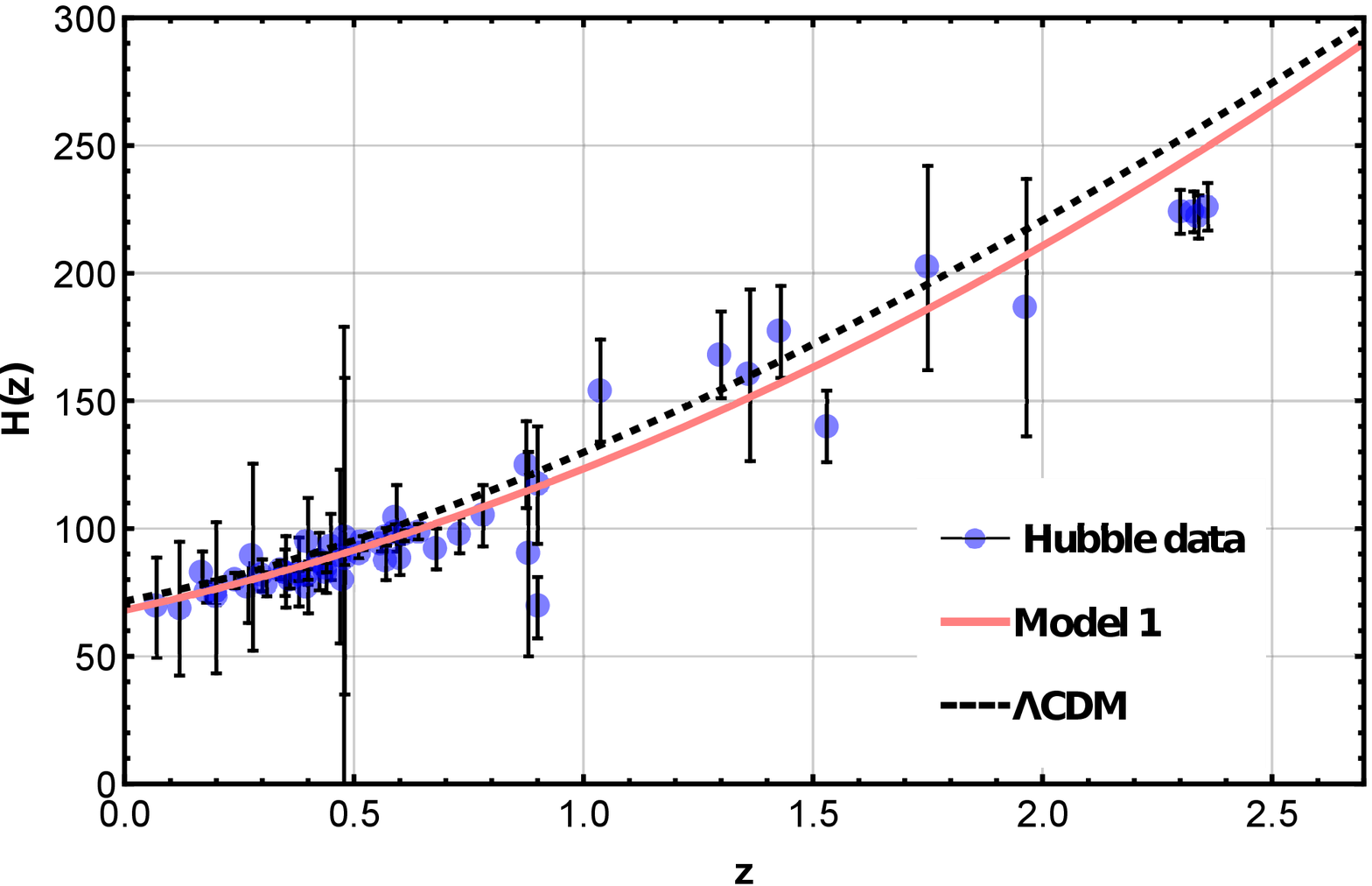}
            \caption{Shows the plot of Hubble function $H(z)$ for Model
                1 (pink line)$, \Lambda$CDM model (black
                dotted line) with $\Omega_{\mathrm{m0}}=$ 0.3 and $\Omega_\Lambda
                =$ 0.7, against Hubble measurements (blue dots) .}\label{HZ1}
        \end{minipage}\hfill
        \begin{minipage}{0.49\textwidth}
            \centering
            \includegraphics[scale=0.4]{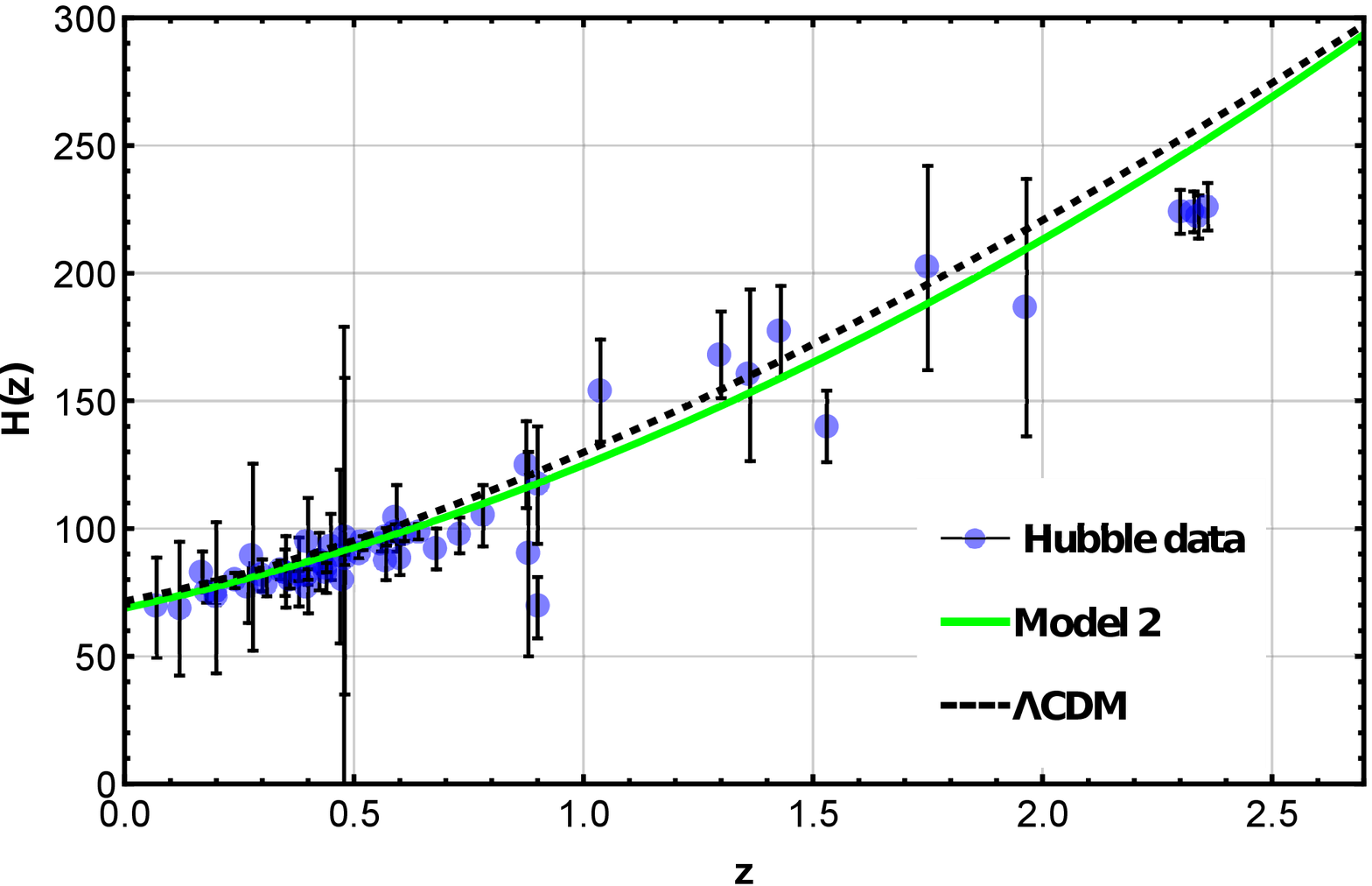}
            \caption{Shows the plot of Hubble function $H(z)$ for Model
                2 (green line), $ \Lambda$CDM model (black
                dotted line) with $\Omega_{\mathrm{m0}}=$ 0.3 and $\Omega_\Lambda
                =$ 0.7, against Hubble measurements (blue dots) .}\label{HZ2}
        \end{minipage}\hfill
        \begin{minipage}{0.49\textwidth}
            \centering
            \includegraphics[scale=0.4]{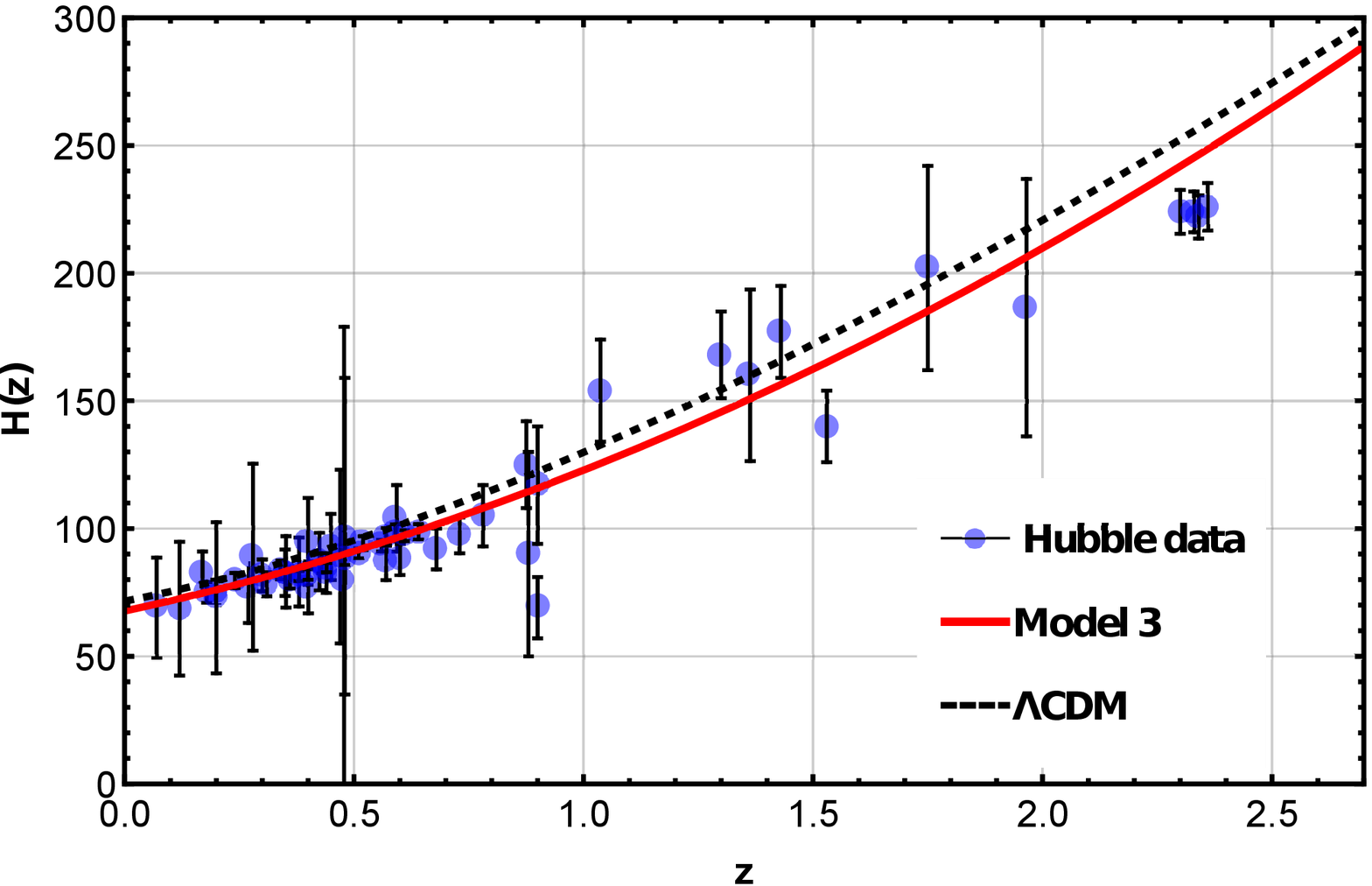}
            \caption{Shows the plot of Hubble function $H(z)$ for Model
                3 (red line), $ \Lambda$CDM model (black dotted
                line) with $\Omega_{\mathrm{m0}}=$ 0.3 and $\Omega_\Lambda =$ 0.7,
                against Hubble measurements (blue dots) .}\label{HZ3}
        \end{minipage}
    \end{figure}

\begin{figure}[!htb]
        \begin{minipage}{0.49\textwidth}
            \centering
            \includegraphics[scale=0.4]{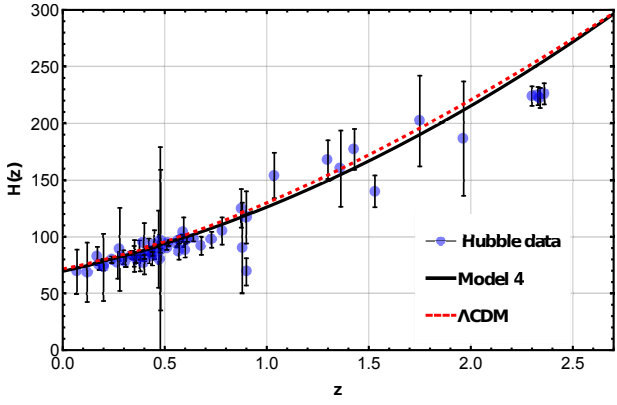}
            \caption{Shows the plot of Hubble function $H(z)$ for Model
                4 (black line), $ \Lambda$CDM model (red dotted
                line) with $\Omega_{\mathrm{m0}}=$ 0.3 and $\Omega_\Lambda =$ 0.7,
                against Hubble measurements (blue dots) .}\label{HZ4}
        \end{minipage}
    \end{figure}

Each model has been compared to the 57 Hubble measurements, $\Lambda$CDM model, and 1$\sigma$ and 2$\sigma$ error bands. The comparison findings are shown in Figs.~\ref{HZ1}, \ref{HZ2}, \ref{HZ3}, \ref{HZ4}. The Figure illustrates that each model accurately fits with the Hubble measurements.

    \subsubsection{Comparison with the Pantheon data.}
\begin{figure}[!htb]
        \begin{minipage}{0.49\textwidth}
            \centering
            \includegraphics[scale=0.56]{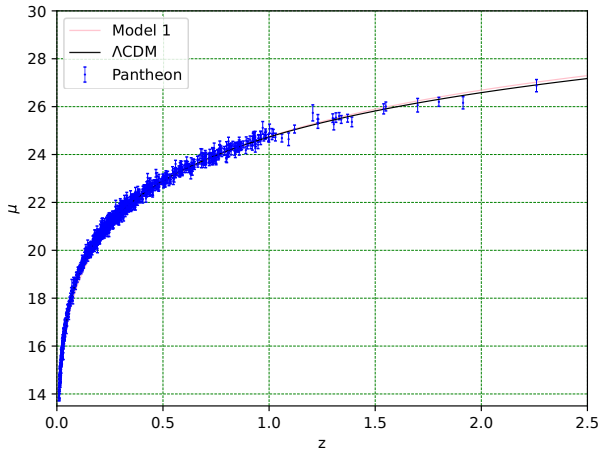}
            \caption{Shows the plot of distance modulus $ \mu(z) $ for
                Model 1 (pink line), $ \Lambda$CDM model (in black line) with $\Omega_{\mathrm{m0}}=$ 0.3 and $\Omega_\Lambda
                =$ 0.7, against Pantheon dataset (blue dots) }\label{mu1}
        \end{minipage}
    \end{figure}

\begin{figure}[!htb]
        \begin{minipage}{0.49\textwidth}
            \centering
            \includegraphics[scale=0.56]{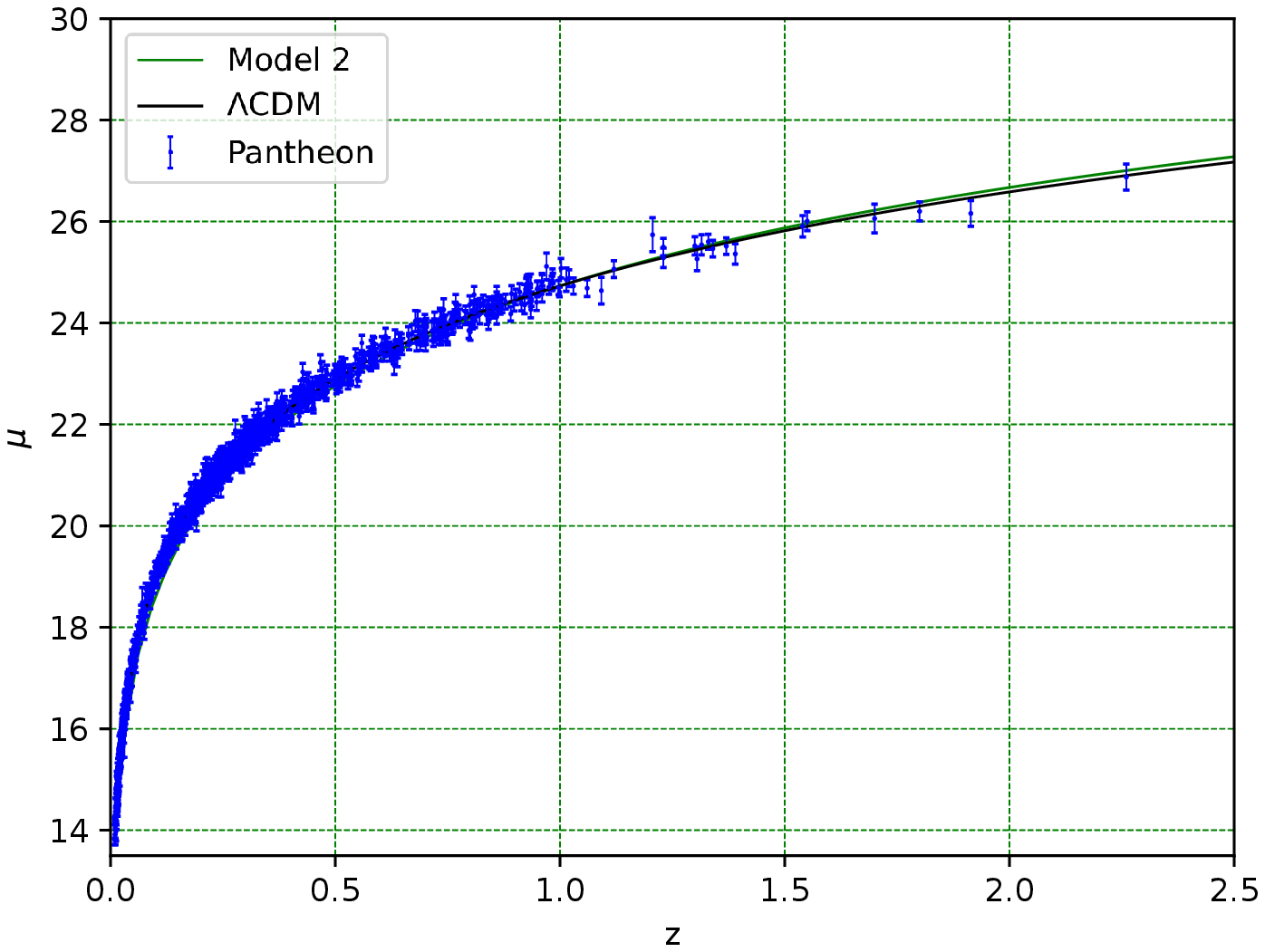}
            \caption{Shows the plot of distance modulus $ \mu(z) $ of the
                model 2 (green line), $ \Lambda$CDM model (black line) with $\Omega_{\mathrm{m0}}=$ 0.3 and $\Omega_\Lambda
                =$ 0.7, against Pantheon dataset (blue dots) .}\label{mu2}
        \end{minipage}\hfill
        \begin{minipage}{0.49\textwidth}
            \centering
            \includegraphics[scale=0.56]{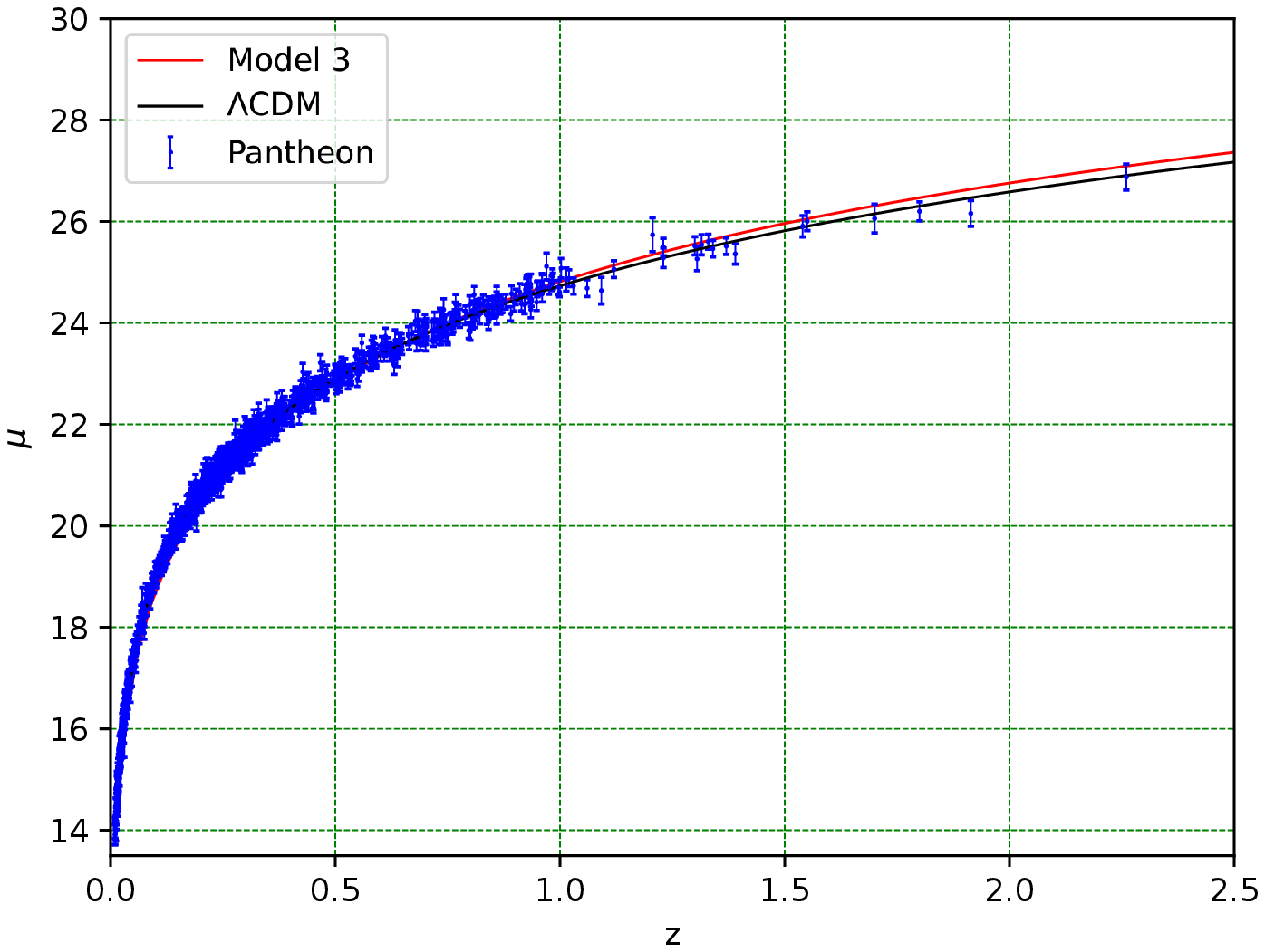}
            \caption{Shows the plot of distance modulus $ \mu(z) $ of the
                model 3 (red line), $ \Lambda$CDM model (black line) with $\Omega_{\mathrm{m0}}=$ 0.3 and $\Omega_\Lambda
                =$ 0.7, against Pantheon dataset (blue dots) .}\label{mu3}
        \end{minipage}\hfill
        \begin{minipage}{0.49\textwidth}
            \centering
            \includegraphics[scale=0.56]{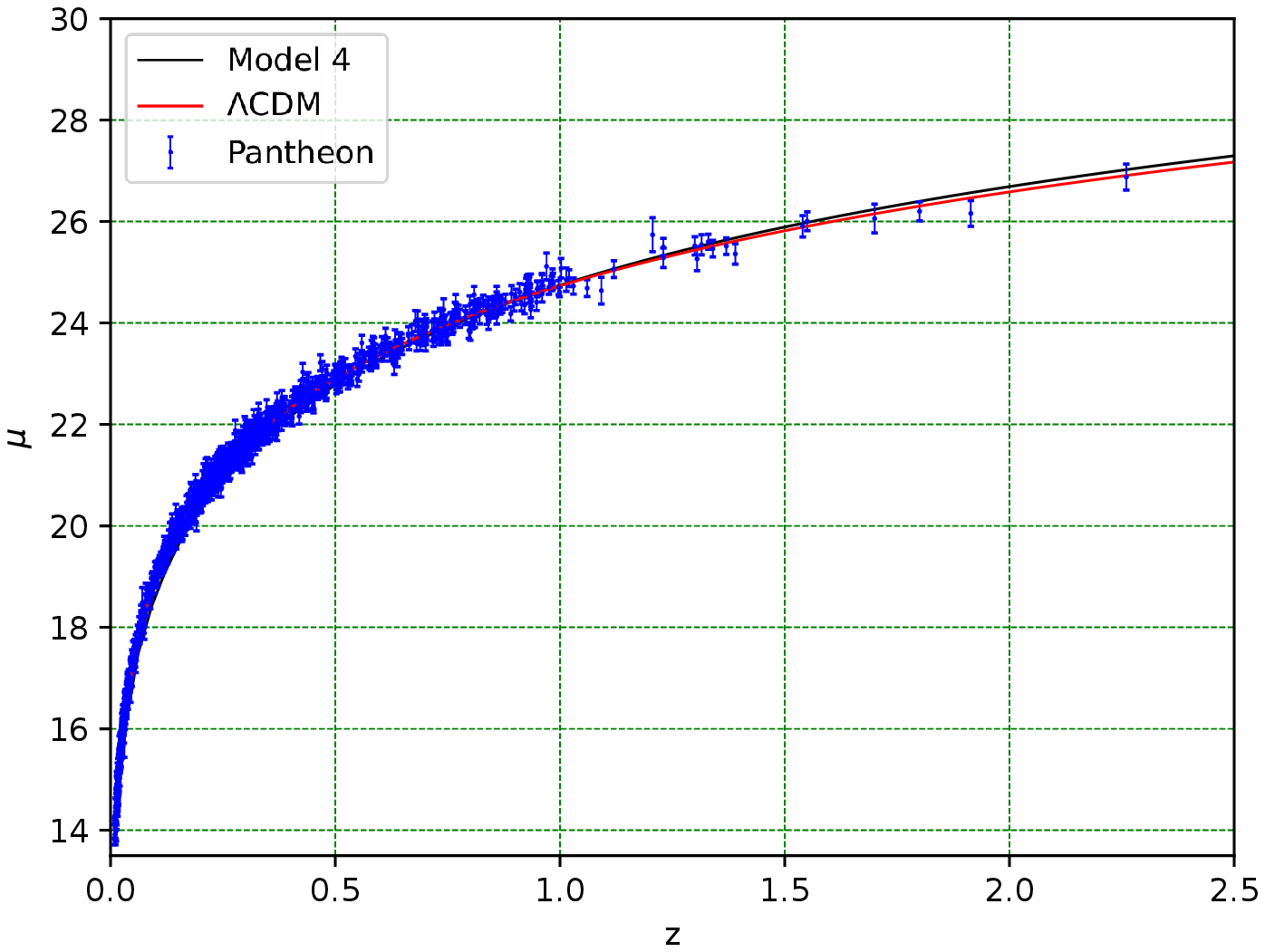}
            \caption{Shows the plot of distance modulus $ \mu(z) $ of the
                model 4 (black line), $ \Lambda$CDM model (black line) with $\Omega_{\mathrm{m0}}=$ 0.3 and $\Omega_\Lambda
                =$ 0.7, against Pantheon dataset (blue dots) .}\label{mu4}
        \end{minipage}
    \end{figure}

    Each model has been compared to the 1048 Pantheon dataset,
    $\Lambda$CDM model, and 1$\sigma$ and 2$\sigma$ error bands. The
    comparison findings are shown in
    Figs.~\ref{mu1}, \ref{mu2}, \ref{mu3}, \ref{mu4}, 0ne could see that
    each model matches the Pantheon dataset quite well.
    \clearpage
    \subsubsection{Relative difference between Each Model and $\Lambda$CDM.}
\begin{figure}[!htb]
        \begin{minipage}{0.49\textwidth}
            \centering
            \includegraphics[scale=0.42]{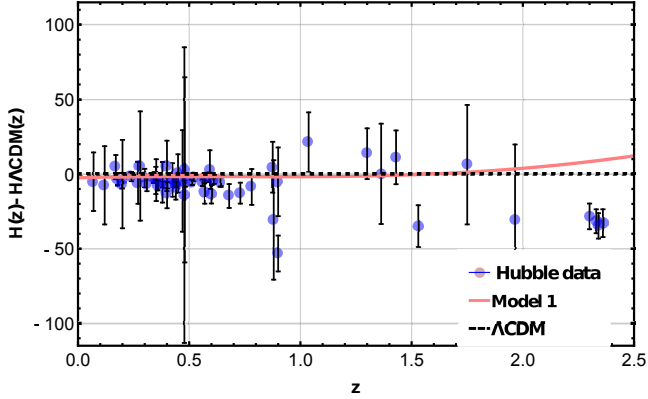}
            \caption{Shows the Relative difference between Model 1 (pink
                line), the $\Lambda$CDM model (black dotted line) with $\Omega_{\mathrm{m0}}=$ 0.3
                and $\Omega_\Lambda =$ 0.7, against
                the 57 Hubble measurements (blue dots), along with their
                corresponding error bars.}\label{1}
        \end{minipage}
    \end{figure}

\begin{figure}[!htb]
        \begin{minipage}{0.49\textwidth}
            \centering
            \includegraphics[scale=0.4]{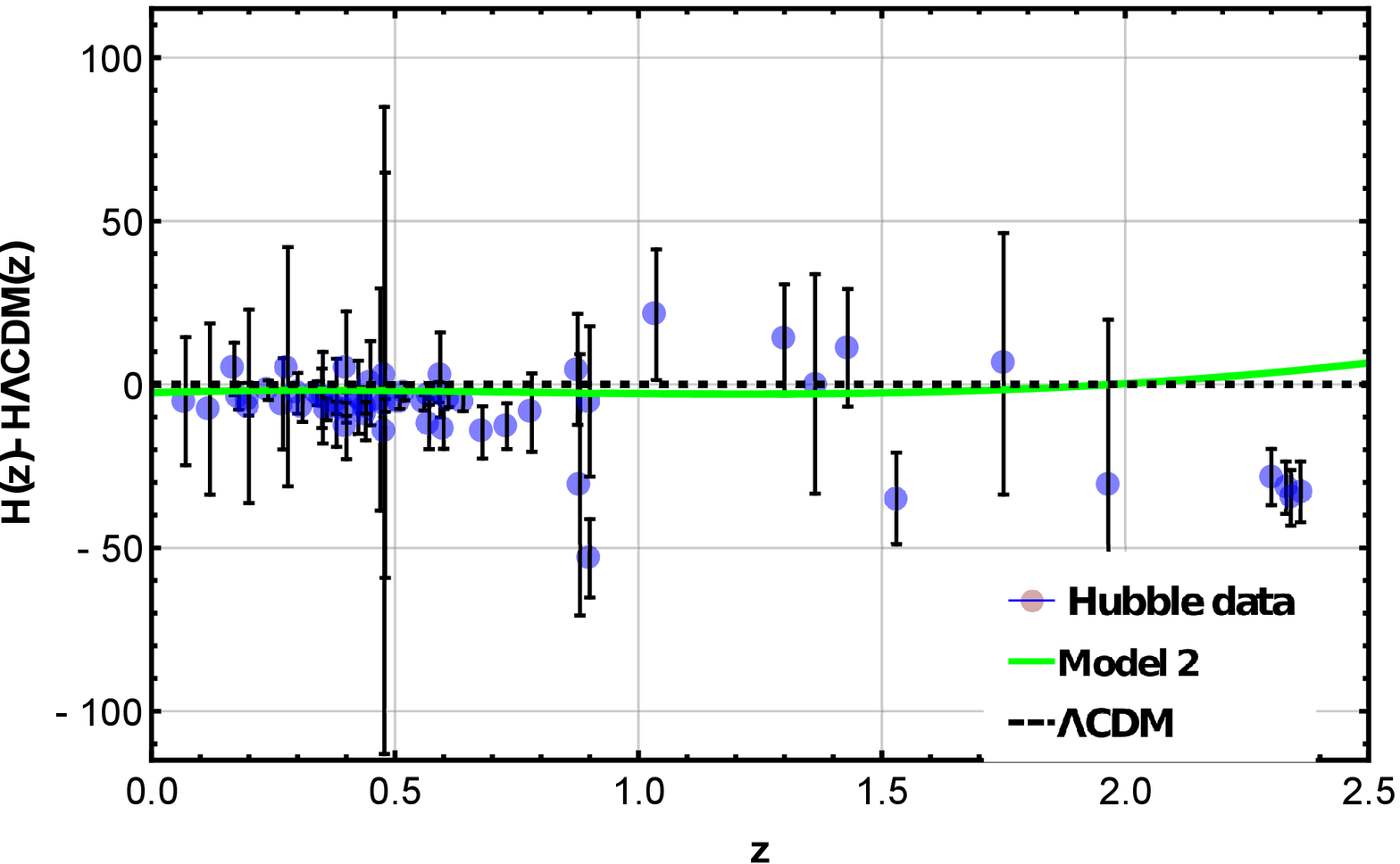}
            \caption{Shows the Relative difference between Model 2 (Green line) and the $\Lambda$CDM model (black dotted line) with
                $\Omega_{\mathrm{m0}}=$ 0.3 and $\Omega_\Lambda =$ 0.7, against the 57 Hubble measurements (blue dots) along with their corresponding error bars.}\label{2}
        \end{minipage}
    \end{figure}

\begin{figure}[!htb]
        \begin{minipage}{0.49\textwidth}
            \centering
            \includegraphics[scale=0.4]{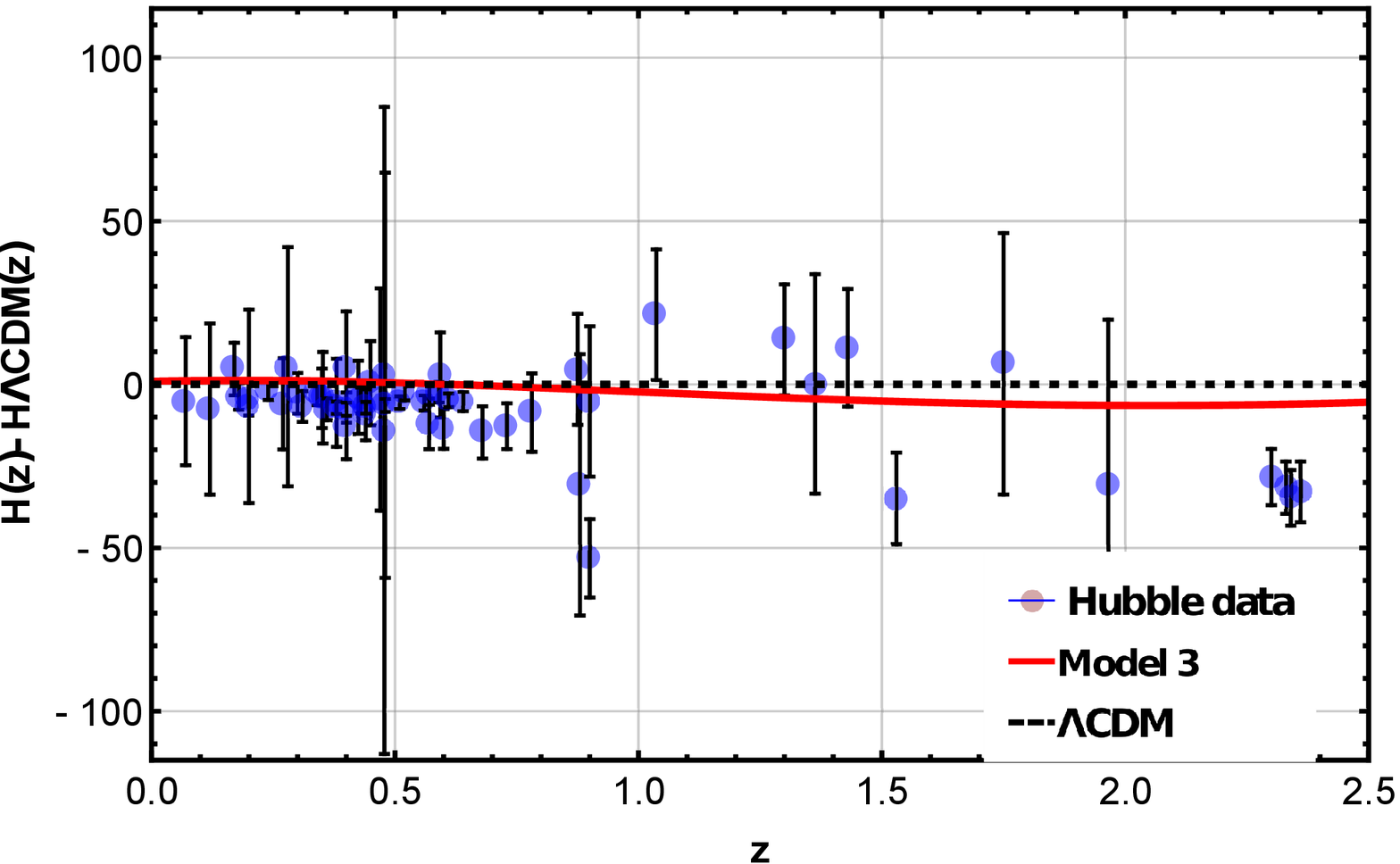}
            \caption{Shows the Relative difference between Model 3 (red
                line), the $\Lambda$CDM model (black dotted line) with $\Omega_{\mathrm{m0}}=$ 0.3
                and $\Omega_\Lambda =$ 0.7, against
                the 57 Hubble measurements (blue dots), along with their
                corresponding error bars.}\label{3}
        \end{minipage}
    \end{figure}

\begin{figure}[!htb]
        \begin{minipage}{0.49\textwidth}
            \centering
            \includegraphics[scale=0.4]{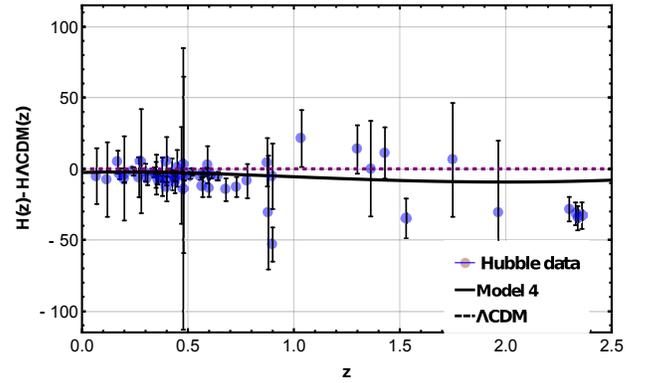}
            \caption{Shows the Relative difference between Model 4 (black line) and the $\Lambda$CDM model (purple dotted line ) with
                $\Omega_{\mathrm{m0}}=$ 0.3 and $\Omega_\Lambda =$ 0.7, against the 57 Hubble measurements (blue dots) along with their corresponding error bars.}\label{4}
        \end{minipage}
    \end{figure}

    Consequently, The relative difference between each model and
    the $\Lambda$CDM standard paradigm has been in
    Figs.~\ref{1}, \ref{2}, \ref{3}, \ref{4}. At low redshift, each model
    behaves substantially identically; however, some distinctions
    between each model and $\Lambda$CDM paradigm emerge at high
    redshift.
    \clearpage
\section{Cosmographic parameters}\label{sec5}
Cosmographic parameters are a set of observables used to describe the behavior and characteristics of the universe's expansion. They provide a way to quantify the dynamics of cosmic expansion beyond the standard cosmological parameters. The cosmographic approach involves approximating the scale factor of the universe as a Taylor series expansion around a reference time. The first three cosmographic parameters are the Hubble constant $H_0$, the deceleration parameter $q_0$, and the jerk parameter $j_0$. The Hubble constant represents the present-day rate of expansion of the universe, while the deceleration parameter describes how the expansion is slowing down or accelerating. The jerk parameter indicates the rate at which the acceleration of the universe's expansion is changing. Additionally, higher-order cosmographic parameters can be considered, such as the snap parameter $s_0$ and the lerk parameter $l_0$. The snap parameter characterizes the fourth derivative of the scale factor, providing insights into the behavior of cosmic expansion beyond acceleration. The lerk parameter represents the fifth derivative of the scale factor, capturing even higher-order effects. Cosmographic parameters offer a phenomenological framework to analyze and compare various cosmological models without assuming specific theories or underlying physical mechanisms. They enable a deeper understanding of the universe's expansion dynamics and can be used to test the consistency of different cosmological scenarios. By incorporating higher-order terms, cosmographic parameters provide a more comprehensive description of the evolution of the universe and its fundamental properties.

\subsection{The deceleration parameter}
The deceleration parameter, denoted as $q_0$, is a fundamental quantity in cosmology that characterizes the changing expansion rate of the universe. It is calculated using the formula:

\begin{equation}
q_0 = -\frac{\ddot{a}a}{{\dot{a}}^2},
\end{equation}

where $a$ represents the size of the universe, and dots denote time derivatives. A positive deceleration parameter ($q_0 > 0$) indicates a decelerating expansion, while a negative value ($q_0 < 0$) suggests an accelerating expansion. A value of zero ($q_0 = 0$) corresponds to a constant expansion rate. The deceleration parameter provides insights into the future fate of the universe and the presence of phenomena such as dark energy. Measurement of $q_0$ from observations contributes to our understanding of the universe's evolution.

    \begin{figure}[!htb]
        \begin{minipage}{0.49\textwidth}
            \centering
            \includegraphics[scale=0.45]{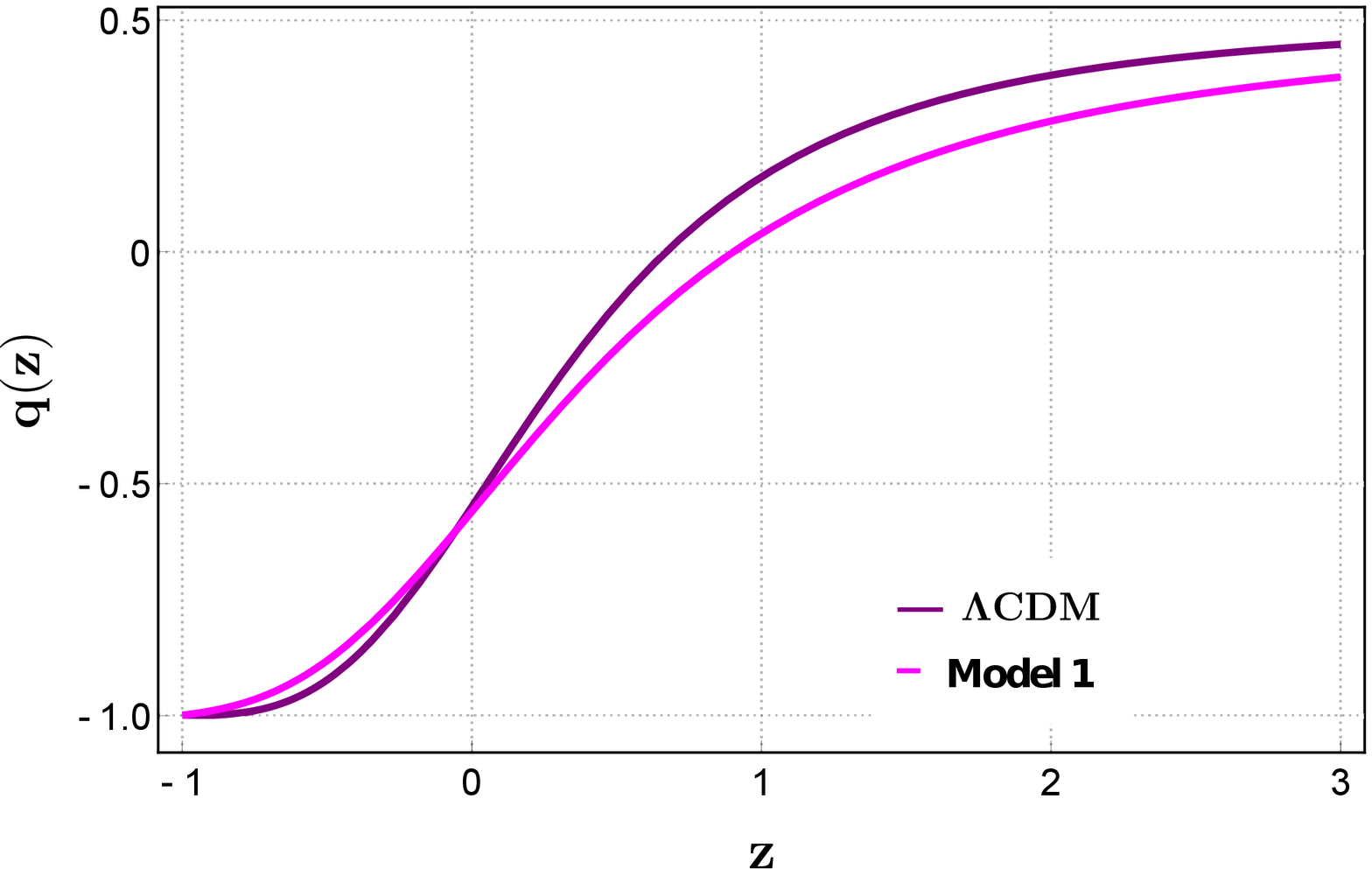}
            \caption{Plot of deceleration parameter with respect to
                redshift.}\label{q(z)1}
        \end{minipage}
    \end{figure}
    \begin{figure}[!htb]
        \begin{minipage}{0.49\textwidth}
            \centering
            \includegraphics[scale=0.45]{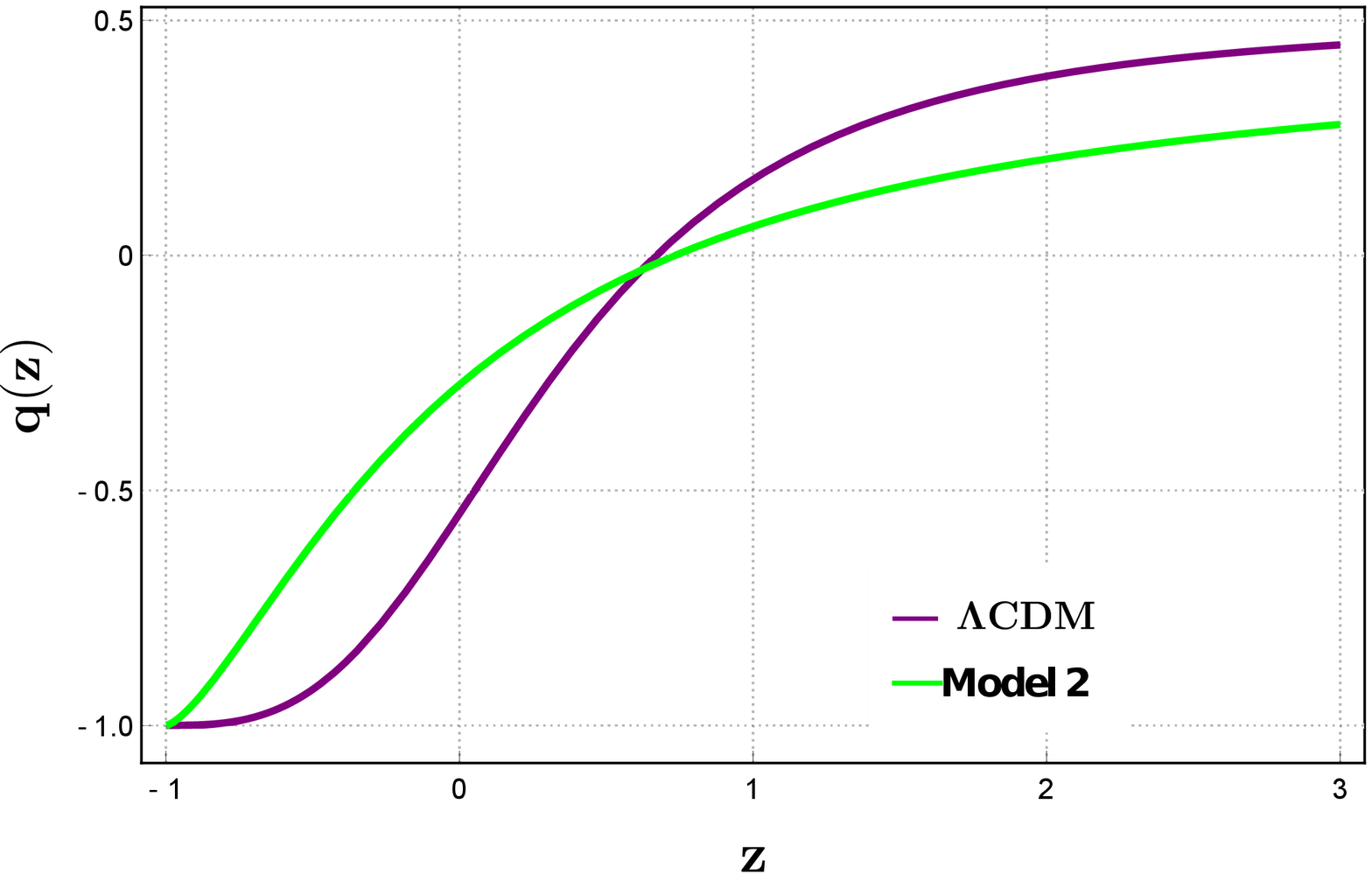}
            \caption{Plot of deceleration parameter with respect to
                redshift.}\label{q(z)2}
        \end{minipage}
    \end{figure}
    \begin{figure}[!htb]
        \begin{minipage}{0.49\textwidth}
            \centering
            \includegraphics[scale=0.45]{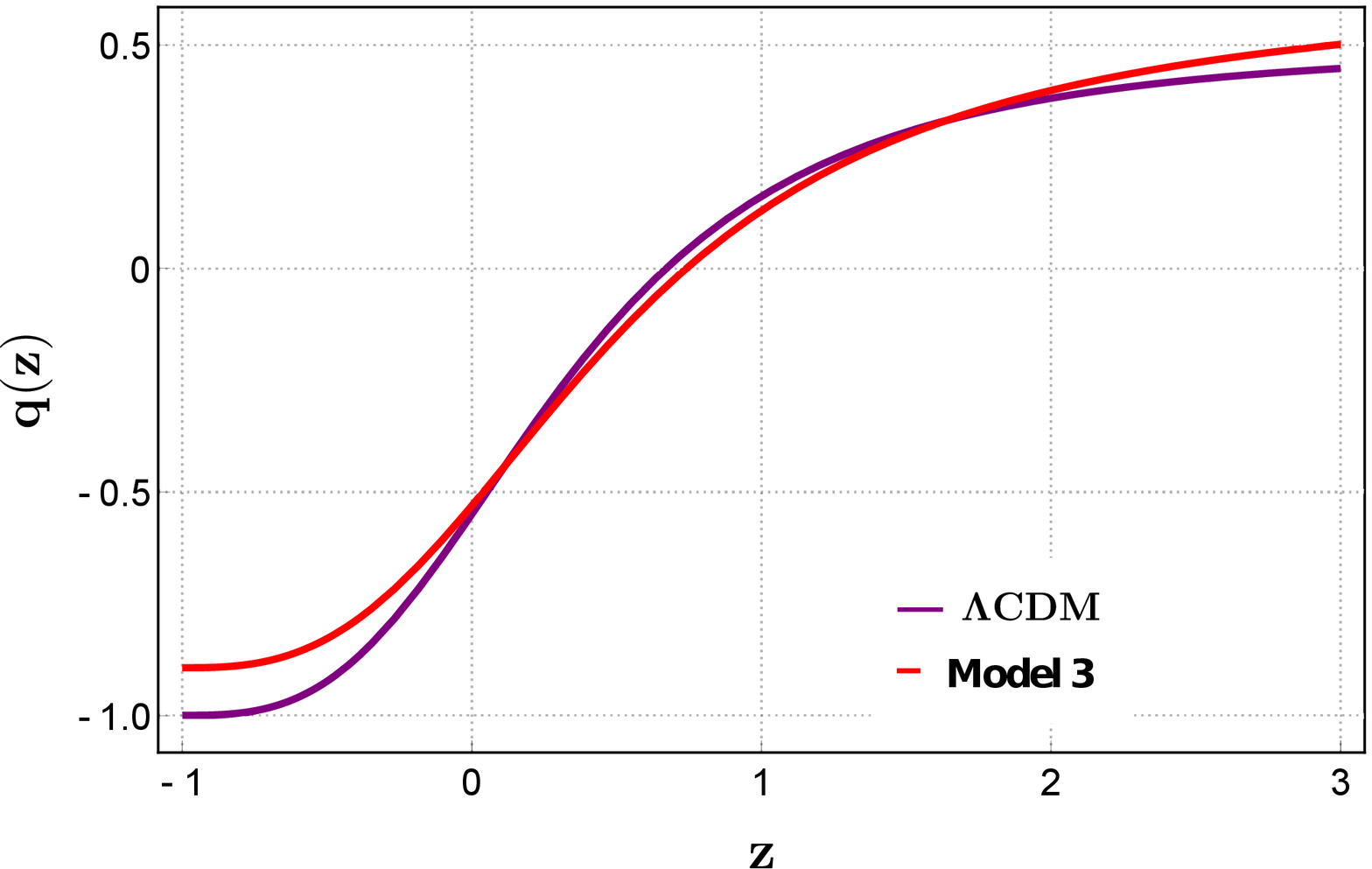}
            \caption{Plot of deceleration parameter with respect to
                redshift.}\label{q(z)3}
        \end{minipage}
    \end{figure}
    \begin{figure}[!htb]
        \begin{minipage}{0.49\textwidth}
            \centering
            \includegraphics[scale=0.45]{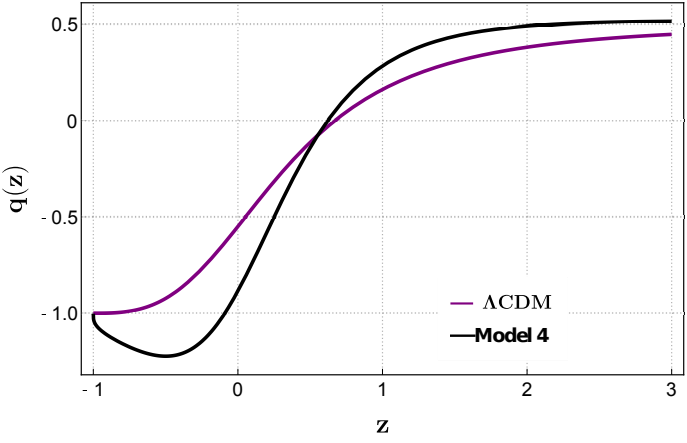}
            \caption{Plot of deceleration parameter with respect to
                redshift.}\label{q(z)4}
        \end{minipage}
    \end{figure}

\subsection{The jerk parameter:}
The jerk parameter, denoted as $j$, is a cosmological quantity that characterizes the behavior of the universe's expansion. It is derived from the fourth term in a Taylor series expansion of the scale factor $a(t)$ around a reference time $t_0$. The formula for the scale factor expansion is given by:

\begin{equation}
\begin{aligned}
\frac{a(t)}{a_0}& = 1 + H_0(t - t_0) - \frac{1}{2}q_0H_0^2(t - t_0)^2 + \frac{1}{6}j_0H_0^3(t - t_0)^3\\
& + O[(t - t_0)^4],
\end{aligned}
\end{equation}

where $H_0$ is the Hubble constant, $q_0$ is the deceleration parameter, and $j_0$ represents the jerk parameter. The jerk parameter $j$ is defined as the third derivative of the scale factor with respect to cosmic time, normalized by the ratio of the first derivative of the scale factor to the scale factor itself. It can be expressed as:

\begin{equation}
j = \frac{1}{a}\frac{d^3a}{d\tau^3}\left(\frac{1}{a}\frac{da}{d\tau}\right)^{-3} = q(2q + 1) + (1 + z)\frac{dq}{dz},
\end{equation}

where $z$ is the redshift, and $\frac{dq}{dz}$ represents the derivative of the deceleration parameter with respect to the redshift. The jerk parameter provides valuable insights into the acceleration and dynamics of the universe's expansion beyond linear and quadratic descriptions. It is a useful tool for exploring different cosmological models and understanding the underlying physics driving the expansion.

    \begin{figure}[!htb]
        \begin{minipage}{0.49\textwidth}
            \centering
            \includegraphics[scale=0.45]{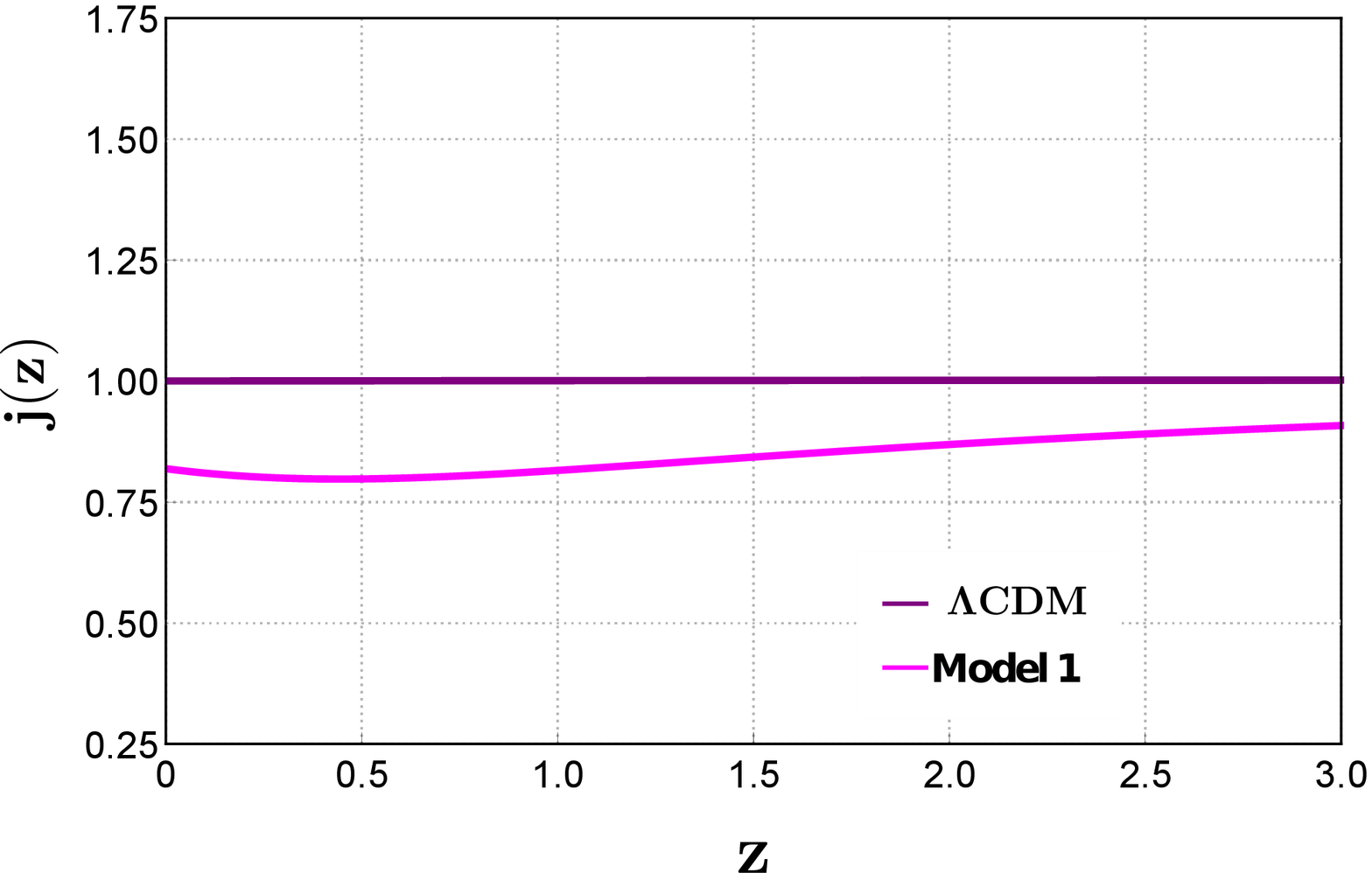}
            \caption{Plot of jerk parameter with respect to
                redshift.}\label{j(z)1}
        \end{minipage}
    \end{figure}
    \begin{figure}[!htb]
        \begin{minipage}{0.49\textwidth}
            \centering
            \includegraphics[scale=0.45]{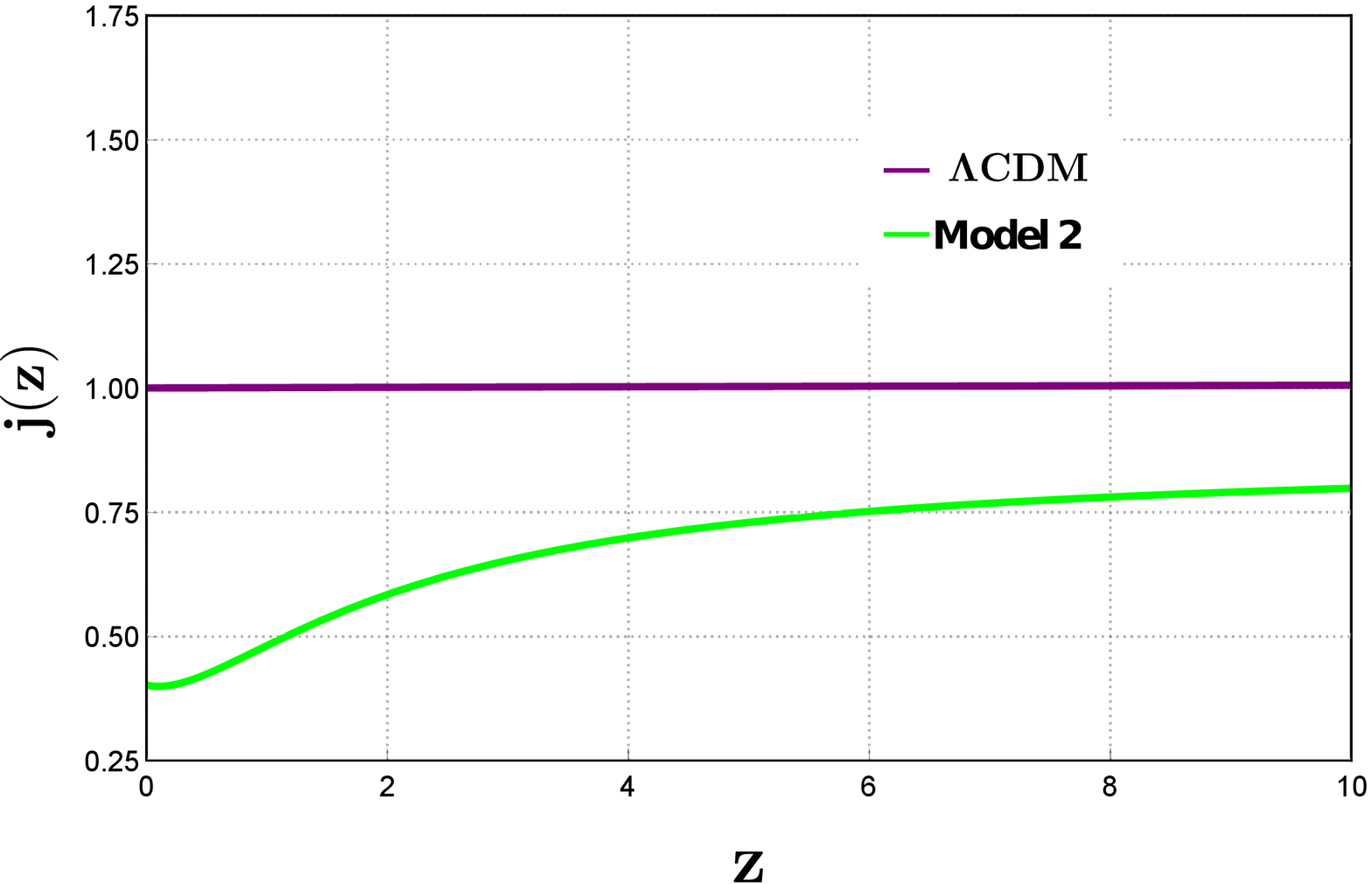}
            \caption{Plot of jerk parameter with respect to
                redshift.}\label{j(z)2}
        \end{minipage}
    \end{figure}
    \begin{figure}[!htb]
        \begin{minipage}{0.49\textwidth}
            \centering
            \includegraphics[scale=0.45]{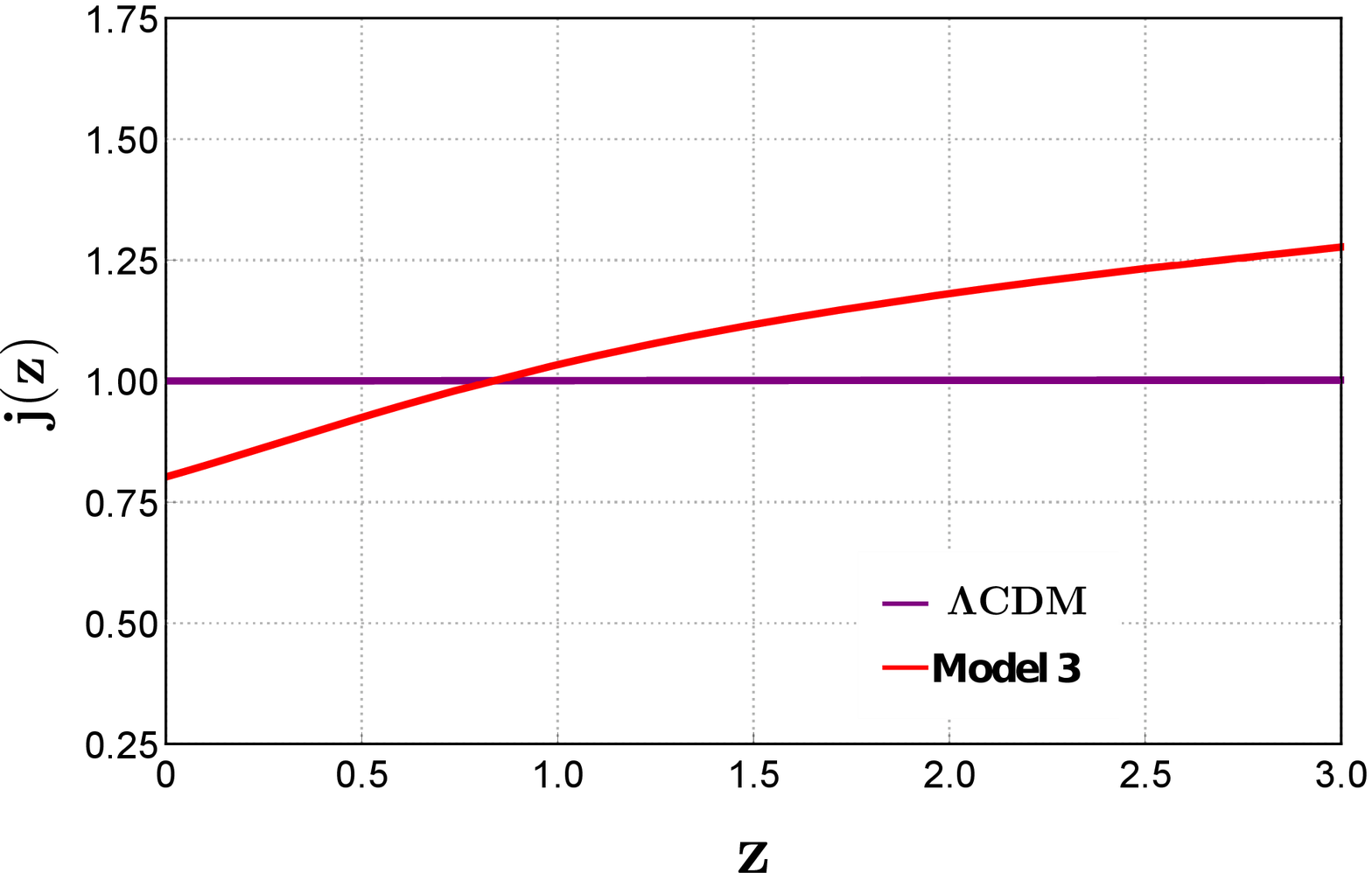}
            \caption{Plot of jerk parameter with respect to
                redshift.}\label{j(z)3}
        \end{minipage}
    \end{figure}
    \begin{figure}[!htb]
        \begin{minipage}{0.49\textwidth}
            \centering
            \includegraphics[scale=0.45]{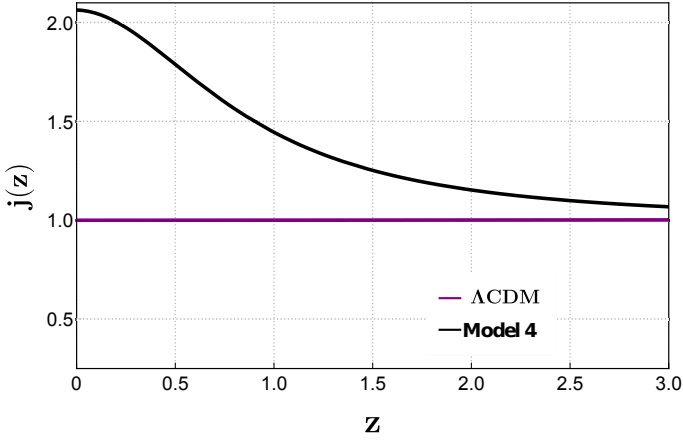}
            \caption{Plot of jerk parameter with respect to
                redshift.}\label{j(z)4}
        \end{minipage}
    \end{figure}
    
\newpage
\subsection{The snap parameter:}
The snap parameter, denoted as $s$, is a cosmological quantity that reveals important insights into the expansion of the universe. It arises from the fourth term of a Taylor series expansion of the scale factor around a reference time $t_0$. The expansion can be expressed as:

\begin{equation}
\begin{aligned}
\frac{a(t)}{a_0} & = 1 + H_0(t - t_0) - \frac{1}{2}q_0H_0^2(t - t_0)^2 + \frac{1}{6}j_0H_0^3(t - t_0)^3 \\ & +\frac{1}{24}s_0H_0^4(t - t_0)^4 + O\left[(t - t_0)^5\right].
\end{aligned}
\end{equation}

Here, $a(t)$ represents the scale factor at cosmic time $t$, $a_0$ is the scale factor at the reference time $t_0$, and $H_0$, $q_0$, $j_0$, and $s_0$ denote the Hubble constant, deceleration parameter, jerk parameter, and snap parameter, respectively. The snap parameter $s$ is defined as the fourth derivative of the scale factor normalized by the ratio of the first derivative of the scale factor to the scale factor itself:

\begin{equation}
s = \frac{1}{a}\frac{d^4a}{d\tau^4}\left(\frac{1}{a}\frac{da}{d\tau}\right)^{-4} = \frac{j - 1}{3\left(q - \frac{1}{2}\right)},
\end{equation}

Understanding the snap parameter provides valuable insights into the higher-order dynamics of cosmic expansion and its interplay with other cosmological parameters.

    \begin{figure}[!htb]
        \begin{minipage}{0.49\textwidth}
            \centering
            \includegraphics[scale=0.45]{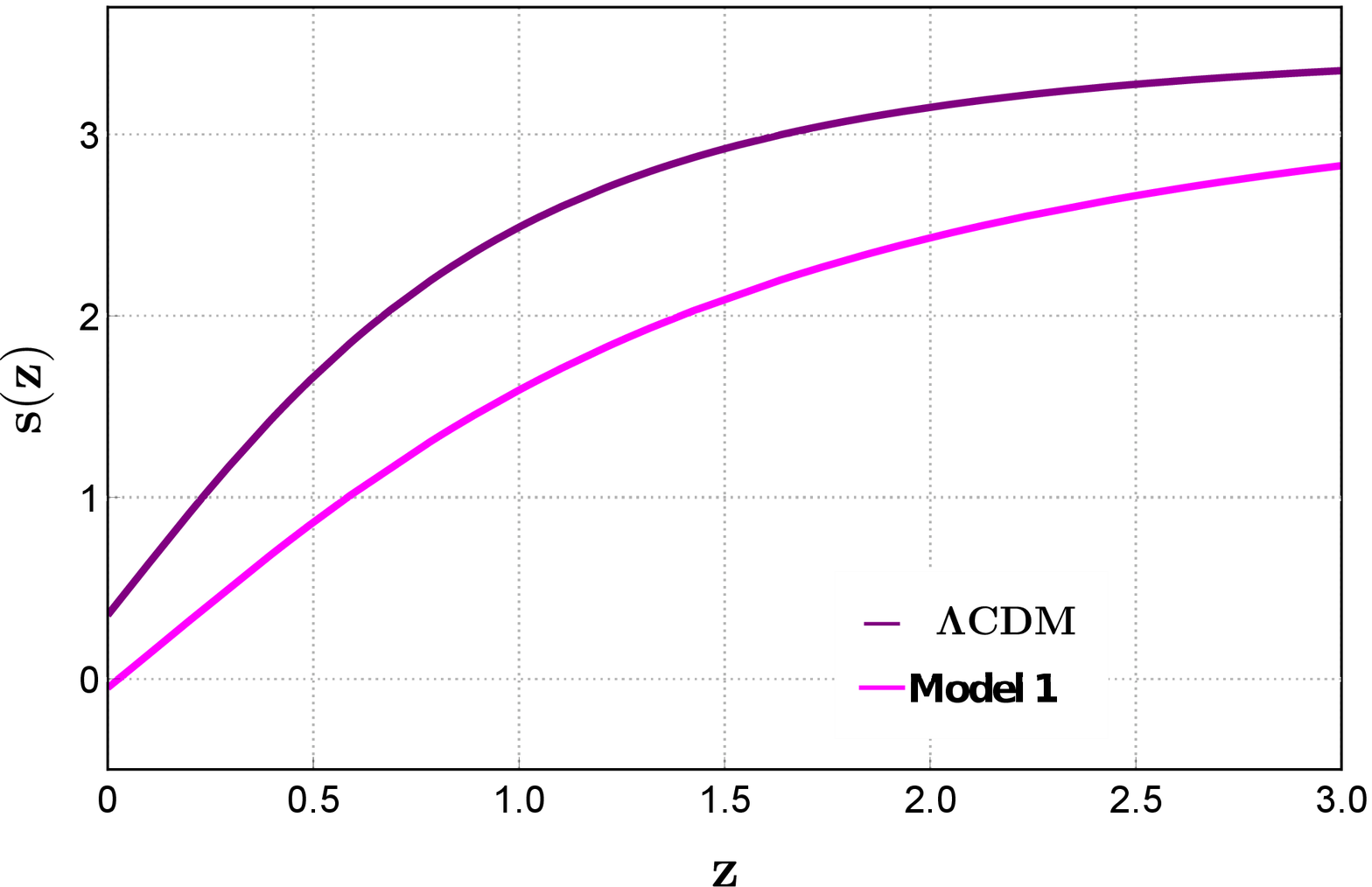}
            \caption{Plot of snap parameter with respect to
                redshift.}\label{s(z)1}
        \end{minipage}
    \end{figure}
    \begin{figure}[!htb]
        \begin{minipage}{0.49\textwidth}
            \centering
            \includegraphics[scale=0.45]{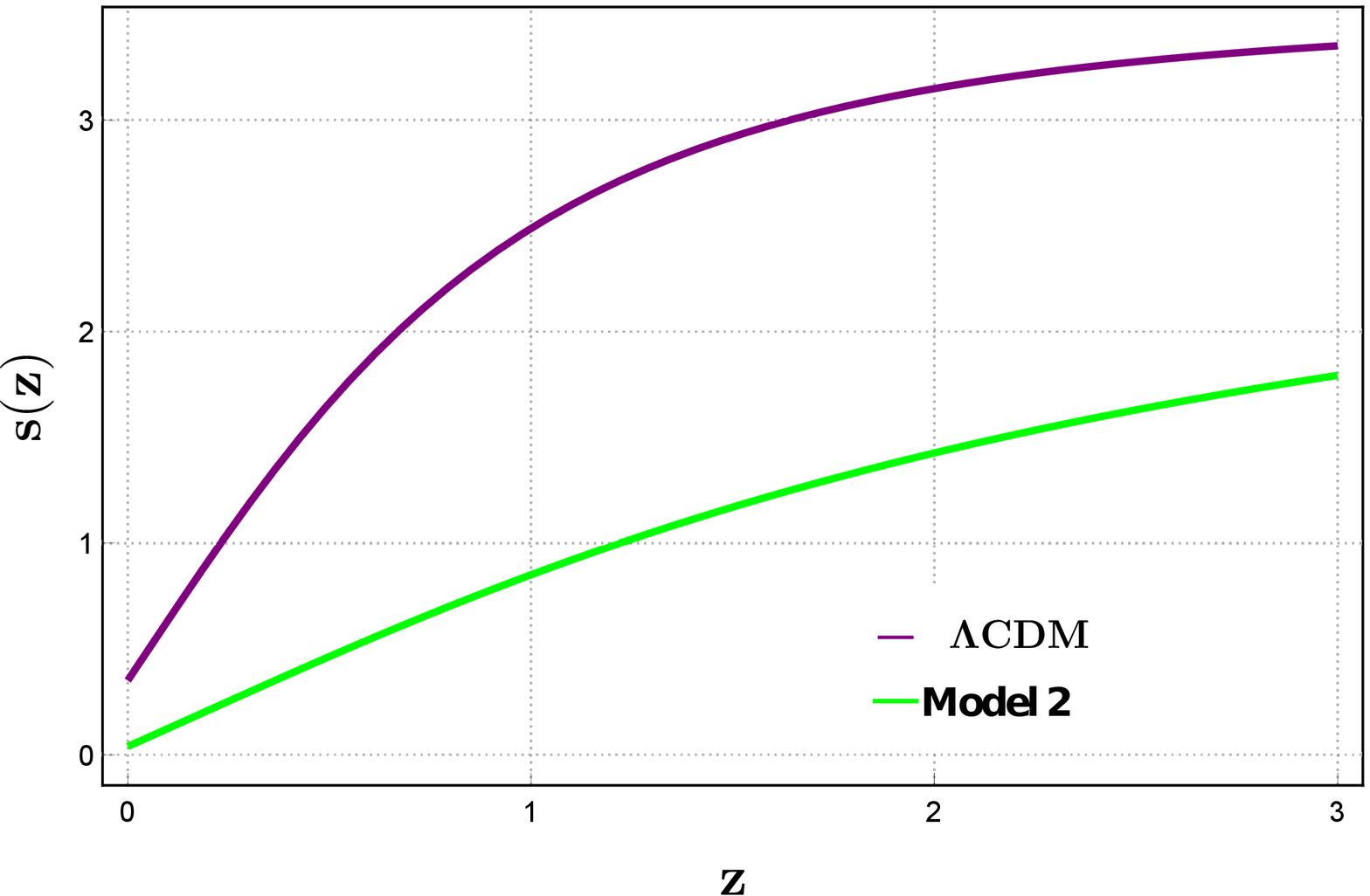}
            \caption{Plot of snap parameter with respect to
                redshift.}\label{s(z)2}
        \end{minipage}
    \end{figure}
    \begin{figure}[!htb]
        \begin{minipage}{0.49\textwidth}
            \centering
            \includegraphics[scale=0.45]{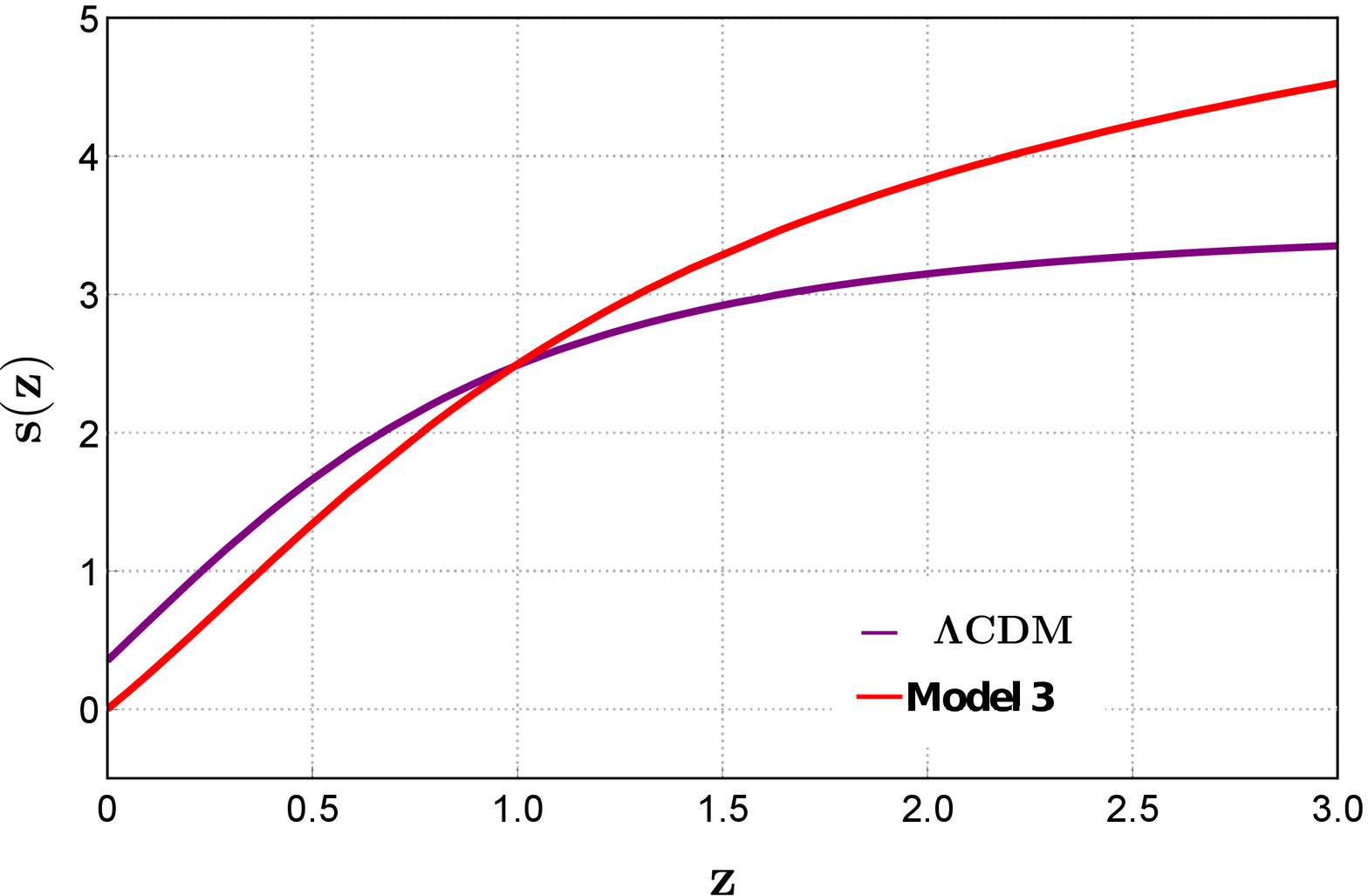}
            \caption{Plot of snap parameter with respect to
                redshift.}\label{s(z)3}
        \end{minipage}
    \end{figure}
    \begin{figure}[!htb]
        \begin{minipage}{0.49\textwidth}
            \centering
            \includegraphics[scale=0.45]{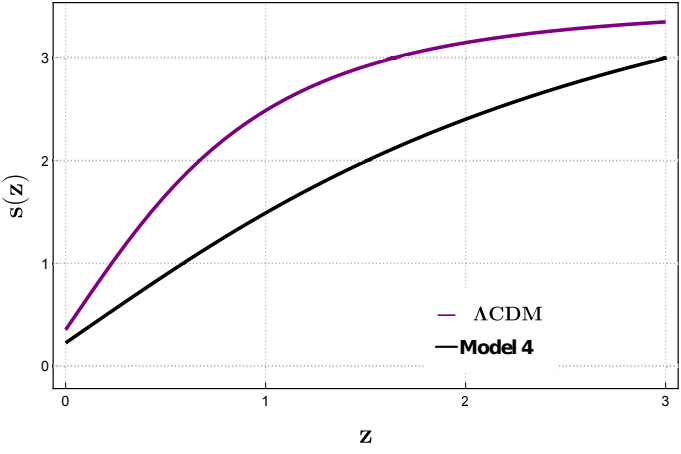}
            \caption{Plot of snap parameter with respect to
                redshift.}\label{s(z)4}
        \end{minipage}
    \end{figure}

    \newpage
    \section{Statefinder Diagnostic}\label{sec6}
    It is a mechanism generally applied to explore different DE models
    and comprehend their nature through higher-order derivatives of
    the scale factor. The statefinder diagnostic pair
    \cite{statefinder1,statefinder2,statefinder3,statefinder4} $\{r,
    s\}$ is dimensionless, which can be used to analyze the cosmic
    features of DE independent of models and can be computed by the
    expressions~~\cite{32:2002fz,33:2008xx}
    \begin{equation}
        r=\left[(1+z)\frac{dq}{dz}+q(2q+1)\right],~
        s=\frac{r-1}{3\left(q-\frac{1}{2}\right)},
    \end{equation}

    where $r$ and $q$ are the usual \textit{jerk} parameter and deceleration parameter respectively. Certain pairs commonly refer to the standard models of DE like $\{r,s\}=\{1,0\}$
    shows $\Lambda$CDM model while $\{r,s\}=\{1,1\}$ corresponds to
    the standard cold dark matter model (SCDM) in FLRW universe.
    Moreover, $(-\infty, \infty)$ indicates the Einstein static
    universe. In the $r-s$ plane, one can obtain quintessence-like and
    phantom-like models of the DE for $s > 0$ and $s < 0$,
    respectively. Also, the evolutionary process occurs (from phantom
    to quintessence) if the value deviates from the standard range
    ${r, s} = {1, 0}$. The value $\{q, r\}=\{-1,1\}$ is associated
    with the $\Lambda \mathrm{CDM}$ model whereas $\{q, r\}=\{0.5,1\}$
    gives SCDM model.

    \begin{figure}[!htb]
        \begin{minipage}{0.49\textwidth}
            \centering
            \includegraphics[scale=0.4]{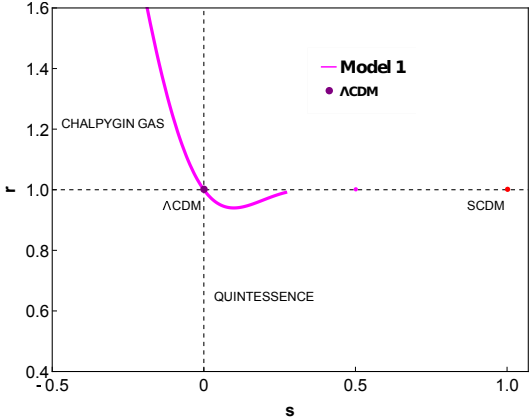}
            \caption{Behavior of $\{r, s\}$ profile of Model 1}\label{rs1}
        \end{minipage}
    \end{figure}
    \begin{figure}[!htb]
        \begin{minipage}{0.49\textwidth}
            \centering
            \includegraphics[scale=0.4]{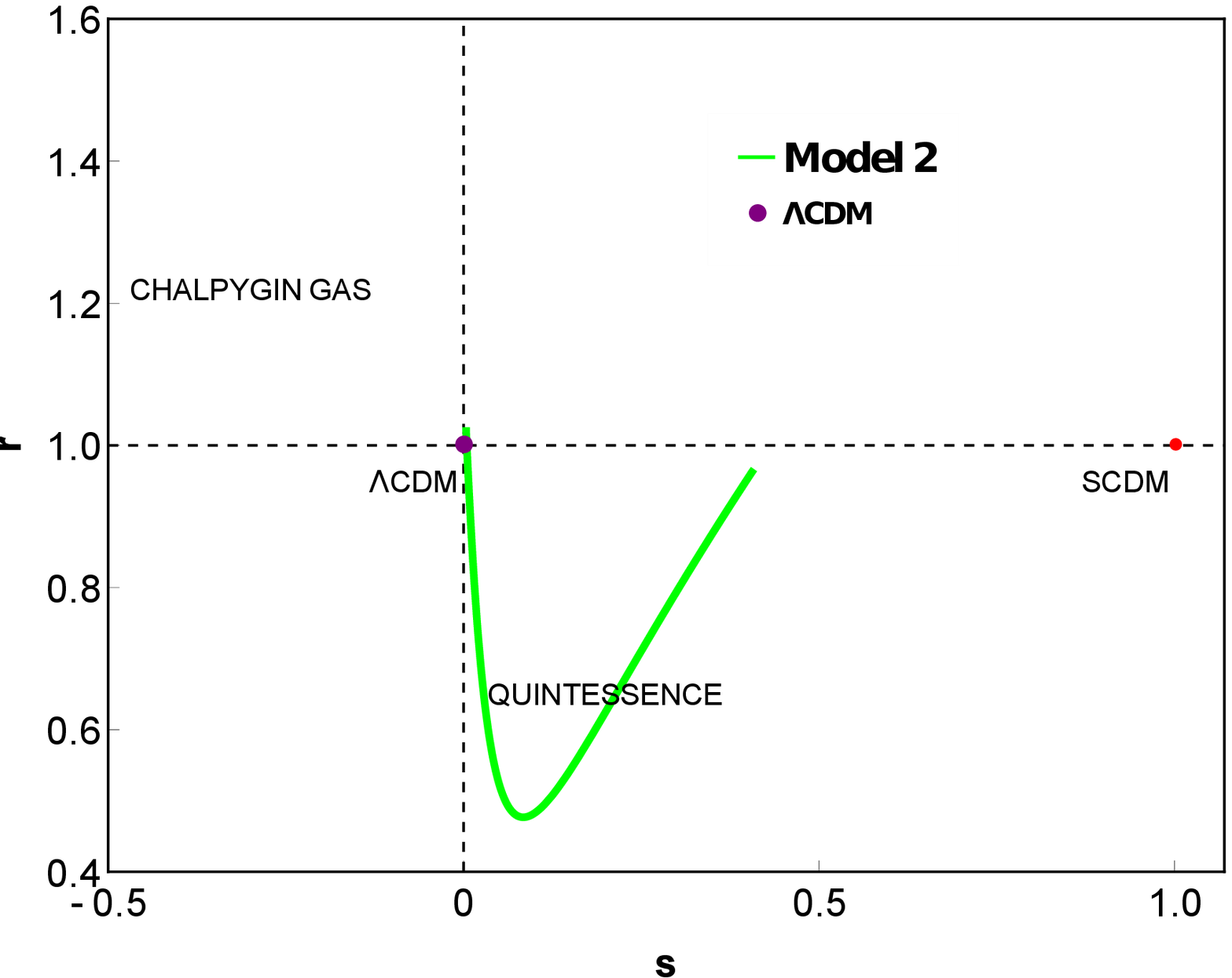}
            \caption{Behavior of $\{r, s\}$ profile of Model 2}\label{rs2}
        \end{minipage}
    \end{figure}
    \begin{figure}[!htb]
        \begin{minipage}{0.49\textwidth}
            \centering
            \includegraphics[scale=0.4]{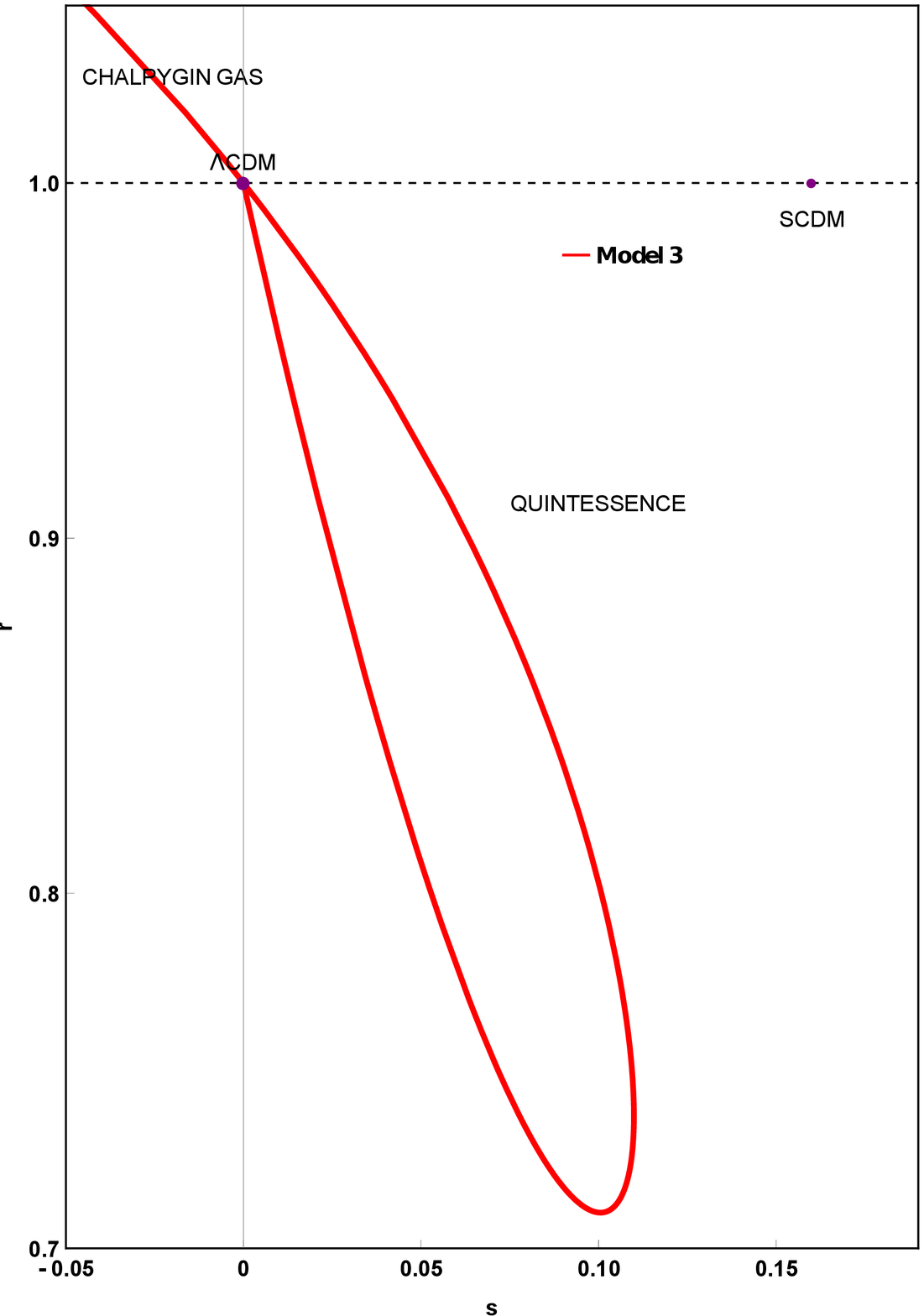}
            \caption{Behavior of $\{r, s\}$ profile of Model 3}\label{rs3}
        \end{minipage}
    \end{figure}
    \begin{figure}[!htb]
        \begin{minipage}{0.49\textwidth}
            \centering
            \includegraphics[scale=0.4]{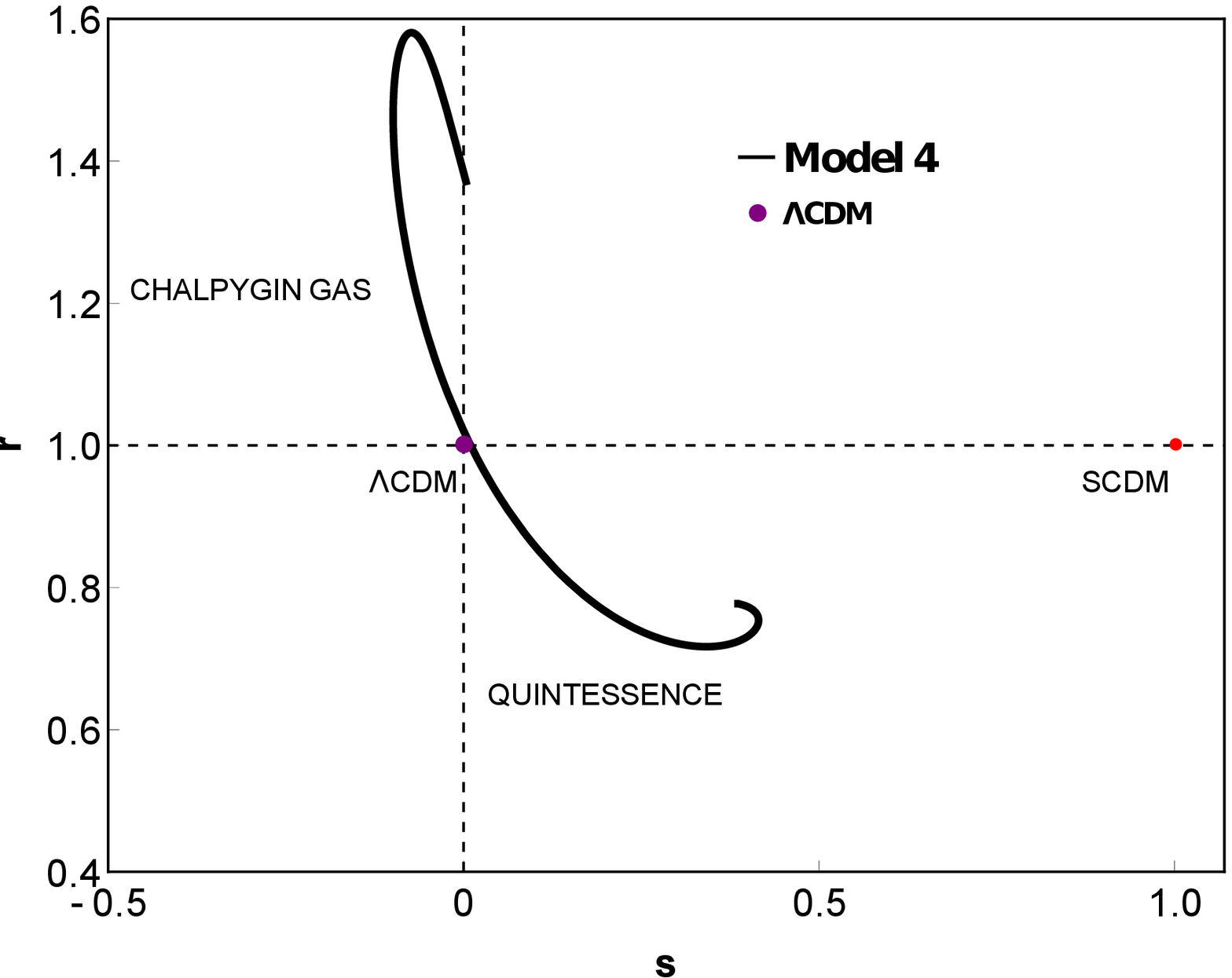}
            \caption{Behavior of $\{r, s\}$ profile of Model 4}\label{rs4}
        \end{minipage}
    \end{figure}
    \begin{figure}[!htb]
        \begin{minipage}{0.49\textwidth}
            \centering
            \includegraphics[scale=0.4]{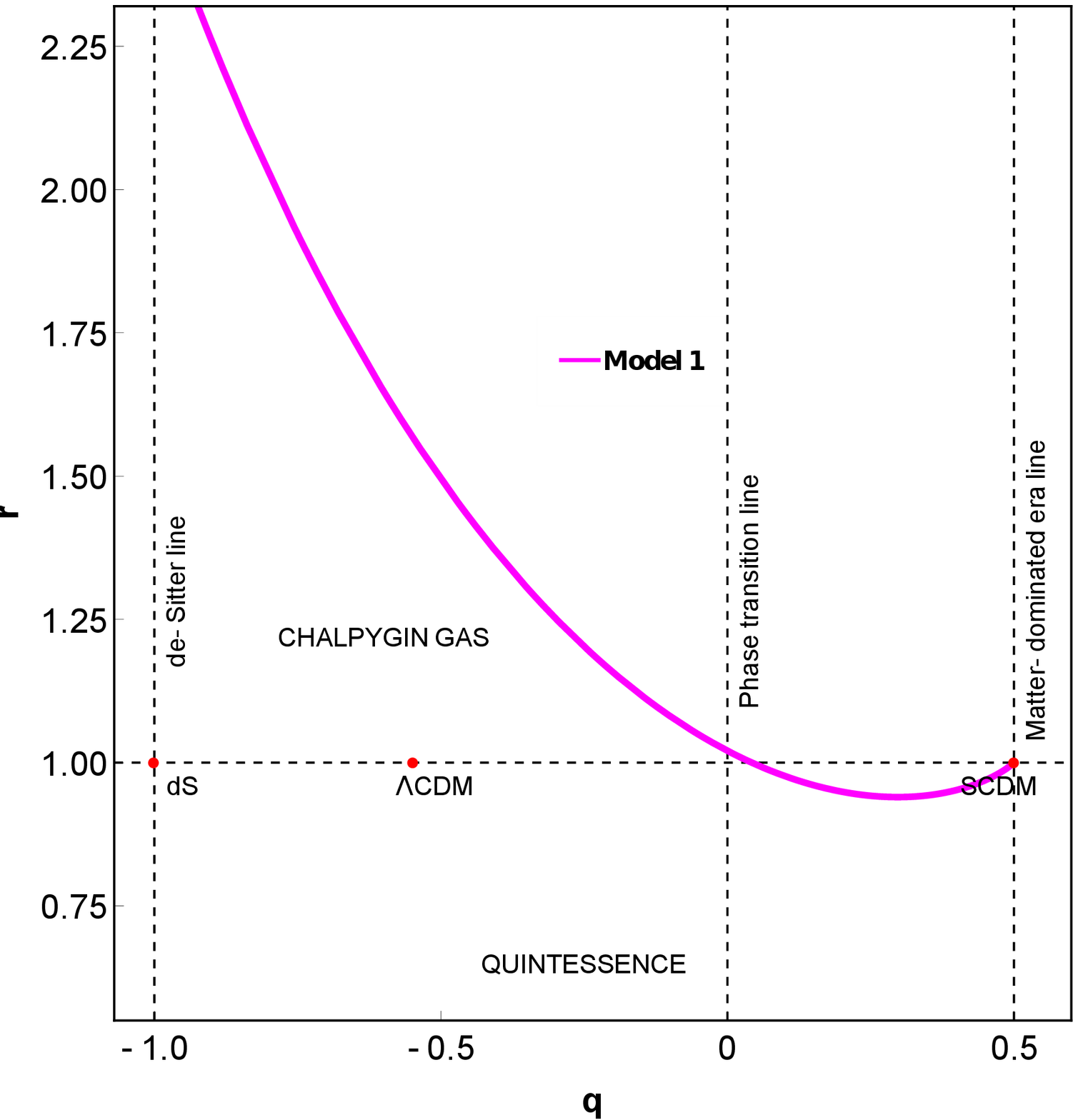}
            \caption{Behavior of $\{r, q\}$ profile of Model 1}\label{qr1}
        \end{minipage}
    \end{figure}
    \begin{figure}[!htb]
        \begin{minipage}{0.49\textwidth}
            \centering
            \includegraphics[scale=0.4]{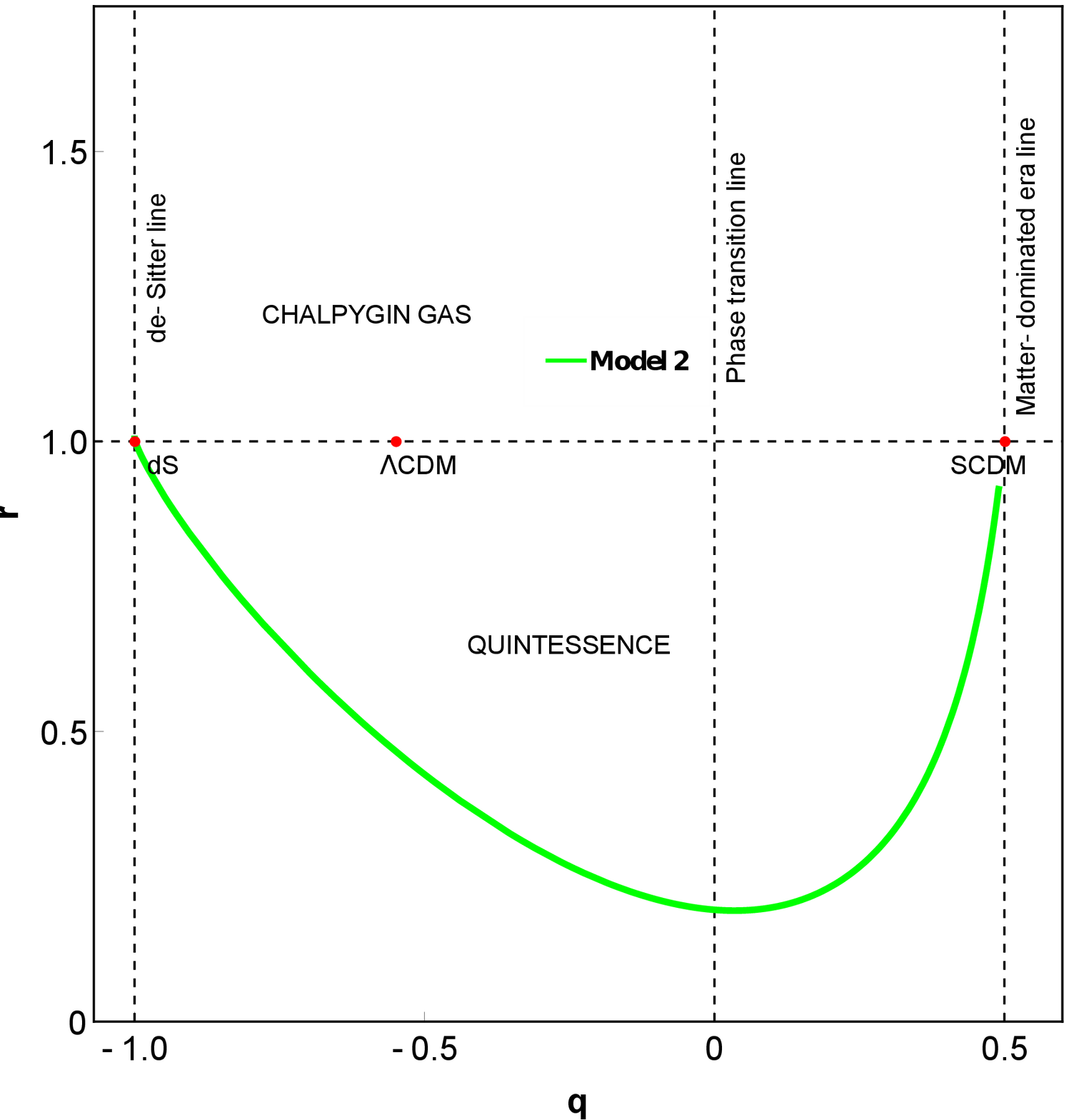}
            \caption{Behavior of $\{r, q\}$ profile of Model 2}\label{qr2}
        \end{minipage}
    \end{figure}
    \begin{figure}[!htb]
        \begin{minipage}{0.49\textwidth}
            \centering
            \includegraphics[scale=0.4]{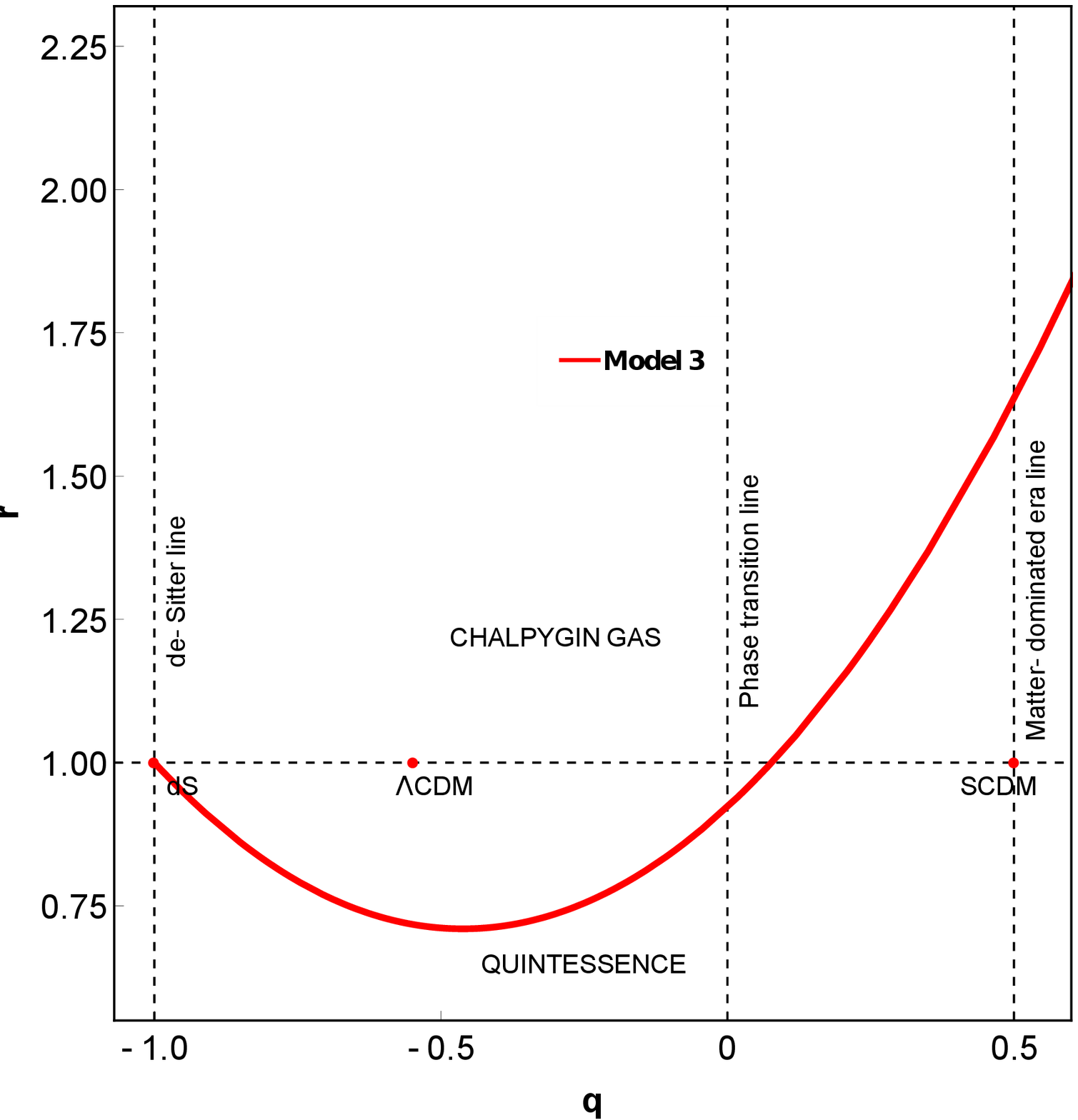}
            \caption{Behavior of $\{r, q\}$ profile of Model 3}\label{qr3}
        \end{minipage}
    \end{figure}
    \begin{figure}[!htb]
        \begin{minipage}{0.49\textwidth}
            \centering
            \includegraphics[scale=0.4]{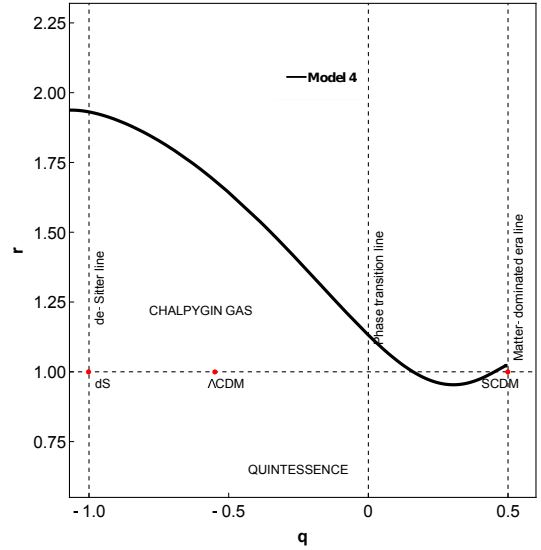}
            \caption{Behavior of $\{r, q\}$ profile of Model 4}\label{qr4}
        \end{minipage}
    \end{figure}

    \clearpage
    \section{$Om$ Diagnostic}\label{sec7}
    It refers to a geometrical formalism in which the Hubble parameter
    yields a null test for the $\Lambda$CDM model
    \cite{om1,om2,om3,om4}. The $Om$ diagnostic also efficiently
    differentiates several DE models from $\Lambda$CDM by the slope
    variation of $Om(z)$. A quintessence or phantom model can be
    acquired through either a positive or negative slope of the
    diagnostic parameter, respectively. Moreover, a constant slope
    with respect to redshift depicts a DE model corresponding to the
    cosmological constant. For a flat universe, one can define $Om(z)$
    as
    \begin{equation}
        Om(z)=\frac{\left(\frac{H(z)}{H_0}\right)^2-1}{(1+z)^3-1}.
    \end{equation}
    Many authors have explored the behavior of various DE models from the viewpoint of statefinder and $Om$ diagnostics parameter~\cite{34:2020wgo,35-Mohammadi:2010you,36:2012wc,37:2010nk}. This diagnostic involves only the first-order temporal derivative
    compared to the statefinder diagnosis \cite{40}. It can also be
    subjected to the Galileons models \cite{41, 42}.
    \begin{figure}[!htb]
        \begin{minipage}{0.49\textwidth}
            \centering
            \includegraphics[scale=0.4]{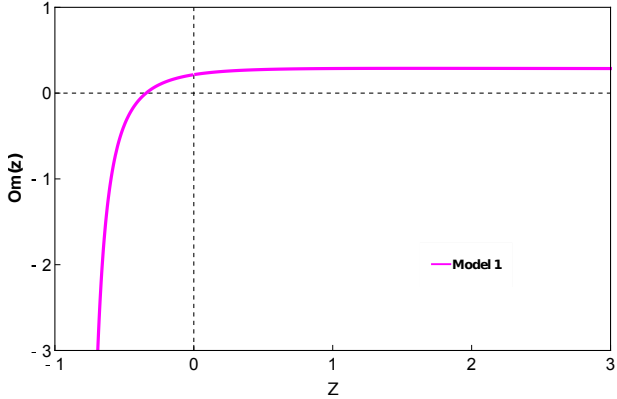}
            \caption{Evolution of $Om$ profile of Model 1}\label{om1}
        \end{minipage}
    \end{figure}
    \begin{figure}[!htb]
        \begin{minipage}{0.49\textwidth}
            \centering
            \includegraphics[scale=0.4]{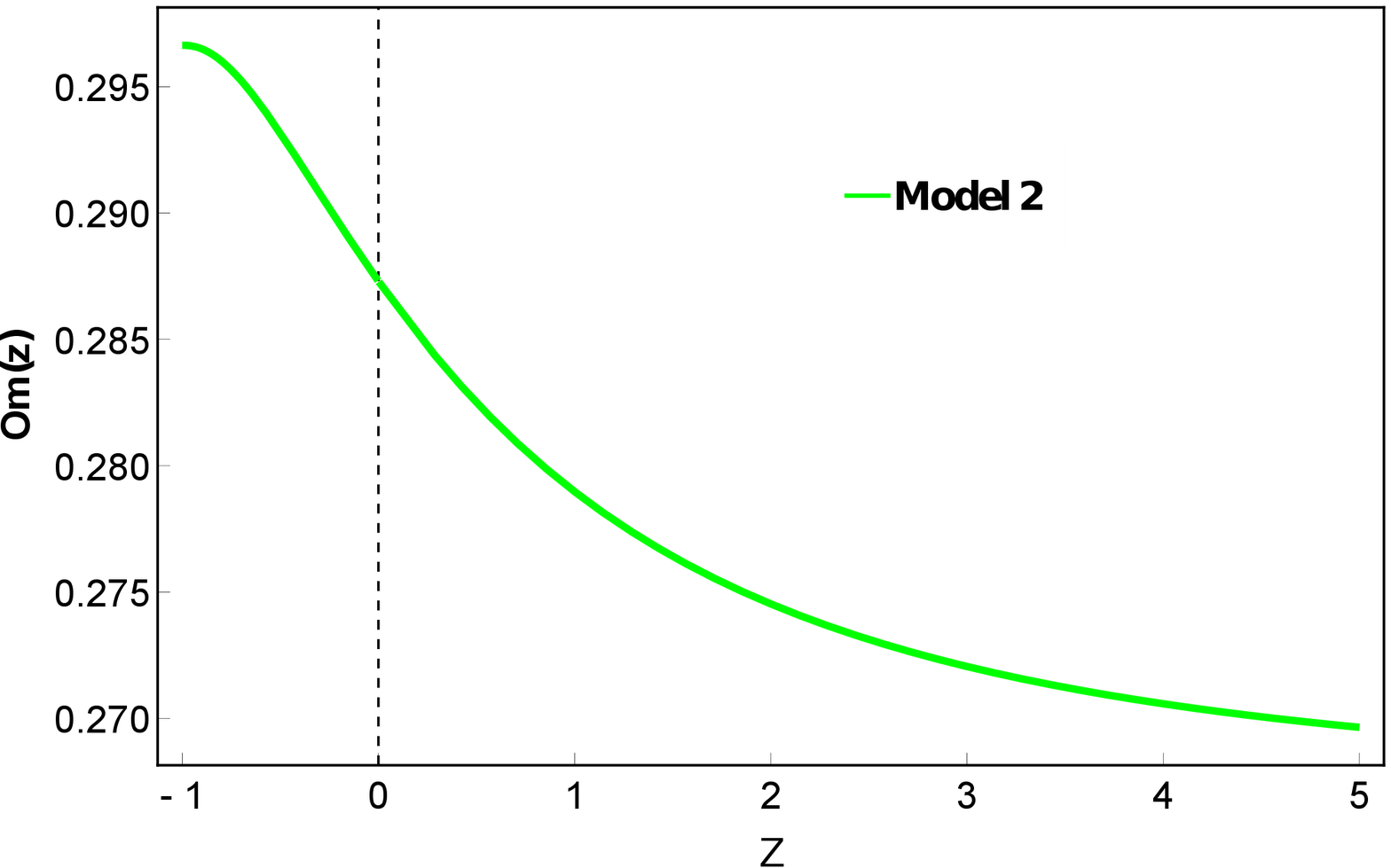}
            \caption{Plot of $Om$ profile of Model 2}\label{om2}
        \end{minipage}
    \end{figure}
    \begin{figure}[!htb]
        \begin{minipage}{0.49\textwidth}
            \centering
            \includegraphics[scale=0.4]{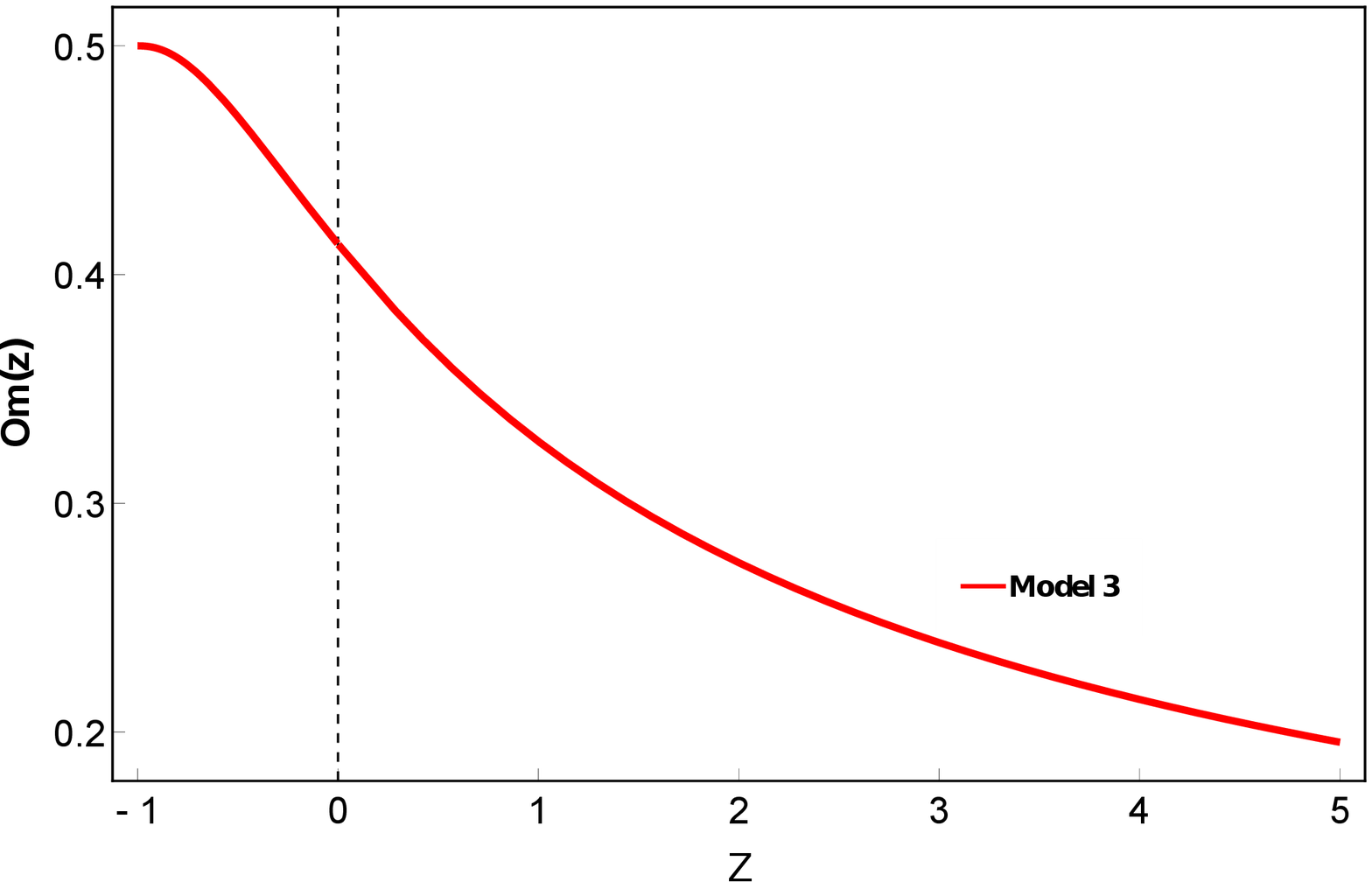}
            \caption{Plot of $Om$ profile of Model 3}\label{om3}
        \end{minipage}
    \end{figure}
    \begin{figure}[!htb]
        \begin{minipage}{0.49\textwidth}
            \centering
            \includegraphics[scale=0.4]{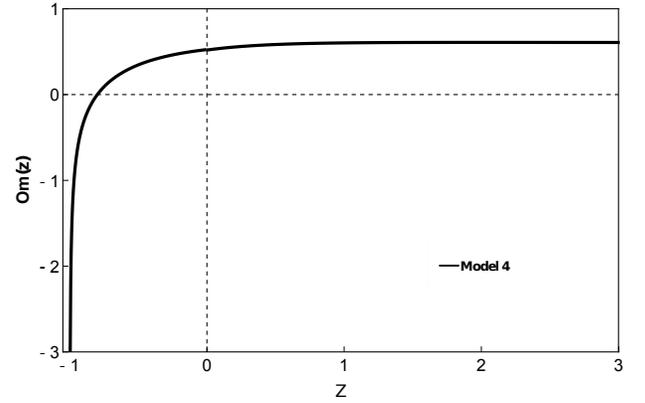}
            \caption{Plot of $Om$ profile of Model 4}\label{om4}
        \end{minipage}
    \end{figure}

\section{Information Criteria}\label{sec8}
To investigate the feasibility of any model, one needs first to comprehend the study of information criteria (IC). The Akaike Information Criteria (AIC) \cite{AIC1} is simply applied to all ICs. Because the AIC is an approximate minimization of the Kullback-Leibler information, it is an asymptotically unbiased estimator of Kullback-Leibler information. The AIC Gaussian estimator could be expressed as \cite{AIC2,AIC3,AIC4,AIC5} $\text{AIC}=-2\ln ({\cal
L}_{max})+2\kappa+\frac{2\kappa(\kappa+1)}{N-\kappa-1}$. where ${\cal L}_{max}$ is the maximum likelihood function, $\kappa$ is the total number of free parameters of any model, and $N$ is the total number of data points utilized. Because $N\gg 1$, is the assumption for the models, the aforementioned formula transforms to the original AIC like $\text{AIC}=-2\ln ({\cal
L}_{max})+2\kappa$. If a set of models is provided, the IC value deviations can be reduced to $\triangle\text{AIC}=\text{AIC}_{model}-\text{AIC}_{min}=\triangle\chi^{2}_{min}+2\triangle\kappa$. The more favorable range of $\triangle\text{AIC}$ are mentioned following $\triangle\text{AIC}$ is $(0,2)$. The low favorable range of
$\triangle\text{AIC}$ is $(4,7)$, while
$\triangle\text{AIC}>10$ provides less support model.\\

\begin{table}[H]
\begin{center}
\begin{tabular}{|c|c|c|c|c|}  
\hline 
Model & $\chi_{min}^{2}$ & $\chi_{red}^{2} $ & $AIC$ & $\Delta AIC $\\ 
\hline $\Lambda$CDM Model & 1755.56 & 0.9476 & 1761.56 & 0  \\
\hline Model 1 & 1751.10  & 0.9462  & 1763.16 & 1.54\\ 

\hline  Model 2 & 1750.24  & 0.9412 & 1762.24 & 0.68  \\

\hline  Model 3 & 1750.02 & 0.9432 & 1762.02 & 0.46 \\ 

\hline Model 4 & 1751.02 & 0.9456 & 1763.02 & 1.46 \\ 
\hline
\end{tabular}
\caption{Summary of the $ {{\chi}^2_{min}}$, ${{\chi}^2_{red}}$,  $AIC$  and $\Delta AIC$.} \label{table}
\end{center}
\end{table}

\section{Results}\label{results}

\paragraph{ \bf Decceleration Parameter}
The behavior of the DP of Model 1 (M1) is shown in Figure~\ref{q(z)1}. The DP closely resembles that of the $\Lambda$CDM model, indicating a similar expansion behavior at both high and low redshifts. At low redshift, M1 transitions into a de Sitter phase, where the universe undergoes accelerated expansion. This behavior suggests that the late-time universe, described by M1, follows a similar cosmic evolution as the standard $\Lambda$CDM model. This behavior can be understood in terms of a dominant cosmological constant driving the accelerated expansion. Moving on to Model 2 (M2) shown in Figure~\ref{q(z)2}, systematic differences in the deceleration parameter are observed at both low and high redshifts compared to $\Lambda$CDM. Despite these differences, M2 also terminates in a de Sitter phase, indicating a late-time universe with accelerated expansion. The physics in the late-time universe described by M2 can be understood in terms of modified gravity theories or additional dark energy components that contribute to the observed deviations from $\Lambda$CDM. For Model 3 (M3) shown in Figure~\ref{q(z)3}, we observe that its deceleration parameter behaves very closely to $\Lambda$CDM at high redshifts. However, at low redshifts, M3 exhibits a slower-accelerated evolution with a deceleration parameter of $q(-1)\approx -0.7$. This slower acceleration implies a delayed transition to the de Sitter phase compared to $\Lambda$CDM. This behavior can be related to the presence of additional matter components or modified gravitational theories that affect the late-time cosmic expansion. Lastly, Model 4 (M4) displays in Figure~\ref{q(z)4}, a distinctive behavior in its deceleration parameter. At low redshift, M4 exhibits super-accelerated evolution for some time before eventually entering the de Sitter phase. This super-acceleration implies an even more rapid expansion compared to $\Lambda$CDM. The late-time universe described by M4 may involve exotic matter or modifications to the gravitational theory that lead to enhanced acceleration.\\\\

\paragraph{ \bf Jerk Parameter}
Starting with Model 1 (M1), as shown in Fig.~\ref{j(z)1}, we observe that its jerk parameter has an approximate value of $0.79$ at $j(0)$. This value suggests a moderate change in acceleration and behavior that closely resembles the $\Lambda$CDM model. Therefore, M1 provides insights into the late-time cosmic evolution that aligns with the standard model. Moving on to Model 2 (M2), depicted in Fig.~\ref{j(z)2}, we can see that its estimated jerk parameter value is $0.42$ at $j(0)$. This lower value indicates a more gradual transition in acceleration compared to the $\Lambda$CDM model. Consequently, M2 introduces distinct physics and dynamics, potentially leading to alternative understandings of the late-time universe. Model 3 (M3) exhibits an approximate value of $0.78$ at $j(0)$, as shown in Fig.~\ref{j(z)3}. At high redshifts, the behavior of M3 closely resembles that of the $\Lambda$CDM model. However, at low redshifts, a slower-accelerated evolution occurs, with $q(-1)\approx -0.7$. This distinct behavior at different redshift ranges offers insights into alternative physics and deviations from the standard model in the late-time universe. Lastly, Fig.~\ref{j(z)4} represents Model 4 (M4), which exhibits super-accelerated evolution at low redshifts. The jerk parameter in M4 has numerical values more than twice as large as those of the $\Lambda$CDM model, with $j(0)\approx 2.1$. This significant difference indicates unique physics governing the late-time cosmic dynamics within this model. The enhanced acceleration observed in M4 can arise from various scenarios, such as the influence of exotic matter or modifications to the theory of gravity.\\\\

\paragraph{\bf The snap Parameter:} 
The progression of each model's snap parameter, $s(z)$, versus redshift is depicted in Figures \ref{s(z)1}, \ref{s(z)2}, \ref{s(z)3}, and \ref{s(z)4}. These figures allow us to compare the behavior of the snap parameter for each model with that of the $\Lambda$CDM model and understand their implications in the cosmological context. At both high and low redshifts, we observe significant deviations between the numerical values of the snap parameter for each model and those of the $\Lambda$CDM model. Specifically, at high redshifts, Models 1, 2, and 4 predict lower values for $s(z)$ compared to $\Lambda$CDM, indicating a different rate of cosmic acceleration. In contrast, Model 3 predicts higher values for $s(z)$ than $\Lambda$CDM, suggesting a more pronounced acceleration at those redshifts. Identifying the present value of the snap parameter, $s(0)$, becomes a critical test for these models. Model 1 predicts a value of $s(0) \approx -0.1$, indicating a small negative deviation from the $\Lambda$CDM behavior. Model 2, on the other hand, estimates $s(0) \approx 0.9$, suggesting a significant positive deviation from $\Lambda$CDM and indicating a distinct late-time cosmic evolution. Model 3 predicts $s(0) \approx 0.9$, again indicating a positive deviation from $\Lambda$CDM but with a slightly lower magnitude. Finally, Model 4 estimates $s(0) \approx 0.7$, small positive deviation from the $\Lambda$CDM. These deviations in the snap parameter have important implications for our understanding of the late-time universe. They signify that the dynamics of cosmic acceleration and the nature of dark energy are not solely determined by the $\Lambda$CDM model but can exhibit significant variations.\\\\

\paragraph{\bf The $\{r, s\}$ Parameter:} 
The evolution of the ${r, s}$ profilr for the four models (M1, M2, M3, and M4) is illustrated in Figs.~\ref{rs1}, \ref{rs2}, \ref{rs3}, and \ref{rs4}. Starting with Model M1, we observe that it initially takes values in the region $r < 1$, $s > 0$, which corresponds to the quintessence domain. However, as the evolution progresses, M1 crosses the fixed point $\{r, s\} = \{1, 0\}$ and transitions into the region where $\Lambda$CDM resides, characterized by $r > 1$ and $s < 0$. This behavior signifies a transition from quintessence to a Chaplygin gas-like domain during the late-time universe. Throughout its evolution, Model M2 consistently exhibits values within the region characterized by $r < 1$ and $s > 0$, indicating the quintessence domain. Notably, this evolution culminates in a fixed point $\{r, s\} = \{1, 0\}$ that aligns with the well-established $\Lambda$CDM paradigm. This suggests a remarkable connection between Model M2's trajectory and the well accepted cosmological framework, potentially providing insights into the compatibility and convergence of these different theoretical descriptions. The behavior exhibited by Model M3 is distinct and intriguing. Initially, it is characterized by values within the region marked by $r > 1$ and $s < 0$, indicative of its association with the domain typically attributed to Chaplygin gas. As the model's evolution unfolds, a noteworthy pattern emerges. Model M3 traverses an intermediate fixed point, precisely $\{r, s\} = \{1, 0\}$, which marks a transition. This transition leads the model into the quintessence region. Remarkably, the trajectory then proceeds to culminate in the fixed point, $\{r, s\} = \{1, 0\}$, which notably corresponds to the established $\Lambda$CDM paradigm. This distinctive behavior provides a unique lens through which to understand the interplay between different cosmological paradigms and their compatibility with one another. Interestingly, the $\{r, s\}$ parameter of Model M4 exhibits identical behavior to that of Model M1. It also starts in the quintessence region, crosses the fixed point $\{r, s\} = \{1, 0\}$, and transitions into the region associated with the Chaplygin gas-like behavior. This similarity in behavior between M1 and M4 suggests comparable dynamics and evolution in terms of the $\{r, s\}$ parameter. The evolution of the $\{r, s\}$ parameter provides valuable information about the nature of dark energy and the underlying physics in the late-time universe. The transition from quintessence to Chaplygin gas-like behavior, as observed in M1 and M4, indicates a change in the dominant components driving the cosmic evolution. The persistent association of M2 with the quintessence domain suggests the prevalence of normal matter-like dynamics throughout the late-time universe. The transformation from Chaplygin gas-like behavior to quintessence, as seen in M3, hints at the interplay between different physical mechanisms governing the universe's expansion.\\\\
 
\paragraph{\bf The $\{r, q\}$ Parameter:}
The $\{q, r\}$ parameter plots (Figures \ref{qr1}, \ref{qr2}, \ref{qr3}, and \ref{qr4}) for models M1, M2, M3, and M4 provide additional valuable information regarding the behavior of these models. Starting with M1, its $\{q, r\}$ profile, it's started with a fixed point $\{r, q\} = \{1, 0.5\}$ corresponds to SCDM. As model evolves it takes values in the range $q > 0$ and $r < 1$, indicating a quintessence region. This signifies a phase of cosmic deceleration dominated by quintessence-like matter. As the evolution progresses, M1 transitions into the range $q < 0$ and $r > 1$, which corresponds to the Chaplygin gas region. Here, the acceleration of the universe is driven by a Chaplygin-like component. Eventually, M1 deviates towards the de Sitter point $\{-1, 1\}$, representing a phase of accelerated expansion with a cosmological constant-like behavior. M2 started with a fixed point $\{r, q\} = \{1, 0.5\}$ corresponds to SCDM then it remains within the quintessence region throughout its evolution, ending at the de Sitter point $\{-1, 1\}$. This indicates that M2 experiences a continuous phase of cosmic acceleration dominated by quintessence-like matter. Moving on to M3, its $\{q, r\}$ parameter initially takes values in the range $q > 0$ and $r > 1$, indicating a Chaplygin gas region. In this phase, the expansion of the universe is driven by a Chaplygin-like component. However, as the evolution proceeds, M3 transitions into the range $q < 0$ and $r < 1$, which corresponds to the quintessence region. Here, the acceleration of the universe is dominated by quintessence-like matter. Similar to M1 and M2, M3 ultimately reaches the de Sitter point $\{-1, 1\}$, signifying a phase of accelerated expansion with a cosmological constant-like behavior. Regarding M4, its $\{q, r\}$ parameter exhibits behavior similar to that of M1. It starts with the fixed SCDM point in the quintessence region, then transitions to the Chaplygin gas region, and finally reaches the de Sitter line . This indicates that M4 undergoes a similar sequence of cosmic acceleration phases, with a super-accelerated phase at low redshifts before entering the de Sitter phase. \\\\

\paragraph{\bf Om Diagnostic Parameter:}
The behavior of $Om(z)$ versus redshift $z$ for each model is depicted in Figures~\ref{om1}, \ref{om2}, \ref{om3}, and \ref{om4}, respectively. Starting with Model 1 (M1) and Model 4 (M2), we observe that they exhibit similar behavior at high redshifts $z > 0$. Both models show a quintessence-like behavior as $Om(z)$ decreases with increasing redshift. However, as the redshift abruptly declines and becomes negative, indicating the transition to the late-time universe, both models enter the phantom region. This behavior signifies a shift towards accelerated expansion and the presence of phantom-like characteristics in cosmic evolution. In contrast, Model 2 (M2) and Model 3 (M2)  maintains positive values of $Om(z)$ throughout both high and low redshifts, indicating consistent support for a quintessence model. This behavior implies the dominance of normal matter and quintessence-like dynamics throughout the late-time universe. Overall, the comparative analysis of the Om diagnostic parameter in the different models sheds light on the cosmological context and the physics involved in the late-time universe. The transition from quintessence-like behavior to the phantom region reflects the dynamics of accelerated expansion, the nature of dark energy, and the overall evolution of the universe at late times.\\\\

The behavior of $Om(z)$ versus redshift $z$ for each model is depicted in Figures~\ref{om1}, \ref{om2}, \ref{om3}, and \ref{om4}, respectively. Starting with Model 1 (M1) and Model 2 (M2), we observe that they exhibit similar behavior at high redshifts $z > 0$. Both models show a quintessence-like behavior as $Om(z)$ decreases with increasing redshift. However, as the redshift abruptly declines and becomes negative, indicating the transition to the late-time universe, both models enter the phantom region. This behavior signifies a shift towards accelerated expansion and the presence of phantom-like characteristics in cosmic evolution. In contrast, Model 2 (M2) maintains positive values of $Om(z)$ throughout both high and low redshifts, indicating consistent support for a quintessence model. This behavior implies the dominance of normal matter and quintessence-like dynamics throughout the late-time universe. Lastly, Model 4 (M4) exhibits similar behavior to M1 and M2. It initially displays quintessence-like behavior at high redshifts, characterized by decreasing values of $Om(z)$. As the redshift decreases, M4 also transitions into the phantom region, signifying accelerated expansion and the presence of phantom-like effects. This behavior aligns with the cosmological context of the late-time universe. Overall, the comparative analysis of the Om diagnostic parameter in the different models sheds light on the cosmological context and the physics involved in the late-time universe. The transition from quintessence-like behavior to the phantom region reflects the dynamics of accelerated expansion, the nature of dark energy, and the overall evolution of the universe at late times.\\\\

\paragraph{ \bf Information Criteria}
From Table: \ref{table} by comparing the $\chi_{min}^{2}$ values, we observe that Models 1-4 have smaller values compared to the $\Lambda$CDM model. This suggests that Models 1-4 provide a better fit to the observed data than the $\Lambda$CDM model. Comparing the $\chi_{red}^{2}$ values, we observe that Models 1-4 have slightly smaller values compared to the $\Lambda$CDM model. The reduced chi-squared statistic is a measure of how well a model fits the data, taking into account the number of degrees of freedom. Smaller $\chi_{red}^{2}$ values indicate better fits to the data. Based on these values, we could say that Models 1-4 have slightly better fits the data compared to the $\Lambda$CDM model, as they have lower $\chi_{red}^{2}$ values. Looking at the AIC values, we can see that Models 1-4 have slightly smaller values compared to the $\Lambda$CDM model. This indicates that Models 1-4 have a better trade-off between goodness-of-fit and model complexity, as AIC penalizes models with a larger number of parameters. The $\Delta AIC$ values represent the difference in AIC between each model and the model with the lowest AIC (in this case, the $\Lambda$CDM model). Smaller $\Delta AIC$ values indicate models that are closer in goodness-of-fit to the best model. In this comparison, Models 1, 2, and 3 have lower $\Delta AIC$ values than Model 4, indicating that Models 1, 2, and 3 are relatively closer in fit to the best model the $\Lambda$CDM model. Based on these values, we can interpret that Models 1, 2, and 3 exhibits better goodness-of-fit compared to both the $\Lambda$CDM model and Model 4. However, it's important to consider additional factors, such as the physical interpretability, robustness, and consistency of these models, before drawing definitive conclusions about their superiority over the $\Lambda$CDM model. These values provide a starting point for model comparison, and further analysis and investigation are necessary to fully evaluate the implications of these models in the context of fundamental physics.\\\\

    \section{Discussions and Conclusions}\label{sec9}

    Herein, we have studied the accelerated expansion
    scenario in a flat FLRW universe composed of DE, radiation, and DM.
    We have attempted to construct a viable dark energy model
    which can describe the whole evolutionary history of the Universe.
    For this purpose, we have taken parametrization of the
    deceleration parameter instead of assuming parameterization of EOS
    parameter. Here, we have considered three parametrizations (model
    1-model 3) and also proposed one new parametrization of the
    deceleration parameter. The advantage of these parametrizations is
    that they can provide finite results without considering any
    particular gravitational theory and provide early deceleration and
    late-time acceleration. We have chosen $q(z)$ containing only
    three parameters $q_0$, $q_1$ and $q_2$. Since the choice of $q(z)$
    is quite arbitrary, so one takes more than three terms for the
    parametrization of $q(z)$. But in that case, it may be difficult
    to constrain the parameters using existing observational data.
    A statistical analysis has been made to get the best-fit values
    of the model parameters by the MCMC method using $H(z)$ datasets,
    Pantheon datasets, BAO datasets, and CMB Distant Prior . We have analyzed the
    evolutionary trajectories of deceleration, jerk, and snap parameters by using the
    best-fit values of the model parameters. Moreover, the best-fit values of the model parameters obtained are used to plot the statefinder and $Om$ diagnostics. The information criteria have been employed to examine the models' viability. We examined all of the models and the $\Lambda$CDM model (the basic reference) to figure out which was more probable than the others. From Table: \ref{table}\\\

The results obtained by analyzing all the geometrical parameters are summarized as follows:

\begin{itemize}

\item The deceleration parameter for Model 1 (M1) closely resembles the behavior of the $\Lambda$CDM model both at high and low redshifts, ultimately transitioning into a de Sitter phase at low redshift. Model 2 (M2) exhibits systematic differences compared to $\Lambda$CDM at both high and low redshifts but also ends in a de Sitter phase. Model 3 (M3) behaves similarly to $\Lambda$CDM at high redshifts, but at low redshifts, it undergoes a slower-accelerated evolution with a deceleration parameter of $q(-1)\approx -0.7$. Finally, Model 4 (M4) displays super-accelerated expansion at low redshift for a certain period before entering the de Sitter phase.\\\\

\item The jerk parameter, We observe that M1 has an approximate value of 0.79 at $j(0)$, M2 has an estimated value of 0.42, and M3 has a value of approximately 0.78 at $j(0)$. In contrast, M4 exhibits jerk values more than twice that of the $\Lambda$CDM model, with $j(0)\approx 2.1$. These variations in the jerk parameter shed light on the dynamics and acceleration patterns of the different models.\\\\

\item The snap parameter, we find that M1, M2, and M3 predict lower values compared to $\Lambda$CDM at high redshifts, while M4 predicts higher values. The present value of the snap parameter, $s(0)$, becomes a critical test. M1 predicts $s(0) \approx -0.1$, M2 estimates $s(0) \approx 0.9$, M3 predicts $s(0) \approx 0.9$, and M4 suggests $s(0) \approx 0.7$. These snap parameter values contribute to our understanding of cosmic acceleration and the underlying physics involved.\\\\

\item The statefinder diagnostic provides additional insights into the nature of the models. For M1 and M4, the statefinder parameters lie in the region $r < 1$, $s > 0$, indicating quintessence-like behavior. As they cross the fixed point ${r,s}={1,0}$, the models enter the region where $\Lambda$CDM resides, characterized by $r < 1$ and $s < 0$, resembling the nature of Chaplygin gas. M2, throughout its evolution, exhibits values in the region $r < 1$, $s > 0$, reflecting quintessence-like behavior. On the other hand, M3 starts with values in the region $r > 1$, $s > 0$, indicating a Chaplygin gas nature initially.\\\\

\item The $Om$ diagnostic, we find that M1 and M4 exhibit a phantom region, suggesting the presence of phantom-like behavior in these models. In contrast, M2 and M3 consistently yield positive values throughout both high and low redshifts, supporting a quintessence model characterized by normal matter. These results provide important information about the nature of dark energy and the overall cosmic evolution in the late-time universe.\\\\

\end{itemize}

Therefore, we have noticed that M2 is more viable than M1 as compared to the $\Lambda$CDM model, and M3 is more feasible than M2 and M4. Finally, the above four considered models are departed from the standard $\Lambda$ CDM limit, supporting other dark energy models, and may be crucial in describing the accelerating universe.\\

Finally, we can say that there exist numerous works aiming to constrain various
dark energy models compared to the $\Lambda$CDM, our study offers several distinctive advantages:
(i) We have utilized new observational data in our data analyses, which helps
in providing more up-to-date and accurate results. (ii) Our investigation involved four
parameterizations of the deceleration parameter, which allow us to gain a deeper understanding
of the late-time cosmic evolution from the deceleration to the acceleration phase.
By exploring a broader range of parameter space, we enhanced the comprehensiveness of the
study. (iii) Our work contributed to the broader theoretical understanding by examining
the positive motivations behind such studies from a theoretical standpoint.
\\


\section*{Data Availability}             
No new data were generated in support of this research. 

\section*{Conflict of Interest} 
The authors declare no conflict of interest.
    
\section*{Acknowledgement:} AS is thankful to CSIR, Govt. of India, for providing a Senior Research Fellowship (No. 08/003(0138)/2019-EMR-I). G. Mustafa is very thankful to Prof. Gao Xianlong from the Department of
Physics, Zhejiang Normal University, for his kind support and help during
this research. Further, G. Mustafa acknowledges Grant No. ZC304022919
to support his Postdoctoral Fellowship at Zhejiang Normal University.\\

\bibliographystyle{elsarticle-num}
\bibliography{mybib}

\begin{thebibliography}{100}
\expandafter\ifx\csname url\endcsname\relax
  \def\url#1{\texttt{#1}}\fi
\expandafter\ifx\csname urlprefix\endcsname\relax\def\urlprefix{URL }\fi
\expandafter\ifx\csname href\endcsname\relax
  \def\href#1#2{#2} \def\path#1{#1}\fi

\bibitem{1:1998fmf}
A.~G. Riess, et~al., {Observational evidence from supernovae for an
  accelerating universe and a cosmological constant}, Astron. J. 116 (1998)
  1009--1038.
\newblock \href {http://arxiv.org/abs/astro-ph/9805201}
  {\path{arXiv:astro-ph/9805201}}, \href {https://doi.org/10.1086/300499}
  {\path{doi:10.1086/300499}}.

\bibitem{2:1996grv}
S.~Perlmutter, et~al., {Measurements of the cosmological parameters Omega and
  Lambda from the first 7 supernovae at z\ensuremath{>}=0.35}, Astrophys. J.
  483 (1997) 565.
\newblock \href {http://arxiv.org/abs/astro-ph/9608192}
  {\path{arXiv:astro-ph/9608192}}, \href {https://doi.org/10.1086/304265}
  {\path{doi:10.1086/304265}}.

\bibitem{3:1998vns}
S.~Perlmutter, et~al., {Measurements of $\Omega$ and $\Lambda$ from 42 high
  redshift supernovae}, Astrophys. J. 517 (1999) 565--586.
\newblock \href {http://arxiv.org/abs/astro-ph/9812133}
  {\path{arXiv:astro-ph/9812133}}, \href {https://doi.org/10.1086/307221}
  {\path{doi:10.1086/307221}}.

\bibitem{4:2003aw}
D.~N. Vollick, {1/R Curvature corrections as the source of the cosmological
  acceleration}, Phys. Rev. D 68 (2003) 063510.
\newblock \href {http://arxiv.org/abs/astro-ph/0306630}
  {\path{arXiv:astro-ph/0306630}}, \href
  {https://doi.org/10.1103/PhysRevD.68.063510}
  {\path{doi:10.1103/PhysRevD.68.063510}}.

\bibitem{5:2003ft}
S.~Nojiri, S.~D. Odintsov, {Modified gravity with negative and positive powers
  of the curvature: Unification of the inflation and of the cosmic
  acceleration}, Phys. Rev. D 68 (2003) 123512.
\newblock \href {http://arxiv.org/abs/hep-th/0307288}
  {\path{arXiv:hep-th/0307288}}, \href
  {https://doi.org/10.1103/PhysRevD.68.123512}
  {\path{doi:10.1103/PhysRevD.68.123512}}.

\bibitem{6:2003cyd}
J.~L. Tonry, et~al., {Cosmological results from high-z supernovae}, Astrophys.
  J. 594 (2003) 1--24.
\newblock \href {http://arxiv.org/abs/astro-ph/0305008}
  {\path{arXiv:astro-ph/0305008}}, \href {https://doi.org/10.1086/376865}
  {\path{doi:10.1086/376865}}.

\bibitem{7:2004lze}
A.~G. Riess, et~al., {Type Ia supernova discoveries at z \ensuremath{>} 1 from
  the Hubble Space Telescope: Evidence for past deceleration and constraints on
  dark energy evolution}, Astrophys. J. 607 (2004) 665--687.
\newblock \href {http://arxiv.org/abs/astro-ph/0402512}
  {\path{arXiv:astro-ph/0402512}}, \href {https://doi.org/10.1086/383612}
  {\path{doi:10.1086/383612}}.

\bibitem{8:2005xhg}
A.~Clocchiatti, et~al., {Hubble Space Telescope and Ground-Based Observations
  of Type Ia Supernovae at Redshift 0.5: Cosmological Implications}, Astrophys.
  J. 642 (2006) 1--21.
\newblock \href {http://arxiv.org/abs/astro-ph/0510155}
  {\path{arXiv:astro-ph/0510155}}, \href {https://doi.org/10.1086/498491}
  {\path{doi:10.1086/498491}}.

\bibitem{9:2005xb}
R.~R. Caldwell, W.~Komp, L.~Parker, D.~A.~T. Vanzella, {A Sudden gravitational
  transition}, Phys. Rev. D 73 (2006) 023513.
\newblock \href {http://arxiv.org/abs/astro-ph/0507622}
  {\path{arXiv:astro-ph/0507622}}, \href
  {https://doi.org/10.1103/PhysRevD.73.023513}
  {\path{doi:10.1103/PhysRevD.73.023513}}.

\bibitem{10:1999gb}
V.~Sahni, A.~A. Starobinsky, {The Case for a positive cosmological Lambda
  term}, Int. J. Mod. Phys. D 9 (2000) 373--444.
\newblock \href {http://arxiv.org/abs/astro-ph/9904398}
  {\path{arXiv:astro-ph/9904398}}, \href
  {https://doi.org/10.1142/S0218271800000542}
  {\path{doi:10.1142/S0218271800000542}}.

\bibitem{11:2002ji}
T.~Padmanabhan, {Cosmological constant: The Weight of the vacuum}, Phys. Rept.
  380 (2003) 235--320.
\newblock \href {http://arxiv.org/abs/hep-th/0212290}
  {\path{arXiv:hep-th/0212290}}, \href
  {https://doi.org/10.1016/S0370-1573(03)00120-0}
  {\path{doi:10.1016/S0370-1573(03)00120-0}}.

\bibitem{12:2002gy}
P.~J.~E. Peebles, B.~Ratra, {The Cosmological Constant and Dark Energy}, Rev.
  Mod. Phys. 75 (2003) 559--606.
\newblock \href {http://arxiv.org/abs/astro-ph/0207347}
  {\path{arXiv:astro-ph/0207347}}, \href
  {https://doi.org/10.1103/RevModPhys.75.559}
  {\path{doi:10.1103/RevModPhys.75.559}}.

\bibitem{13}
E.~J. Copeland, M.~Sami, S.~Tsujikawa, Dynamics of dark energy, International
  Journal of Modern Physics D 15~(11) (2006) 1753--1935.

\bibitem{14}
L.~Amendola, S.~Tsujikawa, Dark energy: theory and observations, Cambridge
  University Press, 2010.

\bibitem{15}
P.~J. Steinhardt, L.~Wang, I.~Zlatev, Cosmological tracking solutions, Physical
  Review D 59~(12) (1999) 123504.

\bibitem{16:1988cp}
S.~Weinberg, {The Cosmological Constant Problem}, Rev. Mod. Phys. 61 (1989)
  1--23.
\newblock \href {https://doi.org/10.1103/RevModPhys.61.1}
  {\path{doi:10.1103/RevModPhys.61.1}}.

\bibitem{17:2005pa}
S.~Capozziello, V.~F. Cardone, E.~Elizalde, S.~Nojiri, S.~D. Odintsov,
  {Observational constraints on dark energy with generalized equations of
  state}, Phys. Rev. D 73 (2006) 043512.
\newblock \href {http://arxiv.org/abs/astro-ph/0508350}
  {\path{arXiv:astro-ph/0508350}}, \href
  {https://doi.org/10.1103/PhysRevD.73.043512}
  {\path{doi:10.1103/PhysRevD.73.043512}}.

\bibitem{18-Rivera:2019aol}
C.~Escamilla-Rivera, S.~Capozziello, {Unveiling cosmography from the dark
  energy equation of state}, Int. J. Mod. Phys. D 28~(12) (2019) 1950154.
\newblock \href {http://arxiv.org/abs/1905.04602} {\path{arXiv:1905.04602}},
  \href {https://doi.org/10.1142/S0218271819501542}
  {\path{doi:10.1142/S0218271819501542}}.

\bibitem{19:2020rho}
U.~Debnath, {Gravitational waves for variable modified Chaplygin gas and some
  parametrizations of dark energy in the background of FRW universe}, Eur.
  Phys. J. Plus 135~(2) (2020) 135.
\newblock \href {https://doi.org/10.1140/epjp/s13360-020-00219-9}
  {\path{doi:10.1140/epjp/s13360-020-00219-9}}.

\bibitem{chaudhary2023constraints}
H.~Chaudhary, U.~Debnath, T.~Roy, S.~Maity, G.~Mustafa, Constraints on the
  parameters of modified chaplygin-jacobi and modified chaplygin-abel gases in
  $ f (t) $ gravity model, arXiv preprint arXiv:2307.14691 (2023).

\bibitem{20:2012ya}
S.~del Campo, I.~Duran, R.~Herrera, D.~Pavon, {Three thermodynamically-based
  parameterizations of the deceleration parameter}, Phys. Rev. D 86 (2012)
  083509.
\newblock \href {http://arxiv.org/abs/1209.3415} {\path{arXiv:1209.3415}},
  \href {https://doi.org/10.1103/PhysRevD.86.083509}
  {\path{doi:10.1103/PhysRevD.86.083509}}.

\bibitem{21}
J.~Cunha, J.~A. S.~d. Lima, Transition redshift: new kinematic constraints from
  supernovae, Monthly Notices of the Royal Astronomical Society 390~(1) (2008)
  210--217.

\bibitem{22:2008mt}
J.~V. Cunha, {Kinematic Constraints to the Transition Redshift from SNe Ia
  Union Data}, Phys. Rev. D 79 (2009) 047301.
\newblock \href {http://arxiv.org/abs/0811.2379} {\path{arXiv:0811.2379}},
  \href {https://doi.org/10.1103/PhysRevD.79.047301}
  {\path{doi:10.1103/PhysRevD.79.047301}}.

\bibitem{23:2004lze}
A.~G. Riess, et~al., {Type Ia supernova discoveries at z \ensuremath{>} 1 from
  the Hubble Space Telescope: Evidence for past deceleration and constraints on
  dark energy evolution}, Astrophys. J. 607 (2004) 665--687.
\newblock \href {http://arxiv.org/abs/astro-ph/0402512}
  {\path{arXiv:astro-ph/0402512}}, \href {https://doi.org/10.1086/383612}
  {\path{doi:10.1086/383612}}.

\bibitem{24:2007gvk}
L.-I. Xu, C.-W. Zhang, B.-R. Chang, H.-Y. Liu, {Constraints to deceleration
  parameters by recent cosmic observations}, Mod. Phys. Lett. A 23 (2008)
  1939--1948.
\newblock \href {http://arxiv.org/abs/astro-ph/0701519}
  {\path{arXiv:astro-ph/0701519}}, \href
  {https://doi.org/10.1142/S0217732308025991}
  {\path{doi:10.1142/S0217732308025991}}.

\bibitem{25:2009zza}
L.~Xu, J.~Lu, {Cosmic constraints on deceleration parameter with Sne Ia and
  CMB}, Mod. Phys. Lett. A 24 (2009) 369--376.
\newblock \href {https://doi.org/10.1142/S0217732309027212}
  {\path{doi:10.1142/S0217732309027212}}.

\bibitem{26}
R.~Nair, S.~Jhingan, D.~Jain, Cosmokinetics: a joint analysis of standard
  candles, rulers and cosmic clocks, Journal of Cosmology and Astroparticle
  Physics 2012~(01) (2012) 018.

\bibitem{27:2013lya}
O.~Akarsu, T.~Dereli, S.~Kumar, L.~Xu, {Probing kinematics and fate of the
  Universe with linearly time-varying deceleration parameter}, Eur. Phys. J.
  Plus 129 (2014) 22.
\newblock \href {http://arxiv.org/abs/1305.5190} {\path{arXiv:1305.5190}},
  \href {https://doi.org/10.1140/epjp/i2014-14022-6}
  {\path{doi:10.1140/epjp/i2014-14022-6}}.

\bibitem{28}
B.~Santos, J.~C. Carvalho, J.~S. Alcaniz, Current constraints on the epoch of
  cosmic acceleration, Astroparticle Physics 35~(1) (2011) 17--20.

\bibitem{29:2006gs}
Y.-G. Gong, A.~Wang, {Reconstruction of the deceleration parameter and the
  equation of state of dark energy}, Phys. Rev. D 75 (2007) 043520.
\newblock \href {http://arxiv.org/abs/astro-ph/0612196}
  {\path{arXiv:astro-ph/0612196}}, \href
  {https://doi.org/10.1103/PhysRevD.75.043520}
  {\path{doi:10.1103/PhysRevD.75.043520}}.

\bibitem{30:2001mx}
M.~S. Turner, A.~G. Riess, {Do SNe Ia provide direct evidence for past
  deceleration of the universe?}, Astrophys. J. 569 (2002) 18.
\newblock \href {http://arxiv.org/abs/astro-ph/0106051}
  {\path{arXiv:astro-ph/0106051}}, \href {https://doi.org/10.1086/338580}
  {\path{doi:10.1086/338580}}.

\bibitem{31:2015ali}
A.~Al~Mamon, S.~Das, {A divergence free parametrization of deceleration
  parameter for scalar field dark energy}, Int. J. Mod. Phys. D 25~(03) (2016)
  1650032.
\newblock \href {http://arxiv.org/abs/1507.00531} {\path{arXiv:1507.00531}},
  \href {https://doi.org/10.1142/S0218271816500322}
  {\path{doi:10.1142/S0218271816500322}}.

\bibitem{39}
G.~N. Gadbail, S.~Mandal, P.~K. Sahoo, Parametrization of deceleration
  parameter in f (q) gravity, Physics 4~(4) (2022) 1403--1412.

\bibitem{bouali}
A.~Bouali, B.~Shukla, H.~Chaudhary, R.~K. Tiwari, M.~Samar, G.~Mustafa,
  Cosmological tests of parametrization q= $\alpha$- $\beta$ h in f (q) flrw
  cosmology, International Journal of Geometric Methods in Modern Physics
  (2023).

\bibitem{bouali2023model}
A.~Bouali, H.~Chaudhary, A.~Mehrotra, S.~Pacif, Model-independent study for a
  quintessence model of dark energy: Analysis and observational constraints,
  arXiv preprint arXiv:2304.02652 (2023).

\bibitem{bouali2023data}
A.~Bouali, H.~Chaudhary, U.~Debnath, A.~Sardar, G.~Mustafa, Data analysis of
  three parameter models of deceleration parameter in frw universe, arXiv
  preprint arXiv:2304.13137 (2023).

\bibitem{ref1}
S.~Shekh, H.~Chaudhary, A.~Bouali, A.~Dixit, Observational constraints on
  teleparallel effective equation of state, General Relativity and Gravitation
  55~(8) (2023) 95.

\bibitem{ref2}
H.~Chaudhary, D.~Arora, U.~Debnath, G.~Mustafa, S.~K. Maurya, A new
  cosmological model: Exploring the evolution of the universe and unveiling
  super-accelerated expansion, arXiv preprint arXiv:2308.07354 (2023).

\bibitem{ref3}
A.~Bouali, B.~Shukla, H.~Chaudhary, R.~K. Tiwari, M.~S. Martin, Cosmographic
  studies of q= $\alpha$- $\beta$ h parametrization in f (t) framework,
  International Journal of Geometric Methods in Modern Physics (2023).

\bibitem{ref4}
H.~Chaudhary, A.~Bouali, U.~Debnath, T.~Roy, G.~Mustafa, Constraints on the
  parameterized deceleration parameter in frw universe, Physica Scripta (2023).

\bibitem{ref5}
A.~Bouali, H.~Chaudhary, S.~Mumtaz, G.~Mustafa, S.~Maurya, Observational
  constraining study of new deceleration parameters in frw universe,
  Fortschritte der Physik (2023) 2300033.

\bibitem{ref6}
H.~Chaudhary, A.~Kaushik, A.~Kohli, Cosmological test of $\sigma$$\theta$ as
  function of scale factor in f (r, t) framework, New Astronomy 103 (2023)
  102044.

\bibitem{ref7}
D.~Arora, H.~Chaudhary, S.~K.~J. PACIF, Diagnostic and comparative analysis of
  dark energy models with $ q (z) $ parametrizations, Available at SSRN
  4543124.

\bibitem{38}
A.~A. Mamon, S.~Das, A parametric reconstruction of the deceleration parameter,
  The European Physical Journal C 77~(7) (2017) 495.

\bibitem{41:2023rdx}
A.~Bouali, H.~Chaudhary, U.~Debnath, T.~Roy, G.~Mustafa, {Constraints on the
  Parameterized Deceleration Parameter in FRW Universe} (1 2023).
\newblock \href {http://arxiv.org/abs/2301.12107} {\path{arXiv:2301.12107}}.

\bibitem{42:2022jbw}
S.~Capozziello, R.~D'Agostino, O.~Luongo, {Thermodynamic parametrization of
  dark energy}, Phys. Dark Univ. 36 (2022) 101045.
\newblock \href {http://arxiv.org/abs/2202.03300} {\path{arXiv:2202.03300}},
  \href {https://doi.org/10.1016/j.dark.2022.101045}
  {\path{doi:10.1016/j.dark.2022.101045}}.

\bibitem{54:2003fg}
U.~Alam, V.~Sahni, T.~D. Saini, A.~A. Starobinsky, {Is there supernova evidence
  for dark energy metamorphosis ?}, Mon. Not. Roy. Astron. Soc. 354 (2004) 275.
\newblock \href {http://arxiv.org/abs/astro-ph/0311364}
  {\path{arXiv:astro-ph/0311364}}, \href
  {https://doi.org/10.1111/j.1365-2966.2004.08189.x}
  {\path{doi:10.1111/j.1365-2966.2004.08189.x}}.

\bibitem{55}
U.~Alam, V.~Sahni, A.~A. Starobinsky, The case for dynamical dark energy
  revisited, Journal of Cosmology and Astroparticle Physics 2004~(06) (2004)
  008.

\bibitem{43}
T.~Bandyopadhyay, U.~Debnath, Fluid accretion upon higher-dimensional wormhole
  and black hole for parameterized deceleration parameter, International
  Journal of Geometric Methods in Modern Physics 19~(12) (2022) 2250182.

\bibitem{44}
R.~Kundu, U.~Debnath, A.~Pradhan, Studying the optical depth behaviour of
  parametrized deceleration parameter in non-flat universe, International
  Journal of Geometric Methods in Modern Physics (2023).

\bibitem{57}
H.~Pad{\'e}, Sur la repr{\'e}sentation approch{\'e}e d'une fonction par des
  fractions rationnelles, in: Annales scientifiques de l'Ecole normale
  sup{\'e}rieure, Vol.~9, 1892, pp. 3--93.

\bibitem{58}
H.~Wei, X.-P. Yan, Y.-N. Zhou, Cosmological applications of pade approximant,
  Journal of Cosmology and Astroparticle Physics 2014~(01) (2014) 045.

\bibitem{59}
M.~Rezaei, M.~Malekjani, S.~Basilakos, A.~Mehrabi, D.~F. Mota, Constraints to
  dark energy using pade parameterizations, The Astrophysical Journal 843~(1)
  (2017) 65.

\bibitem{28emcee}
D.~Foreman-Mackey, D.~W. Hogg, D.~Lang, J.~Goodman, emcee: the mcmc hammer,
  Publications of the Astronomical Society of the Pacific 125~(925) (2013) 306.

\bibitem{29polychord}
W.~Handley, M.~Hobson, A.~Lasenby, Polychord: nested sampling for cosmology,
  Monthly Notices of the Royal Astronomical Society: Letters 450~(1) (2015)
  L61--L65.

\bibitem{30getdist}
A.~Lewis, Getdist: a python package for analysing monte carlo samples, arXiv
  preprint arXiv:1910.13970 (2019).

\bibitem{H(z)}
E.~Gaztanaga, C.~Bonvin, L.~Hui, Measurement of the dipole in the
  cross-correlation function of galaxies, Journal of Cosmology and
  Astroparticle Physics 2017~(01) (2017) 032.

\bibitem{bouali2023cosmological}
A.~Bouali, H.~Chaudhary, R.~Hama, T.~Harko, S.~V. Sabau, M.~S. Mart{\'\i}n,
  Cosmological tests of the osculating barthel--kropina dark energy model, The
  European Physical Journal C 83~(2) (2023) 121.

\bibitem{Pan1}
M.~Kowalski, D.~Rubin, G.~Aldering, R.~Agostinho, A.~Amadon, R.~Amanullah,
  C.~Balland, K.~Barbary, G.~Blanc, P.~Challis, et~al., Improved cosmological
  constraints from new, old, and combined supernova data sets, The
  Astrophysical Journal 686~(2) (2008) 749.

\bibitem{Pan2}
R.~Amanullah, C.~Lidman, D.~Rubin, G.~Aldering, P.~Astier, K.~Barbary,
  M.~Burns, A.~Conley, K.~Dawson, S.~Deustua, et~al., Spectra and hubble space
  telescope light curves of six type ia supernovae at 0.511< z< 1.12 and the
  union2 compilation, The Astrophysical Journal 716~(1) (2010) 712.

\bibitem{Pan3}
N.~Suzuki, D.~Rubin, C.~Lidman, G.~Aldering, R.~Amanullah, K.~Barbary,
  L.~Barrientos, J.~Botyanszki, M.~Brodwin, N.~Connolly, et~al., The hubble
  space telescope cluster supernova survey. v. improving the dark-energy
  constraints above z> 1 and building an early-type-hosted supernova sample,
  The Astrophysical Journal 746~(1) (2012) 85.

\bibitem{Pan4}
M.~Betoule, R.~Kessler, J.~Guy, J.~Mosher, D.~Hardin, R.~Biswas, P.~Astier,
  P.~El-Hage, M.~Konig, S.~Kuhlmann, et~al., Improved cosmological constraints
  from a joint analysis of the sdss-ii and snls supernova samples, Astronomy \&
  Astrophysics 568 (2014) A22.

\bibitem{Pan5}
D.~M. Scolnic, D.~Jones, A.~Rest, Y.~Pan, R.~Chornock, R.~Foley, M.~Huber,
  R.~Kessler, G.~Narayan, A.~Riess, et~al., The complete light-curve sample of
  spectroscopically confirmed sne ia from pan-starrs1 and cosmological
  constraints from the combined pantheon sample, The Astrophysical Journal
  859~(2) (2018) 101.

\bibitem{pantheon+}
D.~Scolnic, D.~Brout, A.~Carr, A.~G. Riess, T.~M. Davis, A.~Dwomoh, D.~O.
  Jones, N.~Ali, P.~Charvu, R.~Chen, et~al., The pantheon+ type ia supernova
  sample: the full dataset and light-curve release, arXiv preprint
  arXiv:2112.03863 (2021).

\bibitem{benisty2021testing}
D.~Benisty, D.~Staicova, Testing late-time cosmic acceleration with
  uncorrelated baryon acoustic oscillation dataset, Astronomy \& Astrophysics
  647 (2021) A38.

\bibitem{bao1}
W.~J. Percival, B.~A. Reid, D.~J. Eisenstein, N.~A. Bahcall, T.~Budavari, J.~A.
  Frieman, M.~Fukugita, J.~E. Gunn, {\v{Z}}.~Ivezi{\'c}, G.~R. Knapp, et~al.,
  Baryon acoustic oscillations in the sloan digital sky survey data release 7
  galaxy sample, Monthly Notices of the Royal Astronomical Society 401~(4)
  (2010) 2148--2168.

\bibitem{bao2}
F.~Beutler, C.~Blake, M.~Colless, D.~H. Jones, L.~Staveley-Smith, L.~Campbell,
  Q.~Parker, W.~Saunders, F.~Watson, The 6df galaxy survey: baryon acoustic
  oscillations and the local hubble constant, Monthly Notices of the Royal
  Astronomical Society 416~(4) (2011) 3017--3032.

\bibitem{bao3}
T.~Delubac, J.~Rich, S.~Bailey, A.~Font-Ribera, D.~Kirkby, J.-M. Le~Goff, M.~M.
  Pieri, A.~Slosar, {\'E}.~Aubourg, J.~E. Bautista, et~al., Baryon acoustic
  oscillations in the ly$\alpha$ forest of boss quasars, Astronomy \&
  Astrophysics 552 (2013) A96.

\bibitem{bao4}
L.~Anderson, E.~Aubourg, S.~Bailey, D.~Bizyaev, M.~Blanton, A.~S. Bolton,
  J.~Brinkmann, J.~R. Brownstein, A.~Burden, A.~J. Cuesta, et~al., The
  clustering of galaxies in the sdss-iii baryon oscillation spectroscopic
  survey: baryon acoustic oscillations in the data release 9 spectroscopic
  galaxy sample, Monthly Notices of the Royal Astronomical Society 427~(4)
  (2012) 3435--3467.

\bibitem{bao5}
H.-J. Seo, S.~Ho, M.~White, A.~J. Cuesta, A.~J. Ross, S.~Saito, B.~Reid,
  N.~Padmanabhan, W.~J. Percival, R.~De~Putter, et~al., Acoustic scale from the
  angular power spectra of sdss-iii dr8 photometric luminous galaxies, The
  Astrophysical Journal 761~(1) (2012) 13.

\bibitem{bao6}
A.~J. Ross, L.~Samushia, C.~Howlett, W.~J. Percival, A.~Burden, M.~Manera, The
  clustering of the sdss dr7 main galaxy sample--i. a 4 per cent distance
  measure at z= 0.15, Monthly Notices of the Royal Astronomical Society 449~(1)
  (2015) 835--847.

\bibitem{bao7}
R.~Tojeiro, A.~J. Ross, A.~Burden, L.~Samushia, M.~Manera, W.~J. Percival,
  F.~Beutler, J.~Brinkmann, J.~R. Brownstein, A.~J. Cuesta, et~al., The
  clustering of galaxies in the sdss-iii baryon oscillation spectroscopic
  survey: galaxy clustering measurements in the low-redshift sample of data
  release 11, Monthly Notices of the Royal Astronomical Society 440~(3) (2014)
  2222--2237.

\bibitem{bao8}
J.~E. Bautista, M.~Vargas-Maga{\~n}a, K.~S. Dawson, W.~J. Percival,
  J.~Brinkmann, J.~Brownstein, B.~Camacho, J.~Comparat, H.~Gil-Mar{\'\i}n,
  E.-M. Mueller, et~al., The sdss-iv extended baryon oscillation spectroscopic
  survey: baryon acoustic oscillations at redshift of 0.72 with the dr14
  luminous red galaxy sample, The Astrophysical Journal 863~(1) (2018) 110.

\bibitem{bao9}
E.~De~Carvalho, A.~Bernui, G.~Carvalho, C.~Novaes, H.~Xavier, Angular baryon
  acoustic oscillation measure at z= 2.225 from the sdss quasar survey, Journal
  of Cosmology and Astroparticle Physics 2018~(04) (2018) 064.

\bibitem{bao10}
M.~Ata, F.~Baumgarten, J.~Bautista, F.~Beutler, D.~Bizyaev, M.~R. Blanton,
  J.~A. Blazek, A.~S. Bolton, J.~Brinkmann, J.~R. Brownstein, et~al., The
  clustering of the sdss-iv extended baryon oscillation spectroscopic survey
  dr14 quasar sample: first measurement of baryon acoustic oscillations between
  redshift 0.8 and 2.2, Monthly Notices of the Royal Astronomical Society
  473~(4) (2018) 4773--4794.

\bibitem{bao11}
T.~Abbott, F.~Abdalla, A.~Alarcon, S.~Allam, F.~Andrade-Oliveira, J.~Annis,
  S.~Avila, M.~Banerji, N.~Banik, K.~Bechtol, et~al., Dark energy survey year 1
  results: Measurement of the baryon acoustic oscillation scale in the
  distribution of galaxies to redshift 1, Monthly Notices of the Royal
  Astronomical Society 483~(4) (2019) 4866--4883.

\bibitem{bao12}
Z.~Molavi, A.~Khodam-Mohammadi, Observational tests of gauss-bonnet like dark
  energy model, The European Physical Journal Plus 134~(6) (2019) 254.

\bibitem{bao13}
N.~B. Hogg, M.~Martinelli, S.~Nesseris, Constraints on the distance duality
  relation with standard sirens, Journal of Cosmology and Astroparticle Physics
  2020~(12) (2020) 019.

\bibitem{bao14}
M.~Martinelli, C.~J. A.~P. Martins, S.~Nesseris, D.~Sapone, I.~Tutusaus,
  A.~Avgoustidis, S.~Camera, C.~Carbone, S.~Casas, S.~Ili{\'c}, et~al., Euclid:
  Forecast constraints on the cosmic distance duality relation with
  complementary external probes, Astronomy \& Astrophysics 644 (2020) A80.

\bibitem{chen2019distance}
L.~Chen, Q.-G. Huang, K.~Wang, Distance priors from planck final release,
  Journal of Cosmology and Astroparticle Physics 2019~(02) (2019) 028.

\bibitem{statefinder1}
V.~Sahni, T.~D. Saini, A.~A. Starobinsky, U.~Alam, Statefinder—a new
  geometrical diagnostic of dark energy, Journal of Experimental and
  Theoretical Physics Letters 77 (2003) 201--206.

\bibitem{statefinder2}
U.~Alam, V.~Sahni, T.~Deep~Saini, A.~Starobinsky, Exploring the expanding
  universe and dark energy using the statefinder diagnostic, Monthly Notices of
  the Royal Astronomical Society 344~(4) (2003) 1057--1074.

\bibitem{statefinder3}
M.~Sami, M.~Shahalam, M.~Skugoreva, A.~Toporensky, Cosmological dynamics of a
  nonminimally coupled scalar field system and its late time cosmic relevance,
  Physical Review D 86~(10) (2012) 103532.

\bibitem{statefinder4}
R.~Myrzakulov, M.~Shahalam, Statefinder hierarchy of bimetric and galileon
  models for concordance cosmology, Journal of Cosmology and Astroparticle
  Physics 2013~(10) (2013) 047.

\bibitem{32:2002fz}
V.~Sahni, T.~D. Saini, A.~A. Starobinsky, U.~Alam, {Statefinder: A New
  geometrical diagnostic of dark energy}, JETP Lett. 77 (2003) 201--206.
\newblock \href {http://arxiv.org/abs/astro-ph/0201498}
  {\path{arXiv:astro-ph/0201498}}, \href {https://doi.org/10.1134/1.1574831}
  {\path{doi:10.1134/1.1574831}}.

\bibitem{33:2008xx}
V.~Sahni, A.~Shafieloo, A.~A. Starobinsky, {Two new diagnostics of dark
  energy}, Phys. Rev. D 78 (2008) 103502.
\newblock \href {http://arxiv.org/abs/0807.3548} {\path{arXiv:0807.3548}},
  \href {https://doi.org/10.1103/PhysRevD.78.103502}
  {\path{doi:10.1103/PhysRevD.78.103502}}.

\bibitem{om1}
V.~Sahni, A.~Shafieloo, A.~A. Starobinsky, Two new diagnostics of dark energy,
  Physical Review D 78~(10) (2008) 103502.

\bibitem{om2}
C.~Zunckel, C.~Clarkson, Consistency tests for the cosmological constant,
  Physical Review Letters 101~(18) (2008) 181301.

\bibitem{om3}
M.~Shahalam, S.~Pathak, M.~Verma, M.~Y. Khlopov, R.~Myrzakulov, Dynamics of
  interacting quintessence, The European Physical Journal C 75 (2015) 1--9.

\bibitem{om4}
A.~Agarwal, R.~Myrzakulov, S.~Pacif, M.~Shahalam, Cosmic acceleration from
  coupling of baryonic and dark matter components: Analysis and diagnostics,
  International Journal of Modern Physics D 28~(06) (2019) 1950083.

\bibitem{34:2020wgo}
A.~Jawad, S.~Qummer, S.~Rani, M.~Younas, {Generalized interaction term inspired
  dark energy model in fractal universe}, Mod. Phys. Lett. A 35~(15) (2020)
  2050126.
\newblock \href {https://doi.org/10.1142/S0217732320501266}
  {\path{doi:10.1142/S0217732320501266}}.

\bibitem{35-Mohammadi:2010you}
A.~Khodam-Mohammadi, M.~Malekjani, {Interacting entropy-corrected holographic
  scalar field models in non-flat universe}, Commun. Theor. Phys. 55 (2011)
  942--948.
\newblock \href {http://arxiv.org/abs/1004.1720} {\path{arXiv:1004.1720}},
  \href {https://doi.org/10.1088/0253-6102/55/5/37}
  {\path{doi:10.1088/0253-6102/55/5/37}}.

\bibitem{36:2012wc}
M.~Malekjani, A.~Khodam-Mohammadi, {Statefinder diagnosis and the interacting
  ghost model of dark energy}, Astrophys. Space Sci. 343 (2013) 451--461.
\newblock \href {http://arxiv.org/abs/1202.4154} {\path{arXiv:1202.4154}},
  \href {https://doi.org/10.1007/s10509-012-1230-3}
  {\path{doi:10.1007/s10509-012-1230-3}}.

\bibitem{37:2010nk}
M.~Malekjani, A.~Khodam-Mohammadi, N.~Nazari-pooya, {Cosmological evolution and
  statefinder diagnostic for new holographic dark energy model in non flat
  universe}, Astrophys. Space Sci. 332 (2011) 515--524.
\newblock \href {http://arxiv.org/abs/1011.4805} {\path{arXiv:1011.4805}},
  \href {https://doi.org/10.1007/s10509-010-0550-4}
  {\path{doi:10.1007/s10509-010-0550-4}}.

\bibitem{40}
M.~Shahalam, S.~Sami, A.~Agarwal, Om diagnostic applied to scalar field models
  and slowing down of cosmic acceleration, Monthly Notices of the Royal
  Astronomical Society 448~(3) (2015) 2948--2959.

\bibitem{41}
M.~Jamil, D.~Momeni, R.~Myrzakulov, Observational constraints on non-minimally
  coupled galileon model, The European Physical Journal C 73~(3) (2013) 2347.

\bibitem{42}
P.~De~Fromont, C.~De~Rham, L.~Heisenberg, A.~Matas, Superluminality in the
  bi-and multi-galileon, Journal of High Energy Physics 2013~(7) (2013) 1--29.

\bibitem{AIC1}
H.~Akaike, A new look at the statistical model identification, IEEE
  transactions on automatic control 19~(6) (1974) 716--723.

\bibitem{AIC2}
M.~Li, X.~Li, X.~Zhang, Comparison of dark energy models: A perspective from
  the latest observational data, Science China Physics, Mechanics and Astronomy
  53~(9) (2010) 1631--1645.

\bibitem{AIC3}
K.~Burnhan, D.~R. Anderson, Model selection and multimodel inference, New York:
  Springer (2002).

\bibitem{AIC4}
K.~P. Burnham, D.~R. Anderson, Multimodel inference: understanding aic and bic
  in model selection, Sociological methods \& research 33~(2) (2004) 261--304.

\bibitem{AIC5}
A.~R. Liddle, Information criteria for astrophysical model selection, Monthly
  Notices of the Royal Astronomical Society: Letters 377~(1) (2007) L74--L78.

\end{thebibliography}

\end{document}